\pdfoutput=1
\documentclass[cernpreprint, atlasdraft=false, UKenglish, texmf, orcidlogo]{atlasdoc}
 
\usepackage{atlaspackage}
\usepackage{atlasbiblatex}
 
\usepackage{atlasphysics}
 
\graphicspath{{logos/}{figures/}}

\usepackage{todonotes}
\usepackage{rotating}
\usepackage{adjustbox}
\usepackage{tabularx}
\usepackage{multirow}
\usepackage{array}
\usepackage{booktabs}
\usepackage{tabularx}
\usepackage{pdflscape}
\usepackage{longtable}
\usepackage{changepage}
\newcommand{\AtlasCoordFootnote}{
ATLAS uses a right-handed coordinate system with its origin at the nominal interaction point (IP)
in the centre of the detector and the \(z\)-axis along the beam pipe.
The \(x\)-axis points from the IP to the centre of the LHC ring,
and the \(y\)-axis points upwards.
Cylindrical coordinates \((r,\phi)\) are used in the transverse plane,
\(\phi\) being the azimuthal angle around the \(z\)-axis.
The pseudorapidity is defined in terms of the polar angle \(\theta\) as \(\eta = -\ln \tan(\theta/2)\).
Angular distance is measured in units of \(\Delta R \equiv \sqrt{(\Delta\eta)^{2} + (\Delta\phi)^{2}}\).}

\AtlasTitle{Inclusive and differential cross-sections for dilepton $t\bar{t}$ production measured in $\sqrt{s}=13\;$TeV \ensuremath{pp}\xspace collisions with the ATLAS detector}

\AtlasJournalRef{\JHEP 07 (2023) 141}
\AtlasDOI{10.1007/JHEP07(2023)141}
 
\AtlasAbstract{
Differential and double-differential distributions of kinematic variables of leptons from decays of
top-quark pairs ($t\bar{t}$) are measured using the full LHC Run 2 data sample collected with the ATLAS detector.
The data were collected at a \ensuremath{pp} collision energy of $\sqrt{s}=13$ TeV and correspond to an integrated
luminosity of 140 fb$^{-1}$.
The measurements use events containing an oppositely charged $e\mu$ pair and $b$-tagged jets.
The results are compared with predictions from several Monte Carlo generators. While no prediction is found to be consistent
with all distributions, a better agreement with measurements of the
lepton \ensuremath{p_{\text{T}}} distributions 
is obtained by reweighting the $t\bar{t}$ sample
so as to reproduce the top-quark \ensuremath{p_{\text{T}}} distribution from an NNLO calculation.
The inclusive top-quark pair production cross-section is measured as well, both in a fiducial region and
in the full phase-space. The total inclusive cross-section is found to be
\[ \sigma_{t\bar{t}} = 829 \pm 1\;(\textrm{stat}) \pm 13\;(\textrm{syst}) \pm 8\;(\textrm{lumi}) \pm 2\; (\textrm{beam})\ \textrm{pb}, \]
where the uncertainties are due to statistics, systematic effects, the integrated luminosity and the beam energy.
This is in excellent agreement with the theoretical expectation.}
 
\author{The ATLAS Collaboration}
 
\AtlasRefCode{TOPQ-2018-26}

\PreprintIdNumber{CERN-EP-2023-016}

\hypersetup{pdftitle={ATLAS document},pdfauthor={The ATLAS Collaboration}}

\begin{document}

\maketitle

\section{Introduction}
\label{sec:intro}
The top quark is the heaviest known elementary particle and studying its properties is a major
goal of the ATLAS experiment~\cite{PERF-2007-01, PIX-2018-001, ATLAS-TDR-19} at the Large Hadron Collider (LHC). Precise measurements of top-quark pair
production in high-energy proton--proton ($pp$) collisions provide sensitive probes of quantum chromodynamics (QCD), particularly parton distribution functions (PDFs).
For the measurement of the top-quark pair production cross-section, the decay channel
$\ttbar \ra \Wplus\Wminus \bbbar$ with subsequent leptonic decays of the \Wboson bosons is of particular interest since,
compared to the hadronic channels, it is minimally affected by QCD modelling uncertainties.
Previous  ATLAS measurements of lepton cross-sections based on events
containing an $e\mu$ pair with opposite electric charges and one or two $b$-tagged jets (jets likely to originate from a $b$-quark) include
measurements using \pp collision events at centre-of-mass energies $\sqrt{s}$~=~7--8~\TeV \ \cite{TOPQ-2013-04, TOPQ-2015-02} and $\sqrt{s}$ = 13 \TeV \ \cite{TOPQ-2018-17}.
The latter measurement was based on data collected during 2015--16, corresponding to an integrated luminosity
of 36~fb$^{-1}$. The same analysis technique is applied here to the entire 13~\TeV\ data sample
from LHC \RunTwo, corresponding to an integrated luminosity of 140~fb$^{-1}$. Similar measurements have also been
performed by the CMS Collaboration at $\sqrt{s}=13\,\TeV$~\cite{CMS-TOP-17-014,CMS-TOP-18-004,CMS-TOP-19-008}.
 
The large integrated luminosity of the \RunTwo data sample allows the lepton differential distributions to be measured
over a wider range and with finer granularity than in Ref.~\cite{TOPQ-2018-17}.
The differential distributions of eight kinematic variables of the two leptons are studied:
\begin{itemize}
\item  $\pT^{\ell}$, the single-lepton transverse momentum\footnote{\AtlasCoordFootnote} ($\ell=e$ or $\mu$);
\item $|\eta^{\ell}|$, the single-lepton pseudorapidity;
\item $m^{e\mu}$, the $e\mu$ system invariant mass;
\item $\pT^{e\mu}$, the $e\mu$ system transverse momentum;
\item $|y^{e\mu}|$, the $e\mu$ system rapidity;
\item $E^e + E^{\mu}$, the sum of lepton energies;
\item $\pT^{e}+\pT^{\mu}$, the scalar sum of lepton transverse momenta;
\item $|\Delta \phi^{e\mu}|$, the azimuthal angular separation of the leptons.
\end{itemize}
 
Both the absolute differential cross-sections and the normalised distributions of these variables, defined at particle level,
are presented in a fiducial region given by $\pT^{\ell}>27\ (25)$~\GeV \ for the leading (sub-leading) lepton and $|\eta^{\ell}|<2.5$ after applying the overlap removal procedure described in Section~\ref{sec:selection}.
Four double-differential distributions are measured as well: $|y^{e\mu}|$ in bins of $m^{e\mu}$,  and $|\Delta \phi^{e\mu}|$ in bins of
$m^{e\mu}$, $\pT^{e\mu}$ and $E^e + E^{\mu}$. The differential and double-differential distributions are compared with predictions from various models of top-quark production in \pp collisions and can later be used to constrain model parameters, such as the strong coupling constant \alphas,
the top-quark mass $m_{t}$ or the PDFs~\cite{TOPQ-2018-17,CMS-TOP-18-004}.
 
The inclusive cross-section for the production of top-quark pairs decaying into an oppositely charged $e\mu$ pair in the fiducial region is also measured, as well as the total inclusive \ttbar cross-section. These measurements make use of recent updates to the luminosity determination and a significant reduction in the luminosity uncertainty at \RunTwo~\cite{ATLAS:2022hro}.

\section{ATLAS detector}
\label{sec:atlas_det}
The ATLAS experiment at the LHC is a multipurpose particle detector
with a forward--backward symmetric cylindrical geometry and a near \(4\pi\) coverage in
solid angle.
It consists of an inner tracking detector surrounded by a thin superconducting solenoid
providing a \SI{2}{\tesla} axial magnetic field, electromagnetic and hadron calorimeters, and a muon spectrometer.
 
The inner tracking detector covers the pseudorapidity range \(|\eta| < 2.5\).
It consists of silicon pixel, silicon microstrip, and transition radiation tracking detectors.
 
Lead/liquid-argon (LAr) sampling calorimeters provide electromagnetic (EM) energy measurements
with high granularity.
A steel/scintillator-tile hadron calorimeter covers the central pseudorapidity range (\(|\eta| < 1.7\)).
The endcap and forward regions are instrumented with LAr calorimeters
for both the EM and hadronic energy measurements up to \(|\eta| = 4.9\).
 
The muon spectrometer surrounds the calorimeters and is based on
three large superconducting air-core toroidal magnets with eight coils each.
The field integral of the toroids ranges between \num{2.0} and \SI{6.0}{\tesla\metre}
across most of the detector.
The muon spectrometer includes a system of precision tracking chambers, and fast detectors for triggering.
 
A two-level trigger system is used to select events.
The first-level trigger is implemented in hardware and uses a subset of the detector information
to accept events at a rate below \SI{100}{\kHz}.
This is followed by a software-based trigger that
reduces the accepted event rate to \SI{1}{\kHz} on average
depending on the data-taking conditions.
An extensive software suite~\cite{ATL-SOFT-PUB-2021-001} is used in data simulation, in the reconstruction and analysis of real
and simulated data, in detector operations, and in the trigger and data acquisition systems of the experiment.

\section{Data and simulated event samples}
\label{sec:samples}

For this analysis, proton--proton collision events collected during \RunTwo of the LHC (2015--2018) with the ATLAS detector are required to
pass the single-electron or single-muon triggers~\cite{TRIG-2016-01,TRIG-2018-01,TRIG-2018-05}, which are highly efficient for leptons with $\pT^{\ell}>27$~\GeV .\
After all quality criteria~\cite{DAPR-2018-01} have been applied, the data recorded in Run 2 correspond to an integrated luminosity of $140$ \si{\per\fb}
with an uncertainty of $0.83\%$~\cite{ATLAS:2022hro}.
 
To aid the analysis, simulated Monte Carlo (MC) samples were produced using either the full ATLAS detector simulation~\cite{SOFT-2010-01} based
on the \GEANT framework~\cite{Agostinelli:2002hh} or, for the estimation of some of the systematic uncertainties, a faster simulation
with parameterised showers in the calorimeters~\cite{ATL-PHYS-PUB-2010-013}.
 
The effect of multiple interactions in the same and neighbouring bunch
crossings (\pileup) was modelled by overlaying each
hard-scattering event with inelastic \pp collisions generated with \PYTHIA[8.186]~\cite{Sjostrand:2007gs} using the NNPDF2.3
set of PDFs~\cite{Ball:2012cx} and the A3 set of tuned parameters~\cite{ATL-PHYS-PUB-2016-017}. The \EVTGEN1.6.0 program~\cite{Lange:2001uf} is used for properties of the bottom and charm hadron decays.
 
\subsection{\ttbar signal samples}
\label{sec:signal-samples}
 
The nominal sample used to model \ttbar events was produced using the next-to-leading-order (NLO) matrix element generator \POWHEGBOX \cite{POWHEG_1, POWHEG_2, POWHEG_3, POWHEG_4} with the NNPDF3.0 PDF set~\cite{Ball:2014uwa}, interfaced to \PYTHIA[8.230] \cite{Sjostrand:2014zea, PYTHIA_2} with the A14 tune~\cite{ATL-PHYS-PUB-2014-021} and the NNPDF2.3 PDF sets~\cite{Ball:2012cx} for the underlying event, parton shower and fragmentation.
The \hdamp parameter was set to $1.5\cdot m_{t}$~\cite{ATL-PHYS-PUB-2016-020}, with $m_{t}$ set to $172.5$~\GeV, and both the renormalisation scale $\muR^\textrm{default}$
and the factorization scale $\muF^\textrm{default}$ were set equal to the top-quark transverse mass.\footnote{$\muR = \muF = \sqrt{(m_t^2+(p_{\mathrm{T},t}^2+p_{\mathrm{T},\bar{t}}^2)/2}$ where $p_{\mathrm{T},t/\bar{t}}$ is the transverse momentum of the top (anti-top) quark.}
 
Several modifications of \POWPY[8.230] are used to assess systematic uncertainties arising from assumptions in the simulation. Variations
in the level of initial-state radiation (ISR) are performed by using the internal \enquote{Var3cUp} (\enquote{Var3cDown}) weight~\cite{ATL-PHYS-PUB-2014-021}
together with the renormalisation ($\muR$) and factorisation ($\muF$) scales set to half (twice) the default values. In two additional samples, the same configurations adopted for the ISR variation sample are used together with a change in the \hdamp parameter (doubled to be $3.0\cdot m_{t}$)~\cite{ATL-PHYS-PUB-2018-009}. These new samples are labelled \enquote{Rad up} and \enquote{Rad down} and used only in the generator--data comparison. Final-state radiation (FSR) is varied by
changing the $\alphas^\textrm{FSR}$ parameter, controlling the FSR emissions in \PYTHIA[8.230]. The PDF uncertainties are estimated with the 30 components of the Hessian PDF4LHC15 error set~\cite{Butterworth:2015oua, META-PDFs, MCH-PDFs}.

The uncertainty associated with the matrix element generation is estimated using \MGNLO \cite{Alwall:2014hca} interfaced with \PYTHIA[8.230] as an alternative generator, with the A14 tune and the NNPDF2.3 set of PDFs for the underlying event, parton shower and fragmentation. Since the \enquote{matrix element correction} (MEC) in \PYTHIA[8.230] is switched off in this simulation~\cite{ATL-PHYS-PUB-2020-023},
a sample of \POWPY[8.230] events with MEC switched off, with the same PDF sets as the nominal \POWPY[8.230] generator, was also produced for comparison with \MGNLO.
In order to estimate the uncertainty associated with the modelling of fragmentation and parton showering,
a sample was generated with \POWHEG interfaced with \HERWIG[7.0.4] \cite{Bahr:2008pv,Bellm:2015jjp} with the H7UE tune~\cite{Gieseke:2012ft} and the NNPDF3.0 PDF set.
 
Additional samples using alternative generators were produced for comparison with data. These include \POWHEG interfaced with \HERWIG[7.1.3]~\cite{Bellm:2017jjp},
\MGNLO interfaced with \HERWIG[7.1.3], and \POWPY[8.230] with the PDF4LHC15\_nnlo\_mc set~\cite{Butterworth:2015oua, CMC-PDFs}. Finally, a reweighted \POWPY[8.230] sample was generated. The reweighting is performed on the top-quark \pT variable, using the
kinematics of the top quarks in the MC sample after initial- and final-state radiation.
The prediction for the top-quark \pT spectrum is calculated to next-to-next-to-leading order (NNLO) in QCD with NLO EW corrections~\cite{Czakon:2017wor, , Catani:2019hip} with the NNPDF3.0 QED PDF set using dynamic renormalisation and factorisation scales $m_{\mathrm{T},t}/2$, i.e.\  half the top-quark transverse mass,\footnote{The transverse mass of the top quark is denoted by $m_{\mathrm{T},t} = \sqrt{m_t^2 + p_{\mathrm{T},t}^2}$.} for the top-quark \pT as proposed in Ref.~\cite{Czakon:2017wor}, with $m_t = 173.3$ \GeV. The reweighting was applied such that at the end of the procedure the reweighted MC sample is in good agreement with the higher-order prediction for the reweighted variable~\cite{Leonid2105.03977}. This sample is referred to as being reweighted to the NNLO prediction in the remainder of the document.

The measurements are sensitive to the fraction of \ttbar events produced together with extra heavy-flavour quarks, which is not well modelled. 
This extra production of heavy flavour,
relative to the prediction, is studied with a modified \POWPY[8.230] 
sample in which the fraction of events with at least three $b$-jets at generator level is increased by 30\% to reproduce the rate of events in data with three $b$-tagged jets, as discussed in Ref.~\cite{TOPQ-2018-17}.

When comparing simulation with data, the \ttbar samples are normalised to the inclusive cross-section prediction calculated at NNLO accuracy in the strong coupling constant \alphas, including the resummation of next-to-next-to-leading logarithmic (NNLL) soft gluon terms, $\sigma_{\ttbar, \textrm{pred}} = 832^{+20}_{-29}$(scale)$^{+35}_{-35}$(PDF+\alphas)~pb, obtained using the \TOPpp[2.0] program~\cite{Czakon:2011xx,Baernreuther:2012ws,Czakon:2012zr,Czakon:2012pz,Czakon:2013goa}.

\subsection{$Wt$ samples}
\label{sec:Wt-samples}
 
In order to describe the dominant background from the single-top $\Wboson t$ channel, samples were produced with \POWPY[8.230] with
the same parameter values as used for the nominal \ttbar sample.
The interference between the \ttbar and $\Wboson t$ amplitudes is modelled using the diagram removal scheme~\cite{White:2009yt,Re:2010bp}.
To estimate the systematic uncertainties from this source, an alternative
sample is used, where the interference is modelled with the diagram subtraction scheme~\cite{White:2009yt}.

The same variations of \POWHEGBOX that were performed for the nominal \ttbar sample were also
carried out for the $\Wboson t$  sample. The same alternative generators are also used to estimate the
hard-scattering matrix element and parton shower plus hadronisation uncertainties in the $\Wboson t$ background.
 
\subsection{Other background samples}
\label{sec:background-samples}
 
The background from diboson events ($WW$, $WZ$ and $ZZ$) was simulated using the \SHERPA[2.2.2]~\cite{Gleisberg:2008ta,Cascioli:2013gfa,Bothmann:2019yzt}
generator with the NNPDF3.0 PDF set. These simulations are accurate to NLO for up to one additional parton and accurate to leading order (LO) for up to three additional parton emissions.

Another background contribution comes from \Zjets with the $Z$ boson decaying into two $\tau$-leptons, which
then decay to an electron and a muon.
Those samples were simulated with the \SHERPA[2.2.1] generator with the NNPDF3.0 PDF set. They are accurate to NLO for up to two additional partons and accurate to LO for up to four additional partons, and so are the
$\Zboson (\ra ee)$\,+\,jets and $\Zboson (\ra \mu\mu)$\,+\,jets samples which are used to extract a factor to scale the
$\Zboson (\ra \tau\tau)$\,+\,jets background to data; see Section~\ref{sec:corrections} for details. To study systematic uncertainties in the \Zjets modelling,
alternative \Zjets samples were generated with \POWPY[8.230].

Backgrounds from $\ttbar W$ and $\ttbar Z$ are described by samples simulated
with the \MGNLO generator at NLO interfaced with \PYTHIA[8.210] with the
A14 tune and the NNPDF2.3 PDF set. The minor background coming from $\ttbar H$ was also simulated with the \MGNLO NLO generator
with the A14 tune and the \NNPDF[2.3lo] PDF set, and the minor single-top contribution from $t$-channel exchange was simulated
with the \POWPY[8.230] generator with the $\NNPDF[3.0nlo]\_4$f PDF set.

For the estimation of the misidentified-lepton backgrounds, the above samples, in which a dileptonic filter is applied, are complemented by top-quark samples and diboson samples containing at least one hadronic top-quark or boson decay, respectively, together with \Wjets samples generated with the same set-up as the \Zjets samples.

\section{Object reconstruction and event selection}
\label{sec:selection}
 
The events used in this analysis must contain a reconstructed electron, a reconstructed muon and either
one or two $b$-tagged jets. All reconstructed objects are required to have $|\eta|<2.5$ and $\pT>25$~\GeV.
For electrons, the pseudorapidity region
is reduced to $|\eta |< 1.37$ and  $1.52 < |\eta | < 2.47$ to exclude the transition
region between the barrel and endcap calorimeters.

Electron candidates are reconstructed from energy clusters in the electromagnetic calorimeter matched to tracks
reconstructed in the inner tracking detector~\cite{EGAM-2018-01}. The candidates are required to satisfy \enquote{tight} selection criteria.
In addition, the candidates are subject to  an isolation requirement
allowing no more than a certain fraction of the electron energy to be carried by particles measured in the vicinity of
the electron candidate. The requirement is passed by 90\% of the electrons from $\Zboson \rightarrow ee$ decays, at $\pT=25$~\GeV. The candidates are also required to
originate from the primary event vertex~\cite{PERF-2015-01}, defined as the reconstructed vertex with the highest sum of $\pT^2$ for the tracks
associated with it. The candidate track must satisfy a requirement on the transverse impact parameter significance of $|d_0|/\sigma_{d_0}<5$
and on the longitudinal impact parameter, $z_0$, of $|\zzsth| <0.5$~mm, where $\theta$ is the polar angle of the track.
 
Muon candidates are reconstructed by combining tracks reconstructed in the inner tracking detector and the muon spectrometer.
They are required to have $|\eta | < 2.5$, to satisfy \enquote{medium} selection criteria~\cite{MUON-2018-03} and
an isolation requirement which has an efficiency of $\sim$85\% for muons with \pT $=$ 25~\GeV , increasing gradually to 98\% for muons with $\pT > 100$~\GeV.
Furthermore, the muon candidate tracks must originate from the primary vertex, ensured by requiring $|d_0|/\sigma_{d_0}<3$
and $|\zzsth| <0.5$~mm.

Jets are reconstructed from topological cell clusters~\cite{PERF-2014-07} in the calorimeters using the anti-$k_t$ algorithm~\cite{Cacciari:2008gp,Fastjet} with a
radius parameter $R=0.4$. After calibration of the jet energy scale~\cite{JETM-2018-05} using information from both data and simulation,
the jets are required to have $\pT> 25$~\GeV \ and $|\eta | < 2.5$. In order to reduce contamination from \pileup, jets with
$\pT< 120$~\GeV \ and $|\eta | < 2.4$ must pass a primary vertex association requirement using the \enquote{jet vertex tagger} (JVT)~\cite{PERF-2014-03}, which
has an efficiency of 87\% for jets with $\pT=25$~\GeV, increasing to 95\% for jets with $\pT=60$~\GeV.
 
Jets likely to contain $b$-hadrons are tagged with the MV2c10 algorithm~\cite{PERF-2016-05} using jet and track variables
sensitive to $b$- and $c$-hadron masses, lifetimes and decay topologies. A working point with an average efficiency of 70\% was used, with rejection factors for $c$-quark jets, $\tau$-leptons and light-quark jets of 8, 13 and 313, respectively. These values are estimated using the \ttbar simulation.

To avoid double counting, an overlap removal procedure is applied. First, any electron candidates that share a track with a muon candidate are removed.
Subsequently, jets within $\Delta R = 0.2$ of an electron are removed, and afterwards, electrons within a region $0.2<\Delta R < 0.4$ around any remaining jet are rejected.
Jets that have fewer than three tracks and are within $\Delta R = 0.2$ of a muon candidate are removed,
and muons within $\Delta R = 0.4$ of any remaining jet are discarded.

Events are retained if they contain exactly one electron and exactly one muon satisfying the selection criteria detailed above,
where at least one of the two leptons is matched to an electron or muon trigger object, which implies a minimum \pT of 27 \GeV.
The  events with opposite-charge $e\mu$ pairs (opposite-sign, OS) are used for the measurement of the \ttbar signal, while the same-charge
$e\mu$ pairs (same-sign, SS) are used to estimate the background from misidentified leptons. Furthermore, the events must contain either exactly one
or exactly two $b$-tagged jets.

\begin{figure}[htbp]
\centering
\includegraphics[scale=0.55]{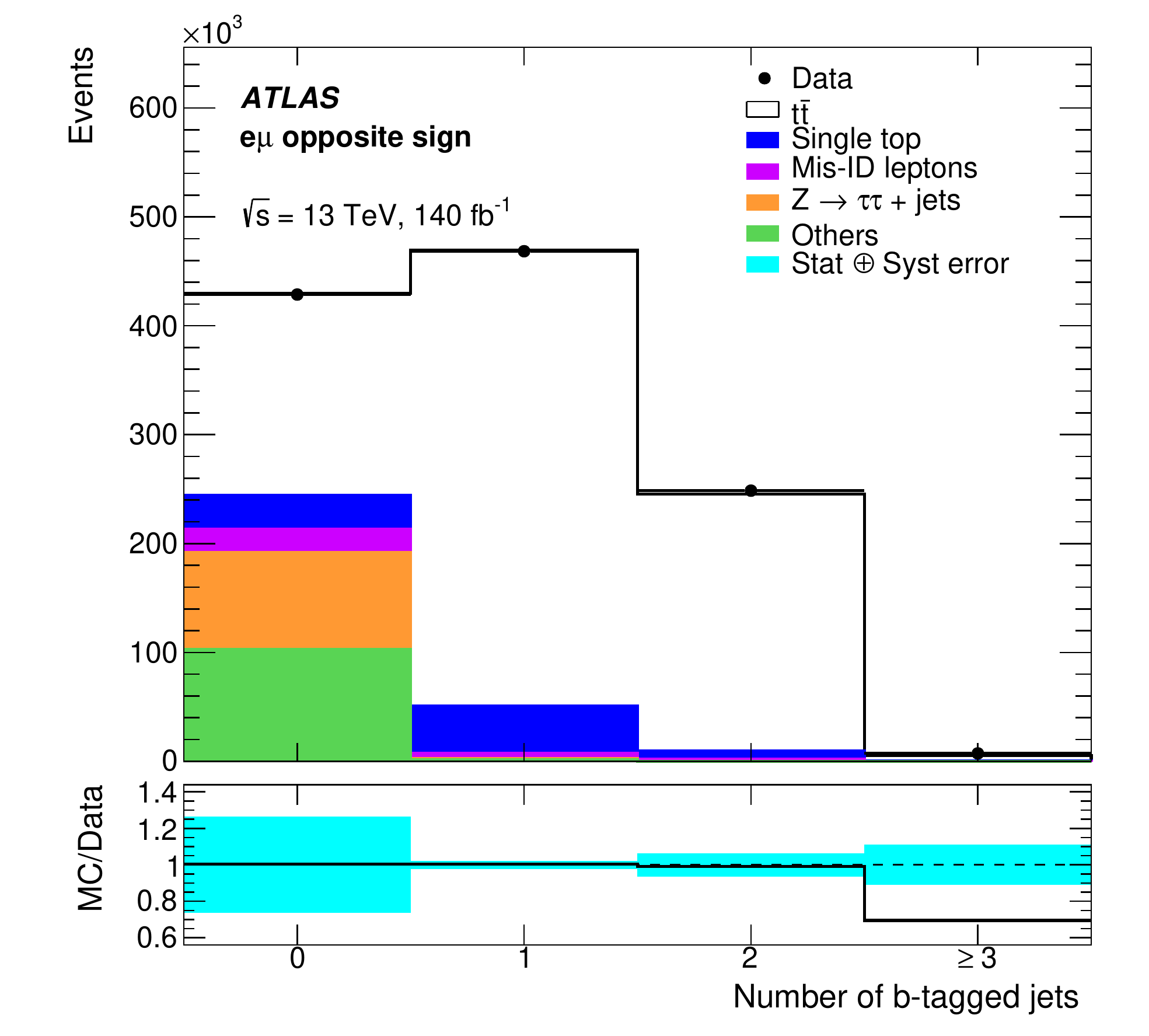}
\caption[Event count for different \bjet multiplicity]{Distribution of the number of $b$-tagged jets in selected opposite-sign $e\mu$ events. The coloured distributions show the breakdown of the predicted background contributions from single top quarks ($\Wboson t$ and $t$-channel), misidentified leptons, $\Zboson (\ra \tau\tau)$\,+\,jets and other sources of background (diboson, $\ttbar W$, $\ttbar Z$, and $\ttbar H$). The bottom panel shows the ratio of the prediction to the data with an uncertainty band covering both the statistical and systematic uncertainties, except for \ttbar generator uncertainties.}
\label{fig:nbjets}
\end{figure}

\begin{figure}[htbp]
\centering
\subfloat[]{\label{fig:lepton_pT_1b} \includegraphics[width=0.49\linewidth]{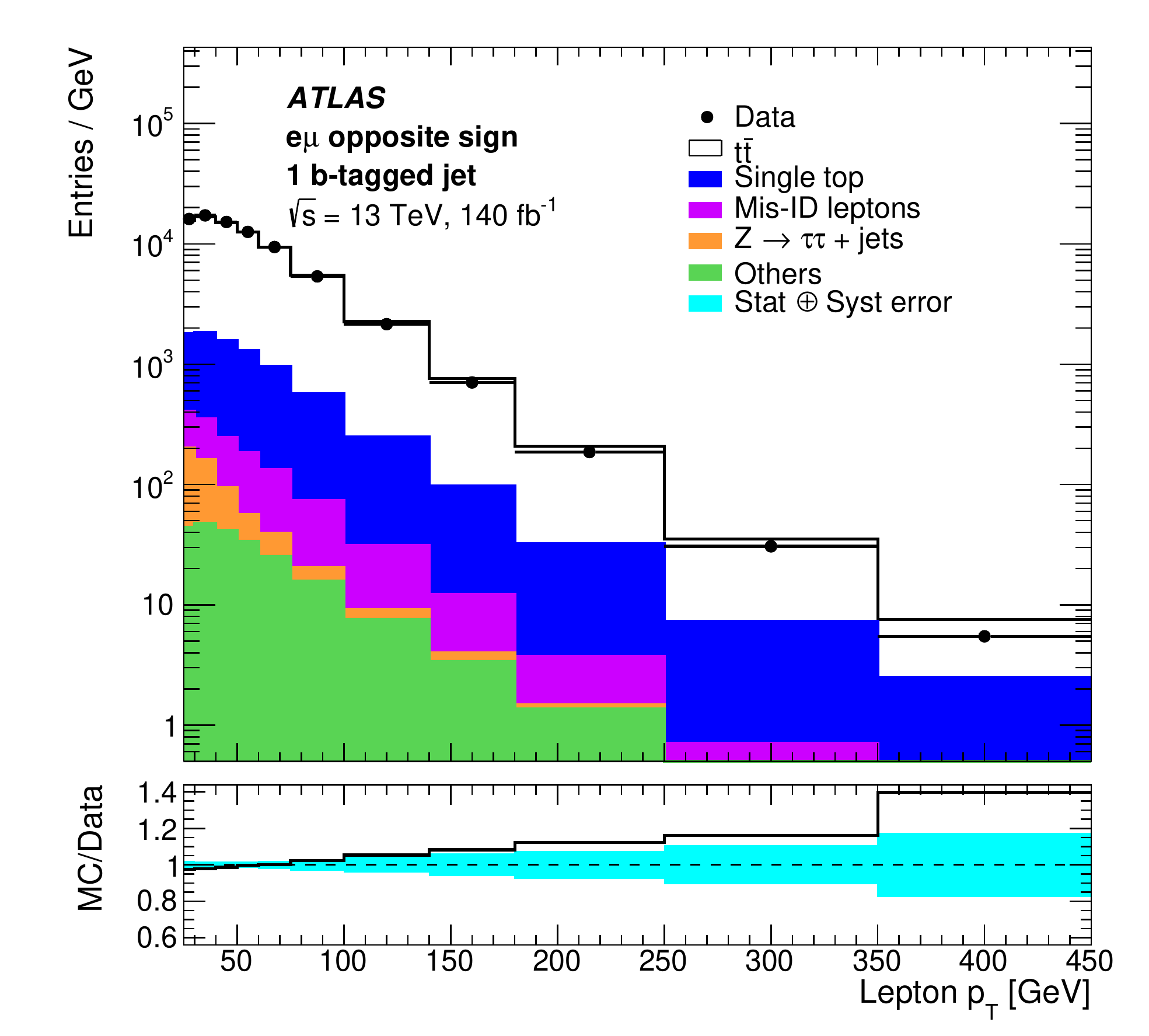}}
\subfloat[]{\label{fig:lepton_eta_1b} \includegraphics[width=0.49\linewidth]{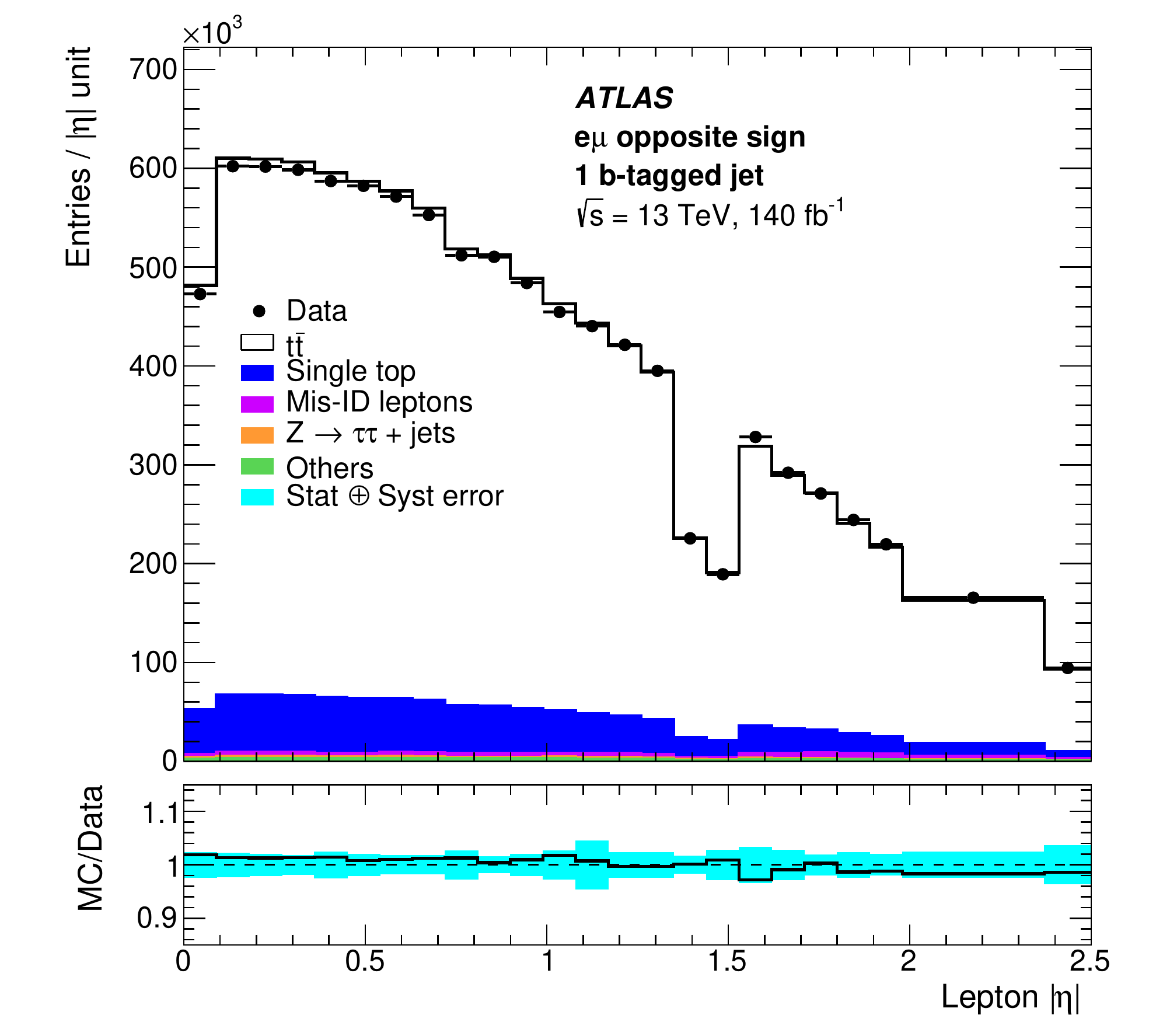}}
\caption[\pT and $|\eta|$ distributions of leptons from opposite sign $e\mu$ events with one \bjet]{Distributions of lepton \pT (left) and $|\eta|$ (right)
in opposite-sign $e\mu$ events with one \btagged jet. The data (dots) are compared with the combined prediction from signal and background processes (line). The simulated event samples are normalised to the integrated luminosity of the data. 
The coloured distributions show the breakdown of the predicted background contributions from single top quarks ($\Wboson t$ and $t$-channel), misidentified leptons, $\Zboson (\ra \tau\tau)$\,+\,jets and other sources of background (diboson, $\ttbar W$, $\ttbar Z$, and $\ttbar H$).
The bottom panel shows the ratio of the prediction to the data with an uncertainty band covering both the statistical and systematic uncertainties, except for \ttbar generator uncertainties.  The last bin of the \pT distribution includes overflow events.}
\label{fig:pT_eta_variables_1b}
\end{figure}
 
\begin{figure}[htbp]
\centering
\subfloat[]{\label{fig:lepton_pT_2b} \includegraphics[width=0.49\linewidth]{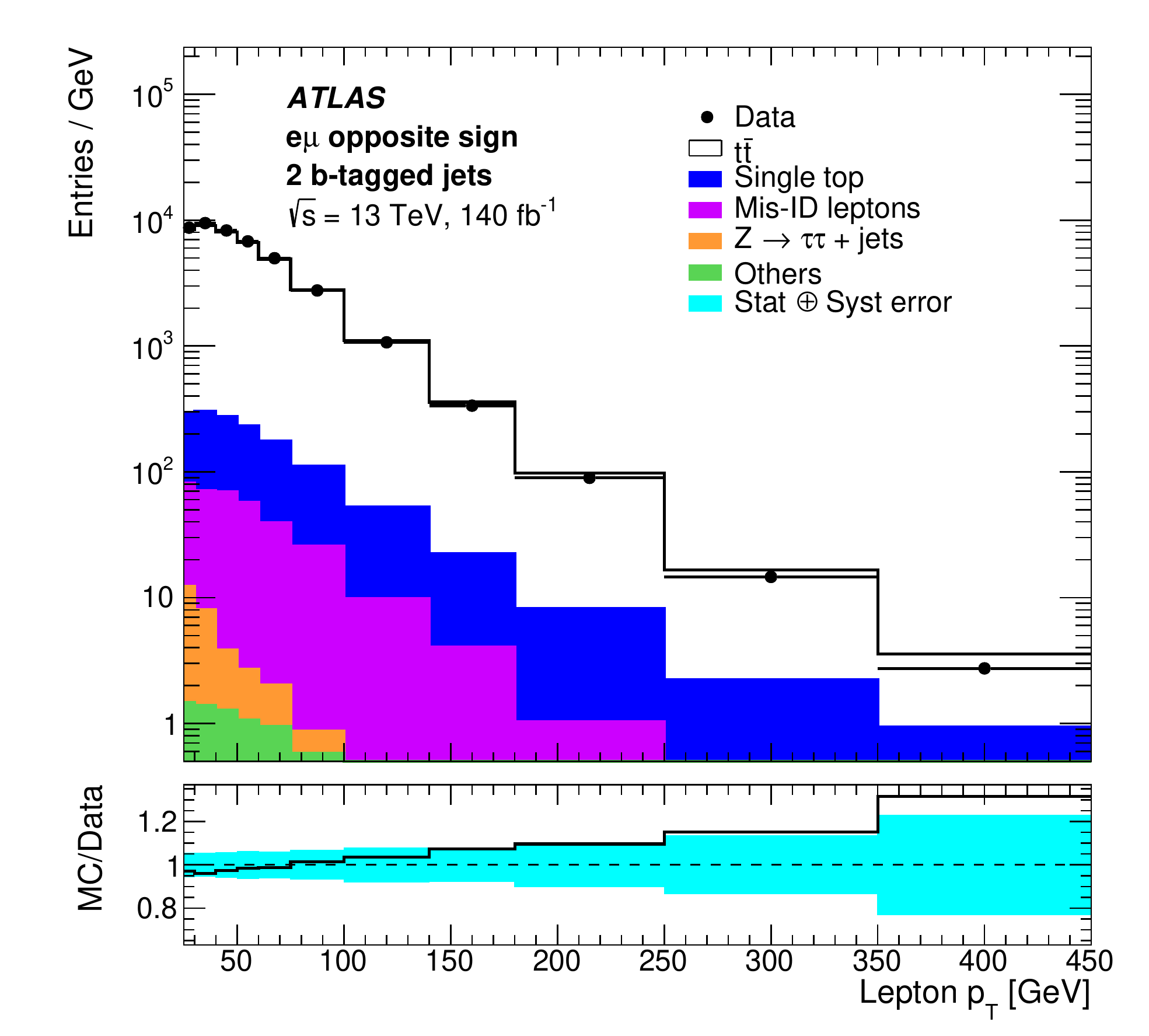}}
\subfloat[]{\label{fig:lepton_eta_2b} \includegraphics[width=0.49\linewidth]{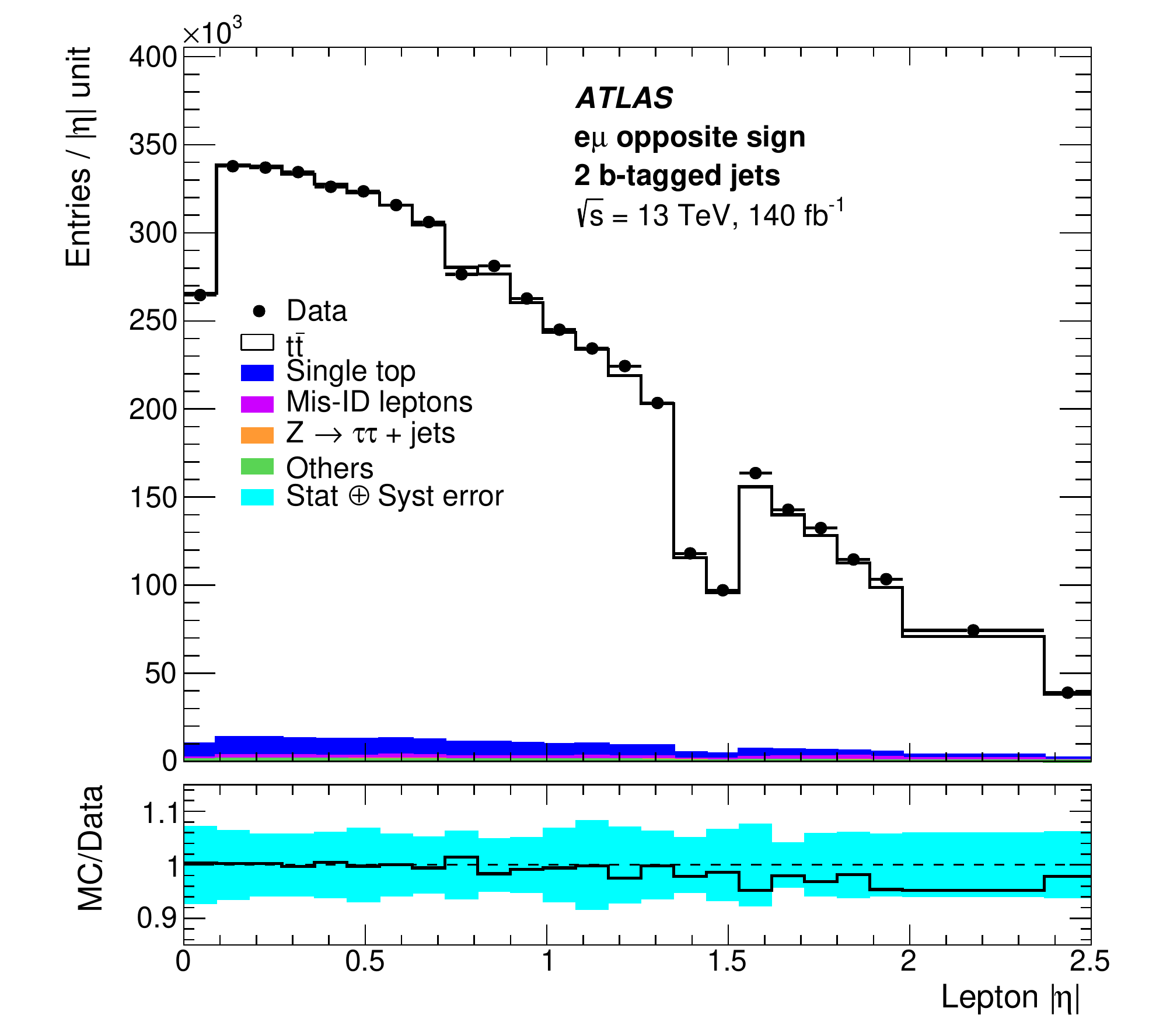}}
\caption[\pT and $\|eta|$ distributions of leptons from opposite sign $e\mu$  events with two \btagged jets]{Distributions of lepton \pT (left) and $|\eta|$ (right)
in opposite-sign $e\mu$ events with two \btagged jets. The data (dots) are compared with the combined prediction from signal and background processes (line). The simulated event samples are normalised to the integrated luminosity of the data. 
The coloured distributions show the breakdown of the predicted background contributions from single top quarks ($\Wboson t$ and $t$-channel), misidentified leptons, $\Zboson (\ra \tau\tau)$\,+\,jets and other sources of background (diboson, $\ttbar W$, $\ttbar Z$, and $\ttbar H$).
The bottom panel shows the ratio of the prediction to the data with an uncertainty band covering both the statistical and systematic uncertainties, except for \ttbar generator uncertainties.  The last bin of the \pT distribution includes overflow events.}
\label{fig:pT_eta_variables_2b}
\end{figure}
 
The selected events containing OS leptons are shown as a function of the number of $b$-tagged jets in Figure~\ref{fig:nbjets}.  The mismodelling of the number of events with three or more \btagged jets is taken into account using the \ttbar sample with an enriched rate of events with at least three $b$-jets at generator level, as described in Section~\ref{sec:signal-samples}.
The reconstructed transverse momentum and $|\eta|$ distributions of the OS leptons in the selected data sample are shown in Figures~\ref{fig:pT_eta_variables_1b}
and \ref{fig:pT_eta_variables_2b} together with signal and background predictions. 
The data and simulated distributions generally agree well, but the lepton
transverse momentum distribution observed in data is softer than in the nominal signal and background simulation,
as has also been observed in previous measurements at $\sqrt{s}=13\,\TeV$~\cite{CMS-TOP-17-014,TOPQ-2018-17}.
 
In addition to the reconstructed objects, \enquote{particle-level} objects are also defined.  These are a collection of stable particles (with lifetime larger than 30~ps) from the full matrix element and parton shower generators, without any simulation of the interaction of these particles with the detector components.
 
Simulated events with an $e\mu$ pair located in a fiducial region, given for both leptons by $\pT^{\ell}>27\ (25)$~\GeV \ for the leading (sub-leading) lepton and $|\eta^{\ell}|<2.5$ at particle-level, are used to extrapolate the observed event rate to a fiducial cross-section. The four-momentum of each charged lepton is taken after final-state radiation and it is summed with the four-momenta of any radiated photons within a cone of size $\Delta R = 0.1$ around the lepton direction.
Particle-level jets are reconstructed using stable particles in the event (excluding charged leptons and neutrinos that do not originate from hadron decays) using the anti-$k_t$ algorithm with a R parameter of $R = 0.4$.  They are required to have $\pT > 25$ \GeV\, and $|\eta| < 2.5$.
Particle-level electrons and muons that overlap with particle jets with $\Delta R < 0.4$ are removed from the event. No electron--muon overlap removal is applied at the particle level.

\section{Data-driven background estimates and efficiency corrections}
\label{sec:corrections}

The simulated backgrounds from misidentified leptons and the backgrounds
from $\Zboson (\ra \tau\tau)$\,+\,jets are corrected using data-driven methods. These two backgrounds
amount to 1\% and 0.2\%, respectively, of the total selected event sample.

\begin{figure}[htbp]
\centering
\subfloat[]{\label{fig:fakes_lepton_pT_1b} \includegraphics[width=0.49\linewidth]{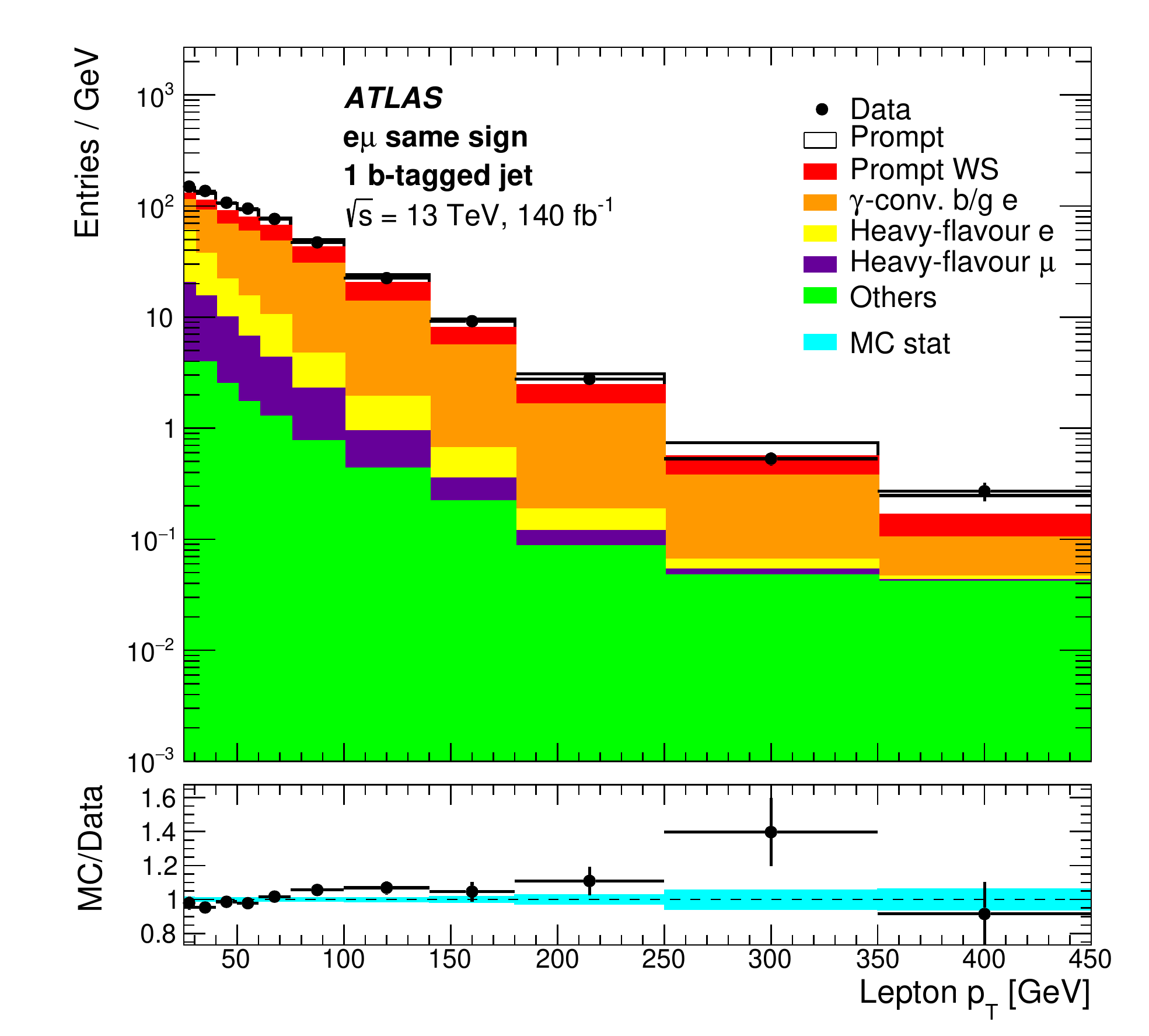}}
\subfloat[]{\label{fig:fakes_lepton_eta_1b} \includegraphics[width=0.49\linewidth]{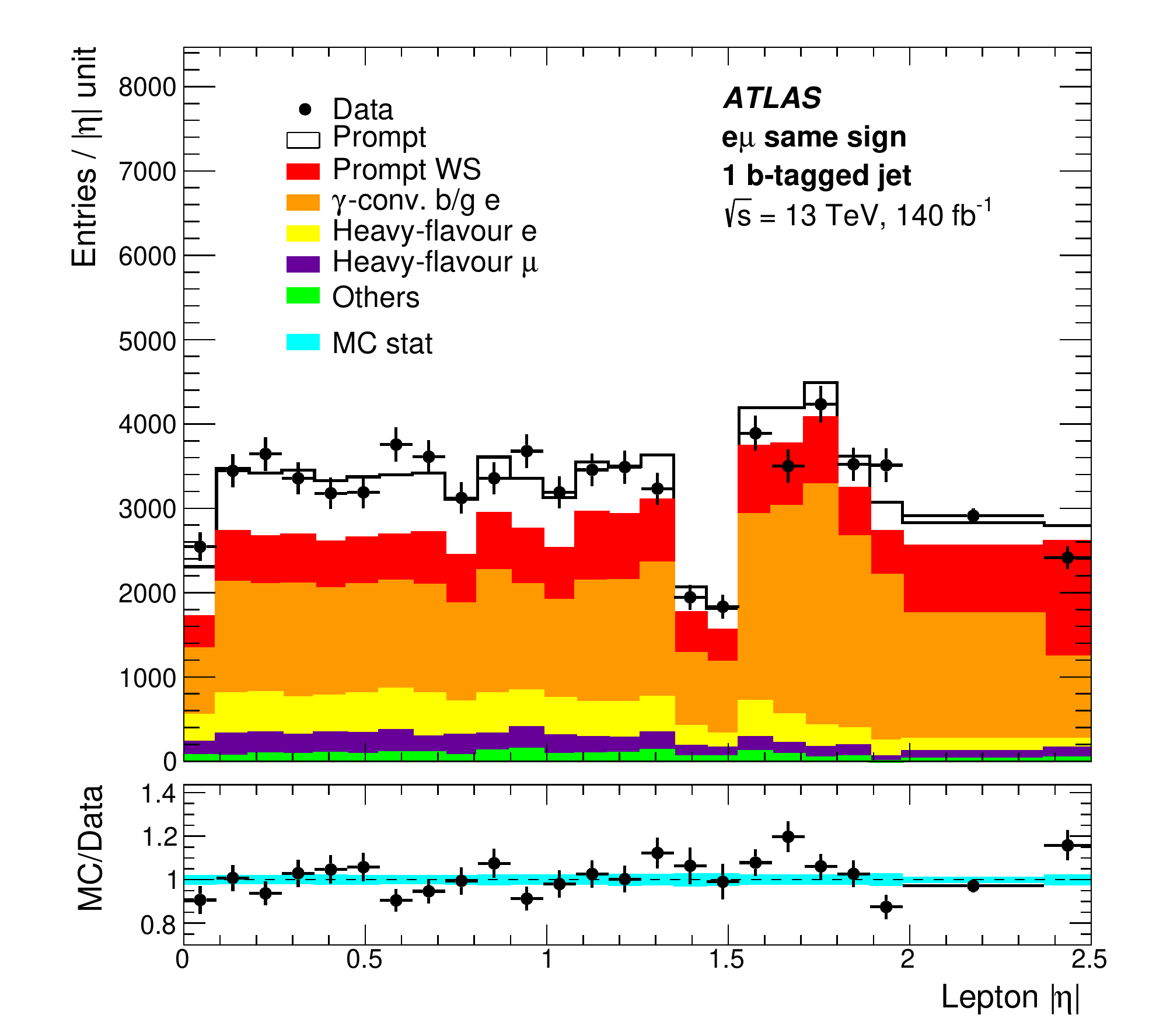}}
\caption[Distribution of lepton \pT (left) and $|\eta|$ (right) in same-sign $e\mu$ events with one \bjet]{Distribution of lepton \pT (left) and $|\eta|$ (right)
in same-sign $e\mu$ events with one $b$-tagged jet. The data (dots) are compared with the combined prediction from signal and background processes
(line). The coloured distributions show the breakdown of the predicted contributions: wrong-sign (WS) prompt leptons,
where the lepton charge is mismeasured; background electrons from photon
conversions; electrons or muons from heavy-flavour decays; and other
sources of fake leptons. The bottom panel shows the ratio of the predictions to the data with an uncertainty band covering the MC statistical uncertainty. The last bin of the \pT distribution includes overflow events.}
\label{fig:pT_eta_SS_1b}
\end{figure}

\begin{figure}[htbp]
\centering
\subfloat[]{\label{fig:fakes_lepton_pT_2b} \includegraphics[width=0.49\linewidth]{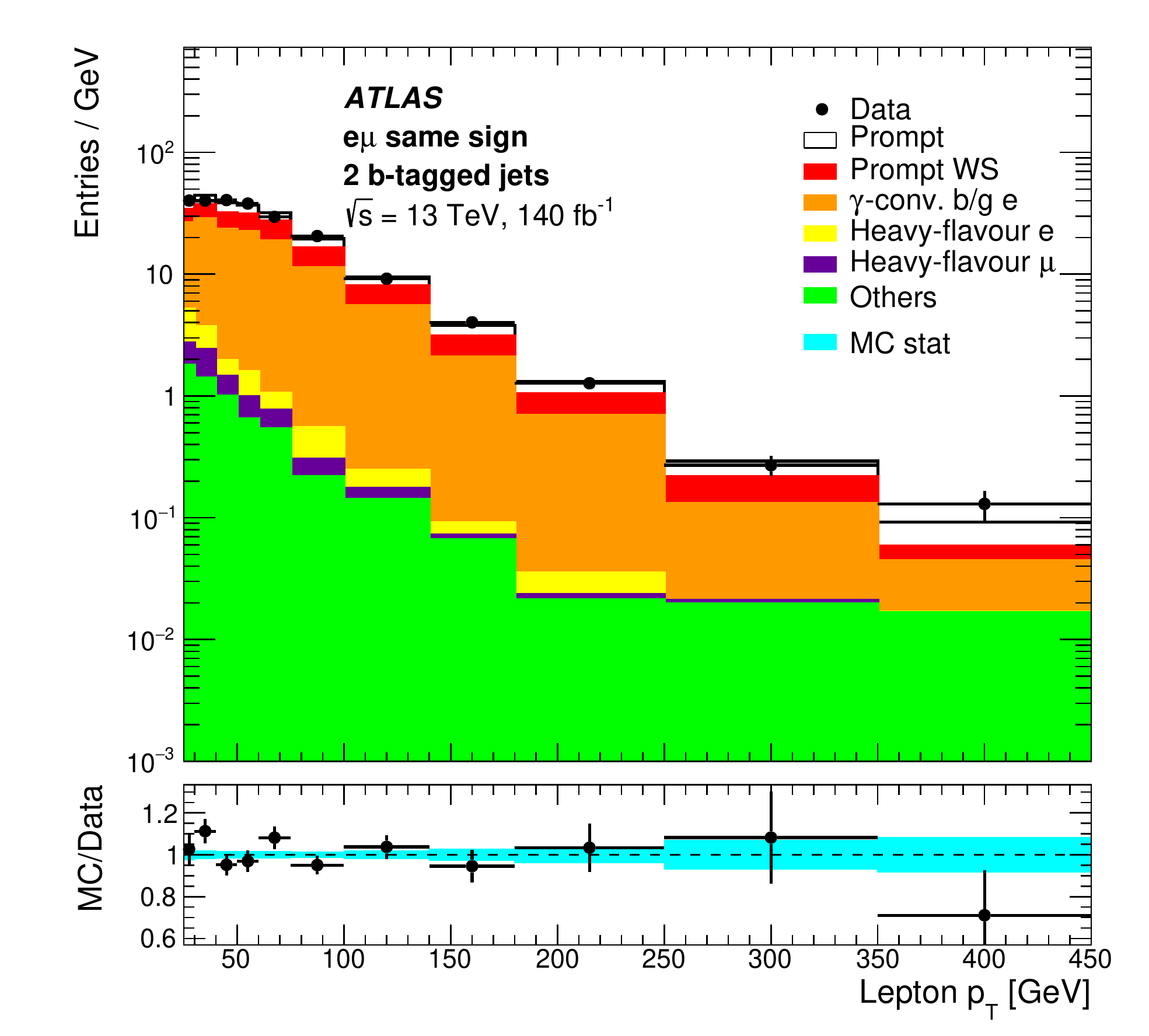}}
\subfloat[]{\label{fig:fakes_lepton_eta_2b} \includegraphics[width=0.49\linewidth]{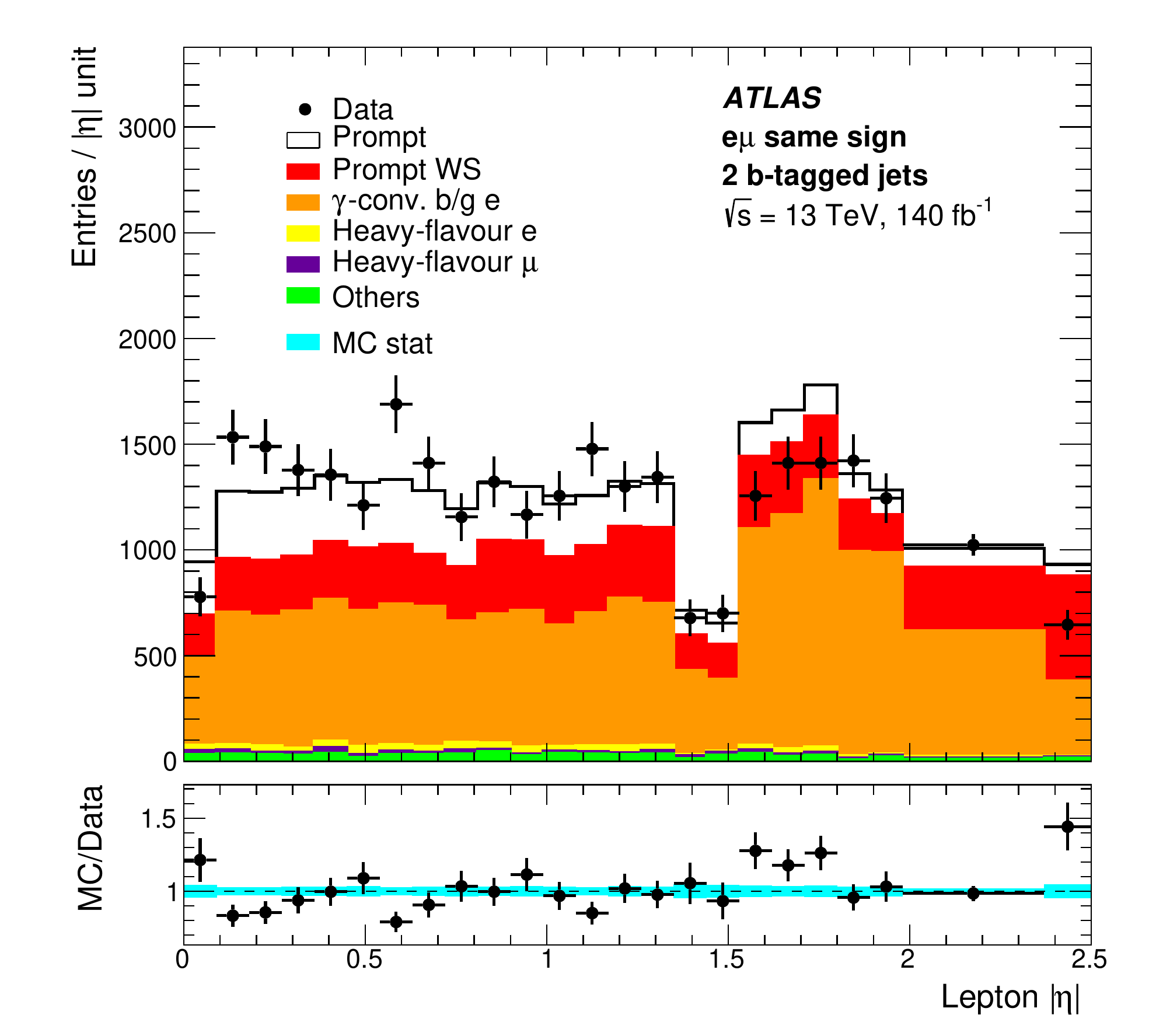}}
\caption[\pT and $|\eta|$ distributions of leptons from same-sign $e\mu$ events with two \bjets]{Distribution of lepton \pT (left) and $|\eta|$ (right)
in same-sign $e\mu$ events with two $b$-tagged jets. The data (dots) are compared with the combined prediction from signal and background processes
(line). The coloured distributions show the breakdown of the predicted contributions: wrong-sign (WS) prompt leptons,
where the lepton charge is mismeasured; background electrons from photon
conversions; electrons or muons from heavy-flavour decays; and other
sources of fake leptons. The bottom panel shows the ratio of the predictions to the data with an uncertainty band covering the MC statistical uncertainty. The last bin of the \pT distribution includes overflow events.}
\label{fig:pT_eta_SS_2b}
\end{figure}

The misidentified-lepton background is composed of five different categories treated together, as shown in the SS regions in Figures~\ref{fig:pT_eta_SS_1b} and \ref{fig:pT_eta_SS_2b}. The major contribution is due to \ttbar dilepton events where the electron stems from the conversion of a photon radiated from a prompt electron. Three more categories are due to an electron or muon coming from the semileptonic decay of heavy-flavour hadrons or one lepton with a wrongly reconstructed charge. The final category, labelled as \enquote{Others}, includes all the other cases, e.g.\ a muon from an in-flight decay of a pion or kaon.

The simulated contribution of prompt leptons to the SS $e\mu$ sample is subtracted and the result is scaled by the
ratio of OS to SS misidentified leptons in the simulation. These data-driven estimates differ from the MC predictions by less than 10\%, as shown in Figure~\ref{fig:pT_eta_SS_1b} and \ref{fig:pT_eta_SS_2b}.
 
The background from $\Zboson (\ra \tau\tau)$\,+\,jets events predicted by the simulation (see Section~\ref{sec:background-samples}) is rescaled by
the ratio of measured to
predicted $\Zboson \ra \mu\mu$ and  $\Zboson \ra ee$ events accompanied by $b$-tagged jets in the \RunTwo sample.
The $\mu^{+} \mu^{-}$ and $e^{+} e^{-}$ invariant mass
spectra in data are each fitted with a linear combination of
two templates, one for leptons from $Z$ boson decays and one for background processes (including misidentified leptons), with both templates taken from simulation.
The two $Z$ scale factors obtained from the events with two electrons or two muons are averaged to obtain a weight to be applied to the $\Zboson$ boson events in the $e\mu$ sample.
The simulated background from $\Zboson (\ra \tau\tau)$\,+\,jets events is found to require a scale factor of 1.180 $\pm$ 0.001 if the $\Zboson$ boson candidate is accompanied by
one $b$-tagged jet and 1.313 $\pm$ 0.006 if the $\Zboson$ boson candidate is accompanied by two $b$-tagged jets. This result agrees within uncertainties with an earlier result~\cite{STDM-2017-38}
based on a data sample corresponding to an integrated luminosity of 36~fb$^{-1}$.
The errors quoted for the scale factors are purely statistical; the systematic uncertainty is discussed in Section~\ref{subsec:backgroundSyst}.
The lepton correction factors described below have been applied before computing these scale factors.

The lepton trigger, reconstruction and selection efficiencies from  simulation receive small corrections derived from
measurements of $\Zboson \ra \ell\ell$ events in the data~\cite{TRIG-2018-05, TRIG-2018-01,EGAM-2018-01,MUON-2018-03}.
However, a dedicated \textit{in situ} measurement of the lepton
isolation efficiencies is made for this analysis. This ensures that the
efficiencies correspond to those in \ttbar events, which have larger
hadronic activity than in  $\Zboson \ra \ell\ell$ events, and it also significantly reduces the dependence
on the \ttbar and \pileup modelling.
Opposite-sign $e\mu$ events are selected with the isolation requirement only applied to one of the two leptons.
The inefficiency of the isolation requirement is given by the fraction of signal events in which the other lepton fails
the requirement. The contribution of the prompt-leptons from background sources is subtracted using simulation,
the contribution from isolated misidentified leptons is determined from same-sign $e\mu$ pairs as described above, and the
contribution from non-isolated misidentified leptons is determined from leptons in data that fail the requirement on the transverse impact parameter significance
$|d_0|/\sigma_{d_0}$.
This yields a scale factor that multiplies the simulated isolation efficiency and is binned in \pT and
$|\eta|$. These scale factors deviate from unity by less than 1\% in all bins and are evaluated individually for each \ttbar generator and applied consistently.

\section{Cross-section determination}
\label{sec:cross-section}
 
The total and differential \ttbar production cross-sections are measured in the $e\mu$ channel of the \ttbar decay in a fiducial
region given by $|\eta^{\ell}|<2.5$ and $\pT^{\ell}>27\ (25)$~\GeV \ for the leading (sub-leading) lepton after applying the overlap removal procedure described in Section~\ref{sec:selection}. The event selection detailed in Section~\ref{sec:selection} targets
$\ttbar \ra \Wplus\Wminus \bbbar$ where one of the \Wboson bosons decays to an electron and the other to a muon, either directly or via the decay
to a $\tau$-lepton which subsequently decays leptonically. When defined in this way, the probability of having an electron and a muon coming from the \ttbar process is around 3\%.
\subsection{Differential fiducial cross-sections}
\label{sec:differential_cross_sections}
 
The events used in the analysis are required to contain either exactly one or exactly two $b$-tagged jets,
thereby reducing backgrounds to the 11\% or 4\% level in the two regions, respectively, and allowing the simultaneous determination of the \ttbar cross-section and the
combined jet selection and $b$-tagging efficiency. In each bin $i$ of the lepton kinematic variables listed in Section~\ref{sec:intro}, these parameters are evaluated by solving the equations

\begin{equation}
\label{eq:double_tagging}
\begin{aligned}
N_1^i &= \mathcal{L} \sigma_{\ttbar}^i G_{e\mu}^i 2 \epsilon_b^i (1 - \epsilon_b^i C_b^i)  + N_{1,\textrm{bkg}}^i \; \\
N_2^i &= \mathcal{L} \sigma_{\ttbar}^i G_{e\mu}^i (\epsilon_b^i)^2 C_b^i  + N_{2,\textrm{bkg}}^i \; 
\end{aligned}
\end{equation}
where
\begin{itemize}
\item $N_1^i$ and $N_2^i$ are the numbers of selected data events with either one $b$-tagged jet or two $b$-tagged jets in the reconstructed bin $i$,
\item $N_{1,\textrm{bkg}}^i$ and $N_{2,\textrm{bkg}}^i$ are the numbers of predicted background events with either one $b$-tagged jet or two $b$-tagged jets in reconstructed bin $i$,
\item $\mathcal{L}$ is the integrated luminosity of the data,
\item $\sigma_{\ttbar}^i$ is the cross-section for \ttbar production resulting in an opposite-sign $e\mu$ pair
in the MC generator-level (i.e.\ particle-level) fiducial region defined by bin $i$,
\item
$G_{e\mu}^i$ is the reconstruction  efficiency, defined in the simulated \ttbar sample as the number of selected $e\mu$ pairs
(without any jet requirements) reconstructed in bin $i$ divided by the total number of $e\mu$ pairs
generated in bin $i$. It has an average value of 0.6 and falls to as low as 0.3 for the lowest-energy $e\mu$ pairs,
\item $\epsilon_b^i$ is the combined probability for a $b$-jet coming from a top-quark decay to be reconstructed as a jet, to fall within the detector and selection acceptance and be tagged as a $b$-jet,
\item $C_b^i$ is the \btag correlation coefficient that corrects the probability of tagging the second jet after having tagged the first one.
This coefficient is determined by simulation and is found to be close to unity in most kinematic bins, differing by a maximum of 2\%.
\end{itemize}
The two unknown variables in these equations, the cross-section and the combined selection and $b$-tagging efficiency, are determined with
a log-likelihood fit, as in Refs.~\cite{TOPQ-2013-04, TOPQ-2018-17}.
 
The binning is chosen so that more than 90\% of $e\mu$ pairs originating in bin $i$ at particle level are usually
measured in bin $i$ at detector level. A bin-by-bin unfolding procedure is used where the $G_{e\mu}^i$ efficiency accounts for both the accuracy of lepton reconstruction and the impact of bin migration, which refers to the situation where events categorized in bin $j$ at the particle level are observed in a different bin $i \neq j$ during reconstruction. The large LHC \RunTwo dataset allows the chosen binning to be finer than in the previous $13\,\TeV$ analysis~\cite{TOPQ-2018-17} and the measurements to be made in an extended energy range. It also allows
the distribution of one leptonic kinematic variable to be measured
in bins of another one. The variable pairs chosen are those deemed most useful
for the testing and tuning of MC generators,
namely the following four double-differential distributions:
\begin{itemize}
\item $|y^{e\mu}|$ in five bins of $m^{e\mu}$;
\item $|\Delta \phi^{e\mu}|$ in five bins of $m^{e\mu}$;
\item $|\Delta \phi^{e\mu}|$ in three bins of $\pT^{e\mu}$;
\item $|\Delta \phi^{e\mu}|$ in five bins of $E^{e} + E^{\mu}$.
\end{itemize}
The cross-section in each two-dimensional bin is determined in the same way as for the one-dimensional distributions and
the binning is again chosen so that at least 90\% of the events populate the diagonal elements of the migration matrix
relating the particle-level variables to the detector-level variables. The chosen binning is sufficiently fine to capture the observed differences between the differential cross-section distributions predicted by a range of event generators.
 
The normalised differential or double-differential cross-sections profit from a large reduction of systematic effects that
are correlated across the distributions. These cross-sections are defined for each bin $i$ as
\begin{equation*}
\label{eq:normalised_cross_section}
\sigma_{\ttbar, \textrm{norm}}^i = \frac{\sigma_{\ttbar}^i}{\sum_j \sigma_{\ttbar}^j}\; ,
\end{equation*}
where the denominator sums the absolute differential cross-sections over the distribution in question.
However, the normalisation introduces new bin-to-bin correlations which are evaluated by \enquote{bootstrapping} pseudo-experiments as explained
in Section~\ref{sec:validation}.

\subsection{Total fiducial cross-section}
\label{sec:fiducial}
The observed number of selected events in data, together with the predicted backgrounds and their associated statistical uncertainties, are presented in Table~\ref{tab:eventcountrun2}.
The MC contributions are normalised to the integrated luminosity of the data and
corrections to the MC predictions are implemented as
explained in Section~\ref{sec:corrections}. The observed and predicted yields agree within 1\%--2\%.
 
\begin{table}[htbp]
\centering
\caption[Observed number of events in data]{Observed number of events in data and expected number of events for each process. $N_{1}$ and  $N_{2}$ are the numbers
of events with one \btagged jet and two \btagged jets, respectively, for the opposite-sign (OS) and same-sign (SS) regions. The \ttbar\;\!+\,$X$ ($X$ = $W$, $Z$, $H$) contributions are included in \enquote{Other}. Misidentified lepton events in this table are divided in two categories: \enquote{Charge-misid.\ lepton} refers to the number of events with one wrong-charge reconstructed lepton while \enquote{Misidentified lepton} refers to the other four categories combined, estimated as explained in Section~\ref{sec:corrections} and shown in Figures~\ref{fig:pT_eta_SS_1b} and~\ref{fig:pT_eta_SS_2b}. The dashes mean that the expected number of events for that process is compatible with zero (given the statistical power of the prediction used). The uncertainties in the ratios of data to MC events are purely statistical. The sum of the individual contributions may differ from the \enquote{Total MC prediction} due to rounding.}
\label{tab:eventcountrun2}
\begin{tabular}{l|cc|cc}
\toprule
&   \multicolumn{2}{c|}{OS} &\multicolumn{2}{c}{SS} \\
& $N_{1}$     & $N_{2}$    & $N_{1}$   & $N_{2}$  \\
\midrule
\ttbar                     		& 418780 $\pm$ 130				& 235937 $\pm$ 95\phantom{0}	& -   						& -    \\
Single $t$                 	& 42944  $\pm$ 77    			& \phantom{0}7295 $\pm$ 31    	& -                 				& -                  \\
\Zjets                     		& \phantom{0}1552 $\pm$ 66		& \phantom{00}96.5 $\pm$ 7.5   	& -                 				& -                  \\
Diboson                    	& 1406.1 $\pm$ 9.5  				& \phantom{00}49.9 $\pm$ 1.1   	& 223.0 $\pm$ 2.4   			& \phantom{0}10.58 $\pm$ 0.30   \\
Charge-misid.\ lepton	& \phantom{000}1.90 $\pm$ 0.14	& \phantom{000}0.614 $\pm$ 0.061	& \phantom{0}858 $\pm$ 11	& 364.0 $\pm$ 7.1      \\
Misidentified lepton		& \phantom{00}4880  $\pm$ 100	& \phantom{0}1990 $\pm$ 67    	& 2550  $\pm$ 57    			& 906 $\pm$ 35      \\
Other                      		& 1192.6 $\pm$ 4.1 				& \phantom{0}807.1 $\pm$ 3.3   	& 407.0 $\pm$ 1.7   			& 238.3 $\pm$ 1.3    \\
\midrule
Total MC prediction		& 470760 $\pm$ 190  			& 246180 $\pm$ 120   			& 4039 $\pm$ 58     			& 1519 $\pm$ 36      \\
\midrule
Data events			& 468450\phantom{ $\pm$ 000}  	& 248560\phantom{ $\pm$ 000}    	& 3995\phantom{ $\pm$ 00}      & 1501\phantom{ $\pm$ 00}       \\
\midrule
Data/MC                    	& $0.995 \pm 0.002$				& $1.010 \pm 0.002$  			& $0.989 \pm 0.021$			& $0.988 \pm 0.035$  \\
\bottomrule
\end{tabular}
\end{table}

In order to determine the total fiducial \ttbar cross-section, Eq.~(\ref{eq:double_tagging}) is solved with bin $i$ taken as the entire fiducial region. An $e\mu$ reconstruction efficiency of  $G_{e\mu} = (57.06 \pm 0.02)\%$
and a \btag correlation coefficient of $C_b = 1.0058 \pm 0.0005$
are calculated with the nominal \ttbar \POWPY[8.230] $e\mu$ sample, where the quoted uncertainties are from MC statistics only.

\subsection{Total inclusive cross-section}
\label{sec:inclusive_total_cross_section}
In order to obtain the inclusive cross-section, the reconstruction efficiency $G_{e\mu}$ in Eq.~(\ref{eq:double_tagging}) is replaced by $E_{e\mu} = A_{e\mu} \cdot G_{e\mu}$.
The acceptance $A_{e\mu}$ is defined as
\begin{equation}
\label{eq:acceptance_factor}
A_{e\mu} = \frac{ N_{e\mu}^{\ttbar,\textrm{fiducial} }}{N^{\ttbar}},
\end{equation}
where $N_{e\mu}^{\ttbar,\textrm{fiducial}}$ is the number of particle-level opposite-sign $e\mu$ events found in the fiducial region
in a simulated \ttbar sample and $N^{\ttbar}$ is the total number of \ttbar pairs produced by the \ttbar generator.
The acceptance factor is taken from the same MC \ttbar sample used to calculate the reconstruction efficiency and
it is corrected for each generator to conform with the \Wboson branching ratio predicted by the Standard Model per lepton flavour, $B(\Wboson \ra \ell \nu) = 10.82\%$~\cite{Zyla:2020zbs}.
The value of $E_{e\mu}$ calculated with the nominal \ttbar \POWPY[8.230] sample is $E_{e\mu} = (0.7251 \pm 0.0003)\%$, while the value of
the acceptance $A_{e\mu}$ calculated with the same sample is $A_{e\mu} = (1.2708 \pm 0.0004)\%$, where the uncertainties are purely statistical.

\subsection{Validation of the analysis method}
\label{sec:validation}
 
The method used to solve Eq.~(\ref{eq:double_tagging}) for each leptonic kinematic bin $i$ is validated by
replacing the data sample by 1000 pseudo-experiments~\cite{Bohm:389738} fluctuating $N_1^i$ and $N_2^i$ within their statistical uncertainties
in various simulated event samples with known \ttbar cross-sections, $\sigma_{\textrm{true}}^i$.
In practice, this is done with a \enquote{bootstrapping method}~\cite{ATL-PHYS-PUB-2018-035}, assigning to each event a set of 1000 weights obtained from fluctuations of a Poisson distribution with a mean value of one.
 
In all such validations, the parameters of the equation, $G_{e\mu}^i$, $C_b^i$ and the background contributions, are taken from the
nominal signal and background samples. The mean of the 1000 cross-sections derived from Eq.~(\ref{eq:double_tagging})
is then compared with the true \ttbar cross-section of the simulated event sample in order to check for possible biases in the method.
The following simulated event samples are used for validation:
\begin{itemize}
\item The nominal \ttbar \POWPY[8.230] sample. The mean measured cross-section equals the true cross-section within the statistical error on the mean in any bin $i$.
\item Half the nominal \ttbar \POWPY[8.230] sample. Here the parameters of Eq.~(\ref{eq:double_tagging}) are evaluated with the other half of the sample.
No bias is seen beyond the expected statistical fluctuations.
\item  Two \ttbar \POWPY[8.230] samples with the top-quark mass changed to $176$~\GeV \ and $169$~\GeV. In the case of $m_t = 169$~\GeV , biases
of 1\%--2\% are seen in some kinematic bins, but in general the biases are smaller than the expected statistical uncertainties of the data and hence neglected.
\item The nominal \ttbar \POWPY[8.230] sample reweighted to produce the same $N_1^i$ and $N_2^i$ as in the data. As in the other tests, no significant bias away from the true cross-section is seen.
\end{itemize}
The bootstrapping method is also applied to both the data and simulated event samples in order to construct the covariance matrices of statistical uncertainties
of the measurements. The matrices
include non-zero off-diagonal elements between different variables and, in the case of normalised differential distributions,
also between bins in the same distribution.
 
\section{Systematic and statistical uncertainties}
\label{sec:errors}
Uncertainties due to theoretical assumptions and detector modelling affect the parameters $\mathcal{L}$, $G_{e\mu}^i$,
$ C_b^i$, $N_{1,\textrm{bkg}}^i$ and $N_{2,\textrm{bkg}}^i$ of Eq.~(\ref{eq:double_tagging}).
Uncertainties associated with generators are
evaluated by changing the values of parameters in the simulations or by using alternative generators. Some background uncertainties
are evaluated by using data-driven uncertainty estimates, as explained in Section~\ref{subsec:backgroundSyst}. For each variation, Eq.~(\ref{eq:double_tagging}) is solved and the change with respect to the baseline sample is assigned as the impact of the uncertainty on the cross-section measurement. 
Effects of finite data and MC sample sizes are evaluated by the bootstrapping method described in Section~\ref{sec:validation} and summarised in a covariance
matrix covering all measured kinematic bins. The individual sources of systematic uncertainty are discussed in the following subsections
and a summary of their impact in the measured fiducial and total inclusive cross-sections is given in Table~\ref{tab:total_fiducial_xs}.
 
\subsection{Detector-related uncertainties}
\label{subsec:detectorSyst}
 
Uncertainties on the trigger~\cite{TRIG-2018-05, TRIG-2018-01}, reconstruction and selection efficiency~\cite{EGAM-2018-01, MUON-2018-03} for the leptons are estimated using $\Zboson \ra ee$ and $\Zboson \ra \mu\mu$ 
events in data. They are expressed
as uncertainties in scale factors for the MC predictions, described in Section~\ref{sec:corrections}. By varying the scale factors, their
uncertainties are propagated to $G_{e\mu}^i$ and  $C_b^i$, as well as the number of background events, and Eq.~(\ref{eq:double_tagging}) is solved for each variation. The \enquote{up} and \enquote{down} variations are applied coherently to all bins of a given kinematic variable, except for the electron efficiency, where
this approach overestimates the uncertainties~\cite{EGAM-2018-01}. The latter variations are carried out separately in
two $|\eta|$ bins and nine $\pT$ bins and their effects are combined, taking measured bin-to-bin correlations into account. The up and down variations of the lepton isolation efficiency scale factors are calculated using $t\bar{t} \rightarrow WbWb \rightarrow e \nu b \mu \nu b$ events.
These variations take into account uncertainties in the isolation efficiencies for data and MC. The uncertainty on the isolation efficiency for data depends on the estimation of events with a misidentified lepton and on the statistical uncertainty, while the uncertainty for the MC depends on the simulated sample size.
 
The uncertainty on the jet energy scale (JES) and the jet energy resolution (JER) affect $\epsilon_b^i$, $C_b^i$ and the number of background events, mainly from $Wt$ production. 
The jet-related uncertainties are evaluated using the \Zjets, $\gamma$\,+\,jet and multijet samples at $\sqrt{s} = 13$~\TeV \ for both real and simulated data~\cite{PERF-2016-04}.
A total of 20 uncorrelated nuisance parameters affecting the JES and five parameters affecting the JER are varied up and down.
The difference between the energy response for reconstructed \bjets and that for other jets is varied separately~\cite{PERF-2016-04} and
the maximum possible variation is applied to the flavour composition of the jets (the mixture of quarks and gluons).
The modelling of \pileup also affects the jet energy, and the associated uncertainty is found by varying the jet energy up and down by
the uncertainty in this effect. The jet vertex association efficiency uncertainty affects the $C_b^i$ coefficient. This is evaluated by changing the JVT scale factor up and down within its uncertainty.
 
A \btag efficiency scale factor for the chosen working point is derived from \ttbar events~\cite{Aaboud:2018xwy} and applied to the Monte Carlo events used in the present analysis. The uncertainties are related to the $b$-jet tagging calibration for $b$-jets, $c$-jets and light-jets, and comprise nine, four and five eigenvector variations to the tagging efficiencies, respectively, and two components for the MC-based extrapolation to jets with very high \pT.

\subsection{Top-quark pair modelling uncertainties}
\label{subsec:ttbarSyst}
 
Uncertainties related to the modelling of \ttbar events have an impact on  $G_{e\mu}^i$ and $C_b^i$ in Eq.~(\ref{eq:double_tagging})
and on the acceptance
factor in Eq.~(\ref{eq:acceptance_factor}). These uncertainties are calculated with alternative \ttbar samples or by reweighting the nominal sample or increasing by 30\% the fraction of events with at least three \bjets, as described in Section~\ref{sec:signal-samples}.
The effect on $E_{e\mu}$, $G_{e\mu}$ and $C_b$ in the selected sample
is summarised in Table~\ref{tab:generators} and the change in the central value is taken as the uncertainty from each source.

As shown in Section~\ref{sec:results_diff_xs},
the \POWPY[8.230] generator does not give a good description of the lepton transverse momentum, which is
believed to be a reflection of the top-quark transverse momentum spectrum. Therefore, the difference with respect to the sample where the top-quark transverse momentum is reweighted to the NNLO predicted spectrum is considered as an additional uncertainty in all measurements. 
This effect is relevant mostly for the extrapolation of the fiducial cross-section to the total phase-space.

\begin{table}[htbp]
\centering
\caption[Difference in \btag correlation coefficient, reconstruction efficiency and preselection efficiency between nominal \ttbar sample and the corresponding \ttbar systematics samples]{Differences in the total \btag correlation coefficient, total reconstruction efficiency and total preselection efficiency between the baseline $e\mu$ \POWPY[8.230] sample and the corresponding \ttbar systematic uncertainty samples. The PDF row refers to the sum in quadrature of the differences derived from the 30 eigenvectors and the baseline. All uncertainties shown are due to the limited MC sample size.}
\label{tab:generators}
\footnotesize
\begin{tabular}{lccc}
\toprule
Systematic uncertainty name 		& $\Delta C_{b}/C_{b}$ [\%] 		& $\Delta G_{e\mu}/G_{e\mu}$ [\%]	& $\Delta E_{e\mu}/E_{e\mu}$ [\%]     \\
\midrule
Matrix element 					& $-$0.10 $\pm$ 0.22\phantom{$-$}  & 0.25 $\pm$ 0.11				& 0.29 $\pm$ 0.12 \\
\midrule
\hdamp 						& $-$0.06 $\pm$ 0.08\phantom{$-$} 	& $-$0.05 $\pm$ 0.04\phantom{$-$} 	& $-$0.05 $\pm$ 0.05\phantom{$-$} \\
\midrule
Parton shower and hadronisation	& 0.16 $\pm$ 0.08 				& $-$0.26 $\pm$ 0.04\phantom{$-$}	& 0.04 $\pm$ 0.05 \\
\midrule
Top \pT reweighting 				& 0.03 $\pm$ 0.08 				& 0.22 $\pm$ 0.04 				& 0.61 $\pm$ 0.05 \\
\midrule
$t\bar{t}$ + heavy flavour 			& $-$0.33 $\pm$ 0.08\phantom{$-$} 	& 0.01 $\pm$ 0.04 				& 0.01 $\pm$ 0.05 \\
\midrule
ISR (high) 					& $-$0.01 $\pm$ 0.08\phantom{$-$}	& 0.06 $\pm$ 0.04 				& 0.35 $\pm$ 0.05 \\
ISR (low) 						& 0.04 $\pm$ 0.08 				& $-$0.13 $\pm$ 0.04\phantom{$-$} 	& $-$0.35 $\pm$ 0.05\phantom{$-$} \\
\midrule
FSR (high) 					& 0.05 $\pm$ 0.09 				& $-$0.07 $\pm$ 0.04\phantom{$-$} 	& $-$0.12 $\pm$ 0.05\phantom{$-$} \\
FSR (low) 					& $-$0.09 $\pm$ 0.15\phantom{$-$} 	& 0.10 $\pm$ 0.07 				& 0.16 $\pm$ 0.09 \\
\midrule
PDF 							& 0.02 $\pm$ 0.08 				& 0.04 $\pm$ 0.04 				& 0.42 $\pm$ 0.05 \\
\bottomrule
\end{tabular}
\end{table}
 
The effect of changing the top-quark mass in the \ttbar and $Wt$ simulations is not included in Table~\ref{tab:generators} since by convention
the inclusive cross-section is quoted at a fixed $m_{t} = 172.5$~\GeV. However,
samples with different top-quark masses (from $m_{t} = 169$~\GeV \ to $m_{t} = 176$~\GeV) are used for validation tests and
to study the combined effect on the fiducial cross-section ($\sigma_{\ttbar}^{\textrm{fid}}$)
and the total inclusive cross-section. This is found to be
\begin{eqnarray*}
\frac{1}{\sigma_{\ttbar}^{\textrm{fid}}} \frac{\dif\sigma_{\ttbar}^{\textrm{fid}}}{\dif m_{t}} & = & -(0.004 \pm 0.003)\%/\GeV \\
\frac{1}{\sigma_{\ttbar}} \frac{\dif\sigma_{\ttbar}}{ \dif m_{\textrm{t}}} & = & -(0.379 \pm 0.005)\%/\GeV \\
\end{eqnarray*}

\subsection{Background modelling uncertainties}
\label{subsec:backgroundSyst}
 
The contribution from $\Wboson t$ is varied up and down by the relative uncertainty in the total cross-section,
which is $5.3\%$, as derived to approximate NNLO for $\sqrt{s} = 13$~\TeV \,using
the calculation in Ref.~\cite{Kidonakis:2010ux}. In addition, the nominal $\Wboson t$ generator is varied as
described in Section~\ref{sec:Wt-samples} and the results are propagated to the cross-section in each lepton kinematic bin.
In particular, the uncertainty due to the interference between the \ttbar and $\Wboson t$ amplitudes is taken as
the change in the result when replacing the diagram removal scheme~\cite{White:2009yt,Re:2010bp} with the diagram subtraction scheme~\cite{White:2009yt}.
The uncertainties in the matrix element, fragmentation and parton showering, and those related to the extra initial- and final-state radiation in the $\Wboson t$ background process are evaluated together with the signal process, considering these uncertainties to be correlated between \ttbar and $\Wboson t$.
 
The uncertainties in the size of the diboson
background are assessed by doubling and halving the factorisation and
renormalisation scales in the \SHERPA[2.2.2] samples
and by comparing the nominal \SHERPA[2.2.2] samples with those generated with
\POWPY[8.210]. A further 40\% normalisation uncertainty is added to
cover the uncertainties due to heavy-flavour jets produced in
association with the diboson pairs as discussed in Ref.~\cite{STDM-2018-03}.
 
An uncertainty of 5\% on the scale factors for the $\Zboson \ra \tau\tau$ contribution derived in Section~\ref{sec:corrections} is considered. 
The uncertainty from the fit amounts to less than 1\%. In order to take into account small differences between the same-flavour $ee$ and $\mu\mu$ control regions and the $e\mu$ measurement region, such as lepton efficiencies, a conservative systematic uncertainty of 5\% is assigned, both in the one $b$-jet and the two $b$-jets regions. 
An additional uncertainty due to the shape of the \Zjets background is estimated by using the \POWPY[8.186] sample
instead of the nominal \SHERPA[2.2.1] sample.
 
The uncertainty assigned to the $\ttbar V$ (where $V$ is \Wboson or \Zboson) cross-sections is $13\%$, following Ref.~\cite{ATLAS:2019fwo}.
 
In order to cover possible mismodelling of the electron charge misidentification and the misidentified-lepton composition in the nominal simulation, the ratio of OS to SS misidentified leptons is varied up and down by 25\% in the 1-$b$-jet region and by 50\% in the 2-$b$-jet region, as discussed in Ref.~\cite{TOPQ-2018-17}. The uncertainties in the predicted cross-sections for the $t\bar{t}V$ and diboson processes also have a non-negligible impact on the number of SS prompt leptons. A conservative uncertainty of 50\% is therefore assigned to those contributions~\cite{TOPQ-2018-17}.

\subsection{Luminosity and beam energy}
\label{subsec:luminositySyst}
 
The uncertainty in the combined 2015--2018 integrated luminosity is 0.83\%~\cite{ATLAS:2022hro}, obtained using the LUCID-2 detector~\cite{LUCID2} for the primary luminosity measurements, complemented by measurements using the inner tracking detector and calorimeters. This is propagated to the cross-sections
via Eq.~(\ref{eq:double_tagging}).
Including the effect of the luminosity uncertainty on the predicted background contribution, the
fiducial and inclusive cross-sections receive a total uncertainty of 0.93\% from the measured luminosity.
 
The uncertainty in the LHC beam energy is evaluated to be $0.1\%$~\cite{PhysRevAccelBeams.20.081003},
which is found to contribute
an uncertainty of $0.23\%$
to the measured total fiducial and inclusive cross-sections at $\sqrt{s} = 13$~\TeV. The effect of a
$0.1$\% uncertainty in the LHC beam energy is also propagated to the
differential measurements by
reweighting the PDFs using the LHAPDF library~\cite{Buckley:2014ana}.
The effect is generally small but increases in the highest-energy kinematic bins, reaching a maximum contribution of 0.1\%

\begin{table}[htbp]
\centering
\caption[Systematic uncertainties in the inclusive]{Breakdown of systematic uncertainties in the measured fiducial and inclusive cross-sections.}
\label{tab:total_fiducial_xs}
\begin{tabular}{lcc}
\toprule
Source of uncertainty                & $\Delta\sigma_{\ttbar}^\textrm{fid}/\sigma_{\ttbar}^\textrm{fid}$ [\%] & $\Delta\sigma_{\ttbar}/\sigma_{\ttbar}$ [\%]  \\
\midrule
Data statistics                   &        0.15    &     0.15 \\
MC statistics                     &        0.04    &     0.04 \\
Matrix element                    &        0.12    &     0.16 \\
\hdamp variation       &        0.01    &     0.01 \\
Parton shower                     &        0.08    &     0.22 \\
\ttbar + heavy flavour                     &        0.34    &     0.34 \\
Top \pT reweighting                     &        0.19    &     0.58 \\
Parton distribution functions     &        0.04  &     0.43 \\
Initial-state radiation           &        0.11  &     0.37 \\
Final-state radiation             &        0.29  &     0.35 \\
Electron energy scale             &        0.10  &     0.10 \\
Electron efficiency               &        0.37  &     0.37 \\
Electron isolation (in situ)      &        0.51  &     0.51 \\
Muon momentum scale               &        0.13  &     0.13 \\
Muon reconstruction efficiency    &        0.35  &     0.35 \\
Muon isolation (in situ)          &        0.33  &     0.33 \\
Lepton trigger efficiency         &        0.05  &     0.05 \\
Vertex association efficiency     &        0.03  &     0.03 \\
Jet energy scale \& resolution      &        0.10  &     0.10 \\
$b$-tagging efficiency              &        0.07  &     0.07 \\
\ttbar /$Wt$ interference           &        0.37  &     0.37 \\
$Wt$ cross-section                  &        0.52  &     0.52 \\
Diboson background                &        0.34  &     0.34 \\
$\ttbar V$ and $\ttbar H$               &        0.03  &     0.03 \\
\Zjets background                 &        0.05  &     0.05 \\
Misidentified leptons             &        0.32  &     0.32 \\
Beam energy                       &        0.23  &     0.23 \\
Luminosity                        &        0.93  &     0.93 \\
\midrule
Total uncertainty                 &        1.6\phantom{0} &     1.8\phantom{0} \\
\bottomrule
\end{tabular}
\end{table}

\FloatBarrier
 
\section{Results}
\label{sec:results}
\subsection{Total inclusive cross-section}
\label{sec:results_total_fid_xs}
 
The cross-section for the fiducial region is measured with the full \RunTwo dataset to be
\begin{equation*}
\sigma_{\ttbar}^{\textrm{fid}} = 10.53 \pm 0.02\; (\textrm{stat}) \pm 0.13\; (\textrm{syst}) \pm 0.10\; (\textrm{lumi}) \pm 0.02\; (\textrm{beam})\ \textrm{pb}
\end{equation*}
and the total inclusive cross-section is
\begin{equation*}
\sigma_{\ttbar} = 829 \pm 1\; (\textrm{stat}) \pm 13\; (\textrm{syst})  \pm 8\; (\textrm{lumi}) \pm 2\; (\textrm{beam})\ \textrm{pb},
\end{equation*}
where the uncertainties are due to statistics, theoretical and experimental systematic effects, the integrated luminosity and the beam energy.
The fiducial region is defined as $\pT^{\ell}>27\ (25)$~\GeV \ for the leading (sub-leading) lepton and $|\eta^{\ell} |<2.5$ after applying the
overlap removal procedure described in Section~\ref{sec:selection} for both leptons from \ttbar decays producing an $e\mu$ pair.
The total relative uncertainties in $\sigma_{\ttbar}^{\textrm{fid}}$
and $\sigma_{\ttbar}$  are 1.6\% and 1.8\%, respectively, where the breakdown of the various sources is shown in Table~\ref{tab:total_fiducial_xs}.

This measurement is compatible with the earlier ATLAS result at $\sqrt{s}=13$~\TeV\ using an integrated luminosity of 36~fb$^{-1}$~\cite{TOPQ-2018-17} but is significantly more precise, due to a reduction in the luminosity uncertainty~\cite{ATLAS:2022hro}. It is the most precise measurement of the inclusive \ttbar cross-section to date.
 
The predicted NNLO+NNLL value of the total inclusive cross-section at $\sqrt{s}=13$~\TeV, for a top-quark mass of $172.5$~\GeV, is
$\sigma_{\ttbar, \textrm{pred}} = 832^{+20}_{-29}$ (scale)$^{+23}_{-23}$ ($m_t$)$^{+35}_{-35}$ (PDF+\alphas)~pb
\cite{Czakon:2011xx,Baernreuther:2012ws,Czakon:2012zr,Czakon:2012pz,Czakon:2013goa}, which is in excellent agreement
with this measurement.

\subsection{Differential cross-section}
\label{sec:results_diff_xs}
 
\begin{figure}[htbp]
\centering
\subfloat[]{\label{fig:abs_xs_pT} \includegraphics[width=0.49\linewidth]{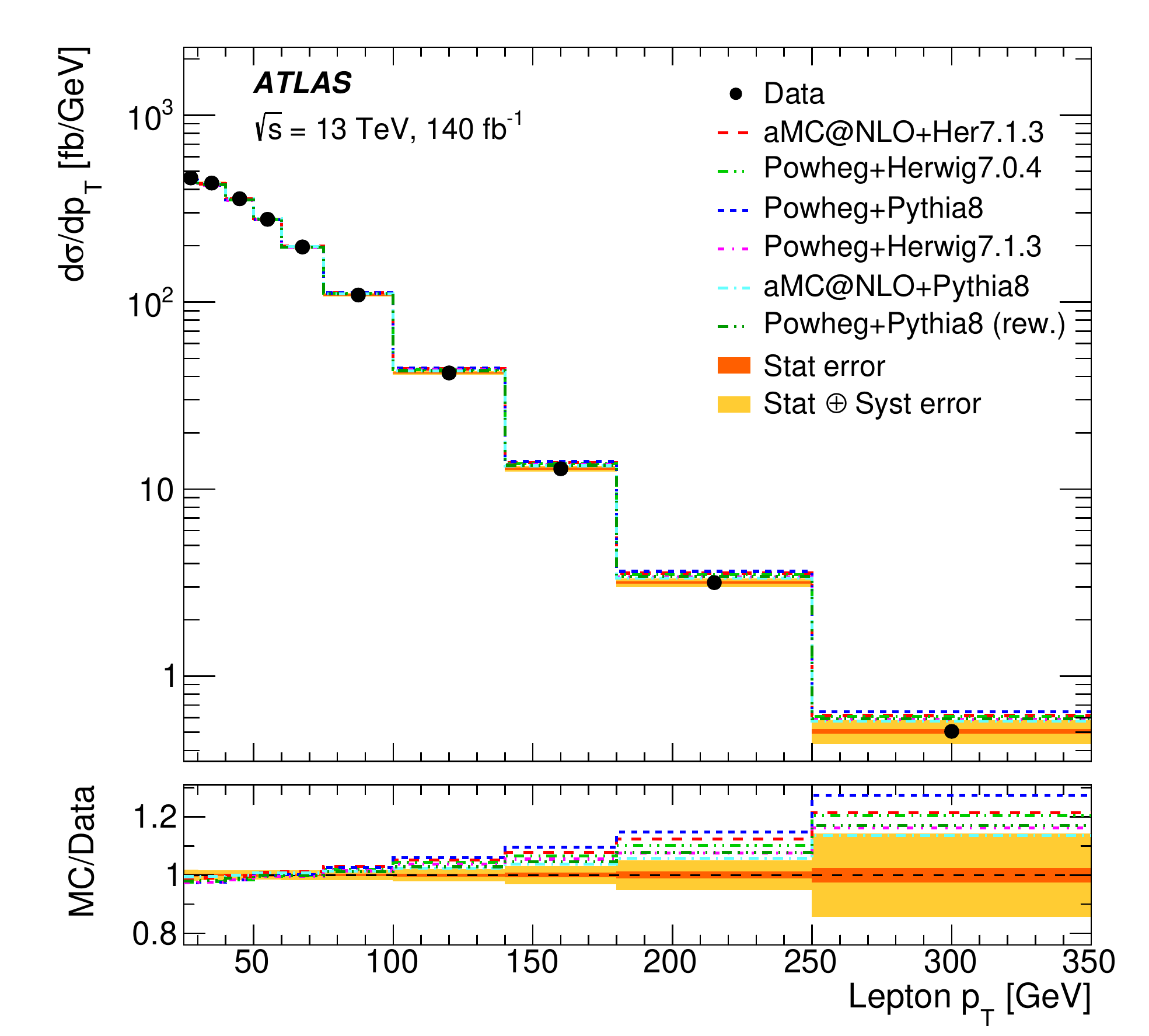}}
\subfloat[]{\label{fig:abs_xs_eta} \includegraphics[width=0.49\linewidth]{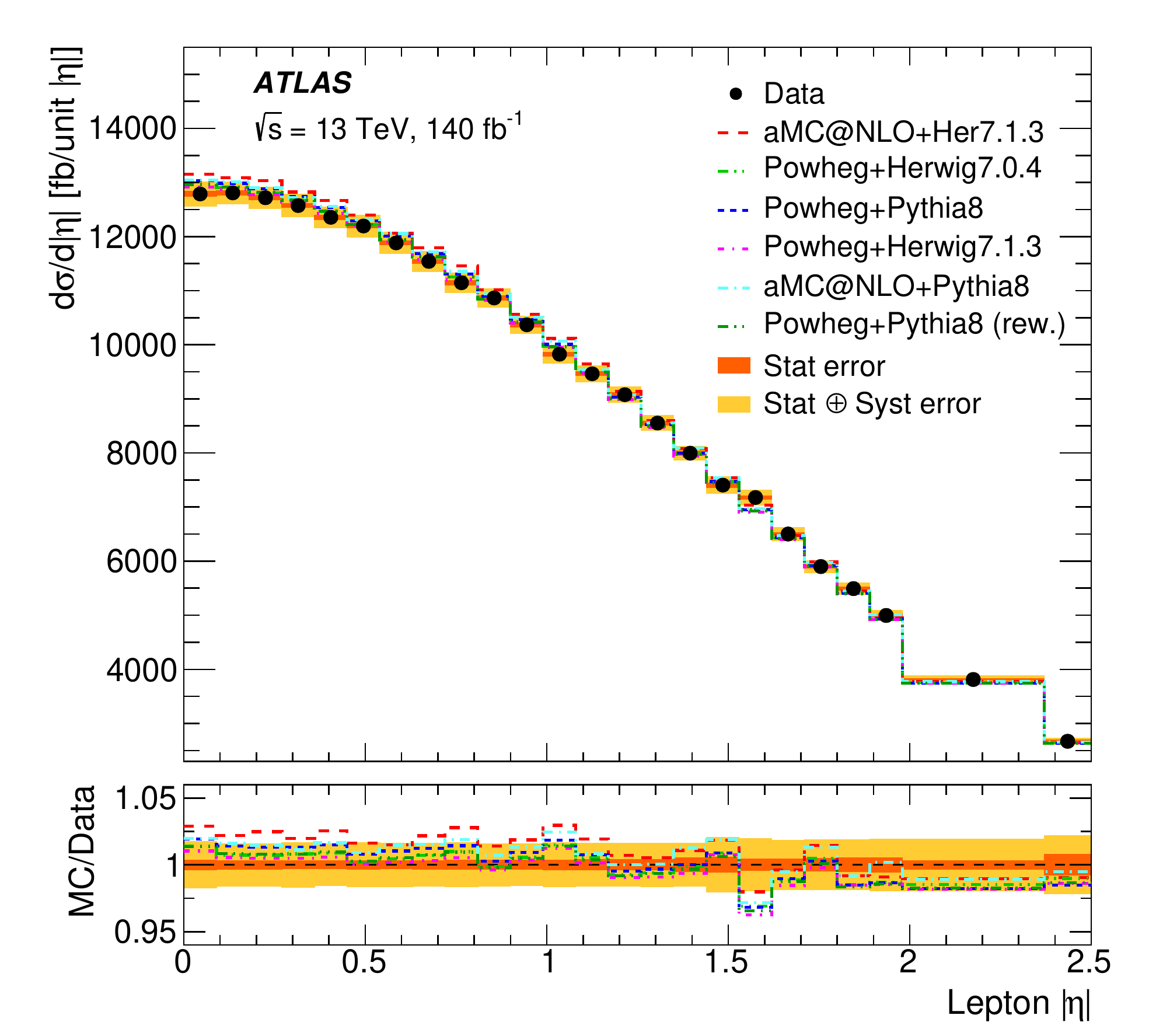}} \\
\subfloat[]{\label{fig:abs_xs_sumE} \includegraphics[width=0.49\linewidth]{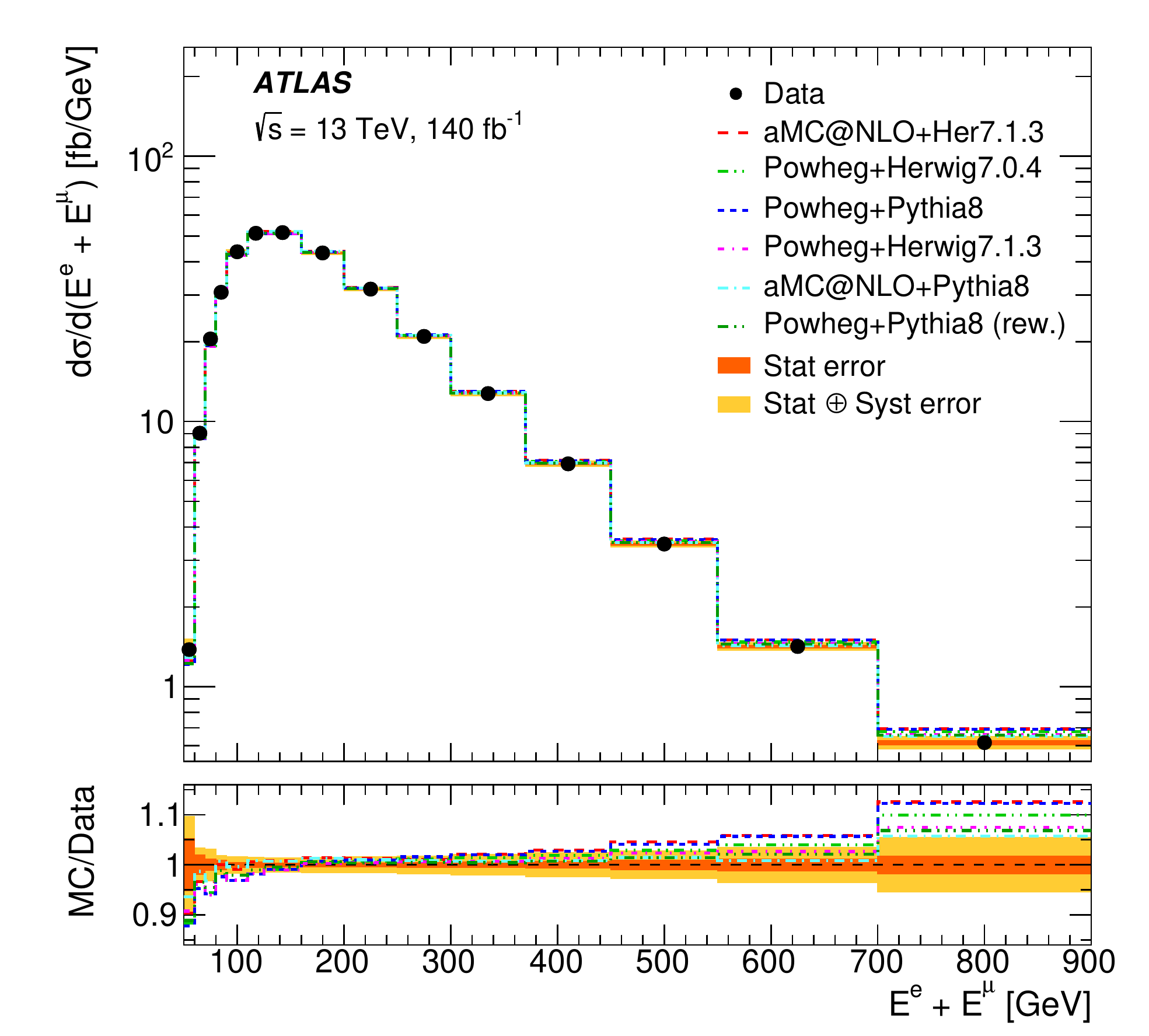}}
\subfloat[]{\label{fig:abs_xs_combinedM} \includegraphics[width=0.49\linewidth]{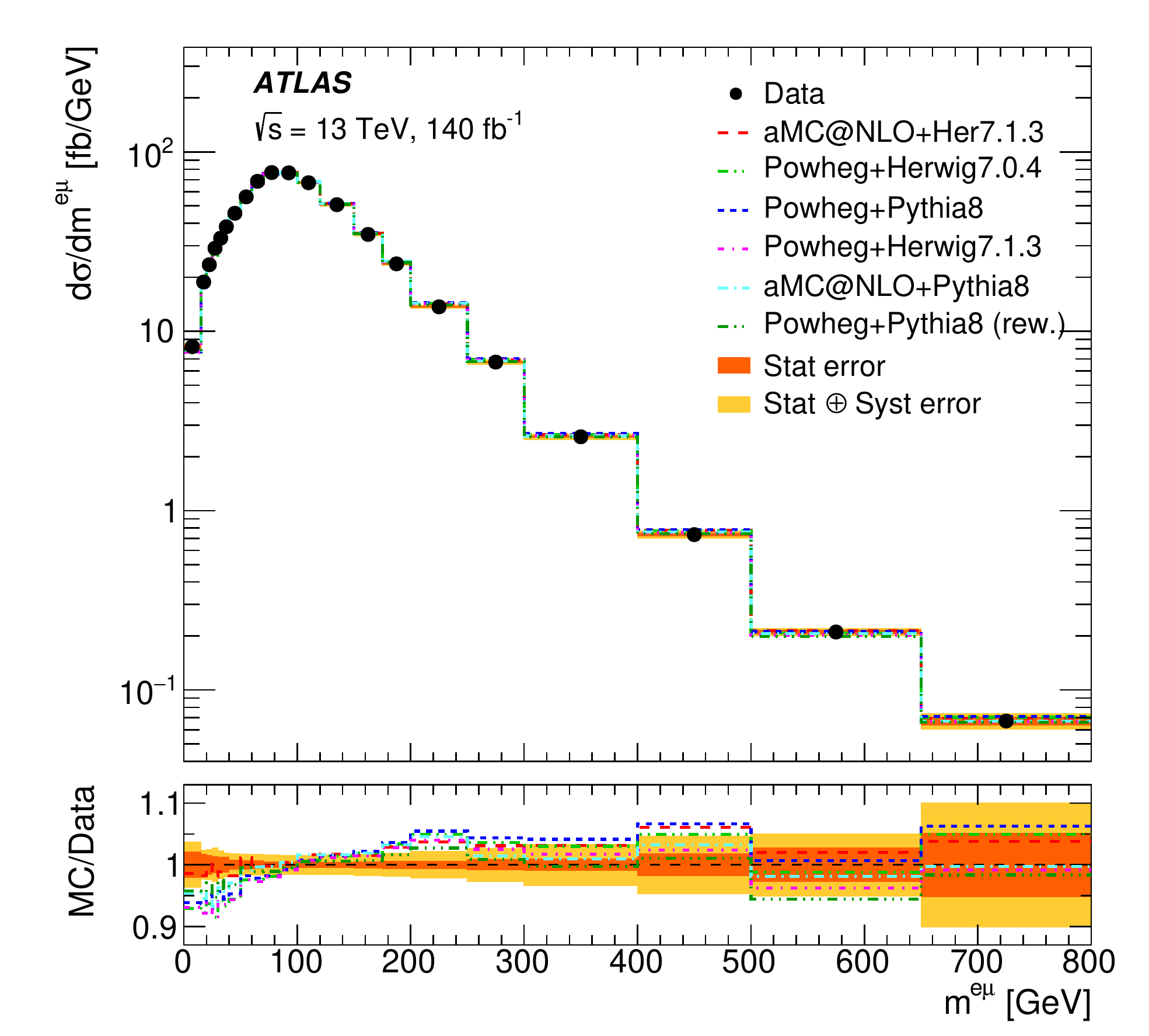}} \\
\caption[Absolute differential cross-sections with statistical and systematic uncertainties]{Absolute differential cross-sections as a function of (a) $\pT^{\ell}$, (b) $|\eta^{\ell}|$, (c) $E^e + E^{\mu}$ and (d) $m^{e\mu}$ with statistical (orange) and statistical plus systematic uncertainties (yellow). The data points are placed at the centre of each bin. The results are compared with the predictions from different Monte Carlo generators normalised to the \TOPpp prediction: the baseline \POWPY[8.230] \ttbar sample (blue), \MGNLOHER[7.1.3] (red), \POWHER[7.0.4] (green), \POWHER[7.1.3] (purple), \MGNLOPY[8.230] (cyan) and \POWPY[8.230] rew.\ (dark green), which refers to \POWPY[8.230] reweighted according to the top-quark \pT. The lower panel shows the ratios of the predictions to data, with the bands indicating the statistical and systematic uncertainties. The last bin in (a), (c) and (d) also contains overflow events.}
\label{fig:xs_abs_predictions}
\end{figure}

\begin{figure}[htbp]
\centering
\subfloat[]{\label{fig:double_abs_xs_double_YvsM} \includegraphics[width=0.85\linewidth]{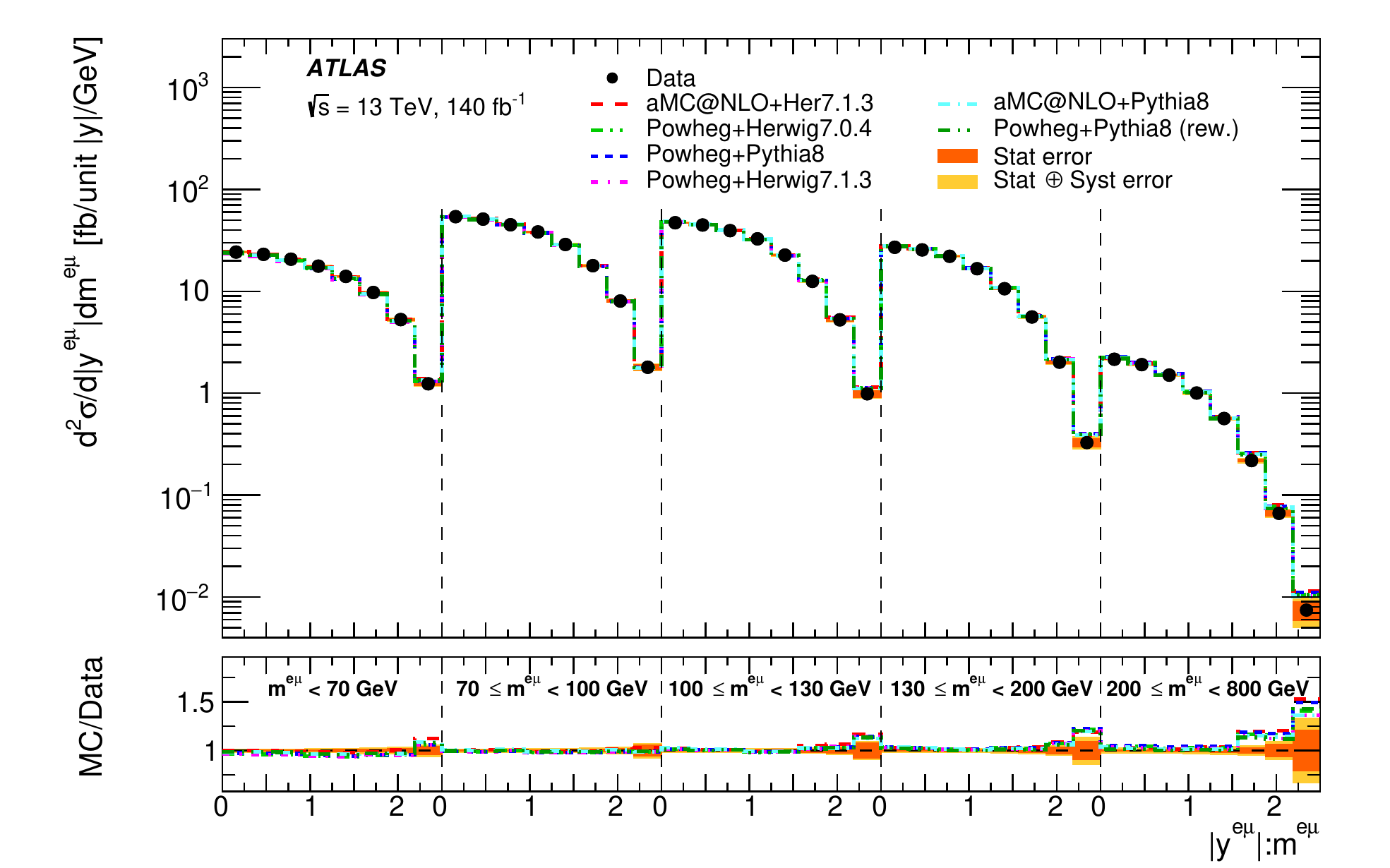}} \\
\subfloat[]{\label{fig:double_abs_xs_double_PhivsM} \includegraphics[width=0.85\linewidth]{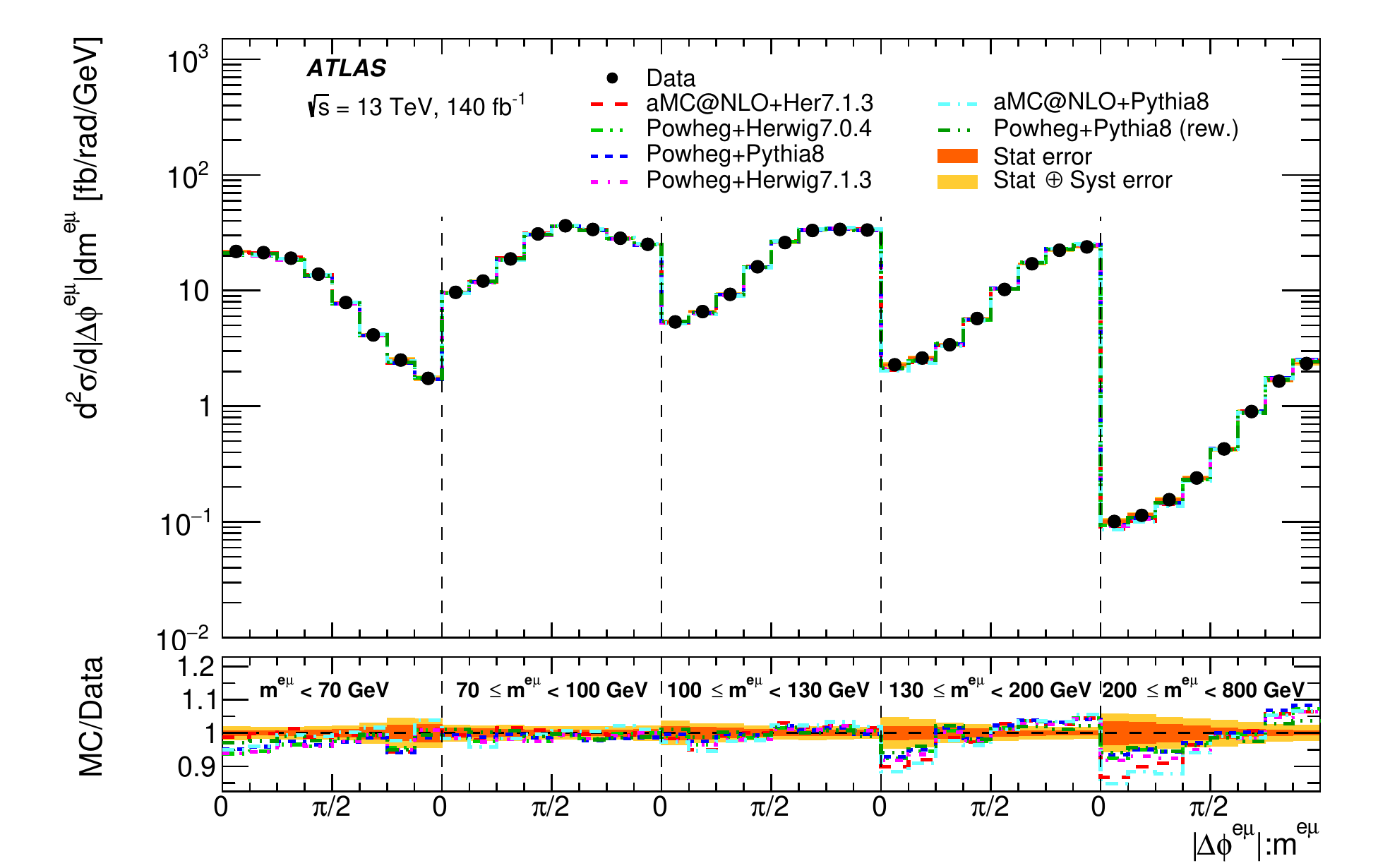}}
\caption[Absolute double-differential cross-sections with statistical and systematic uncertainties]{Absolute double-differential cross-sections as a function of (a) $|y^{e\mu}|$ in bins of $m^{e\mu}$ and (b) $|\Delta \phi^{e\mu}|$ in bins of $m^{e\mu}$ with statistical (orange) and statistical plus systematic uncertainties (yellow). The data points are placed at the centre of each bin. The results are compared with the predictions from different Monte Carlo generators normalised to the \TOPpp prediction: the baseline \POWPY[8.230] \ttbar sample (blue), \MGNLOHER[7.1.3] (red), \POWHER[7.0.4] (green), \POWHER[7.1.3] (purple), \MGNLOPY[8.230] (cyan) and \POWPY[8.230] rew.\ (dark green), which refers to \POWPY[8.230] reweighted according to the top-quark \pT. The lower panel shows the ratios of the predictions to data, with the bands indicating the statistical and systematic uncertainties. The highest invariant mass bin of the two distributions contains the overflows.}
\label{fig:double_xs_abs_predictions1}
\end{figure}

\begin{figure}[htbp]
\centering
\subfloat[]{\label{fig:double_abs_xs_double_PhivspTll} \includegraphics[width=0.85\linewidth]{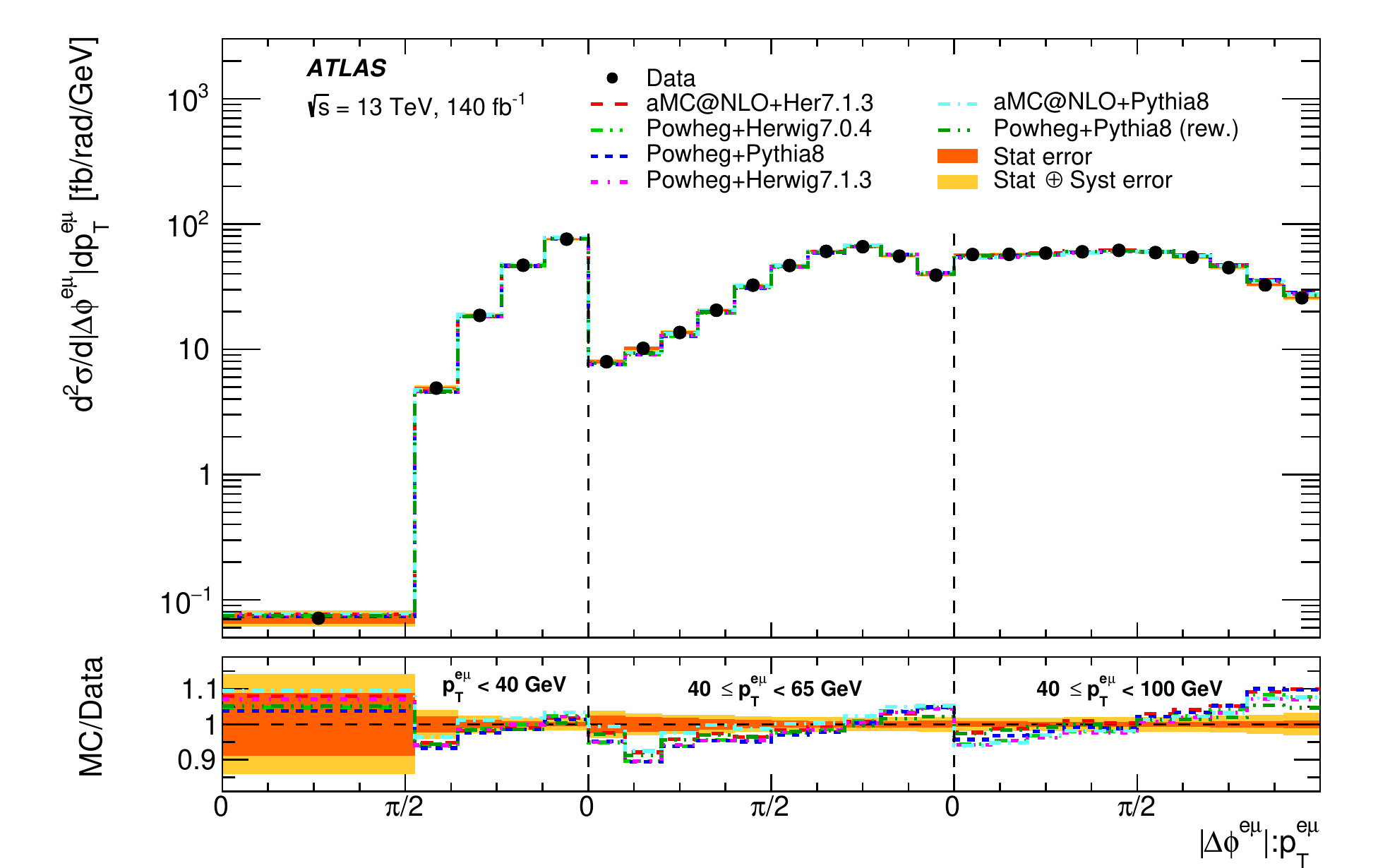}} \\
\subfloat[]{\label{fig:double_abs_xs_double_PhivsE+E} \includegraphics[width=0.85\linewidth]{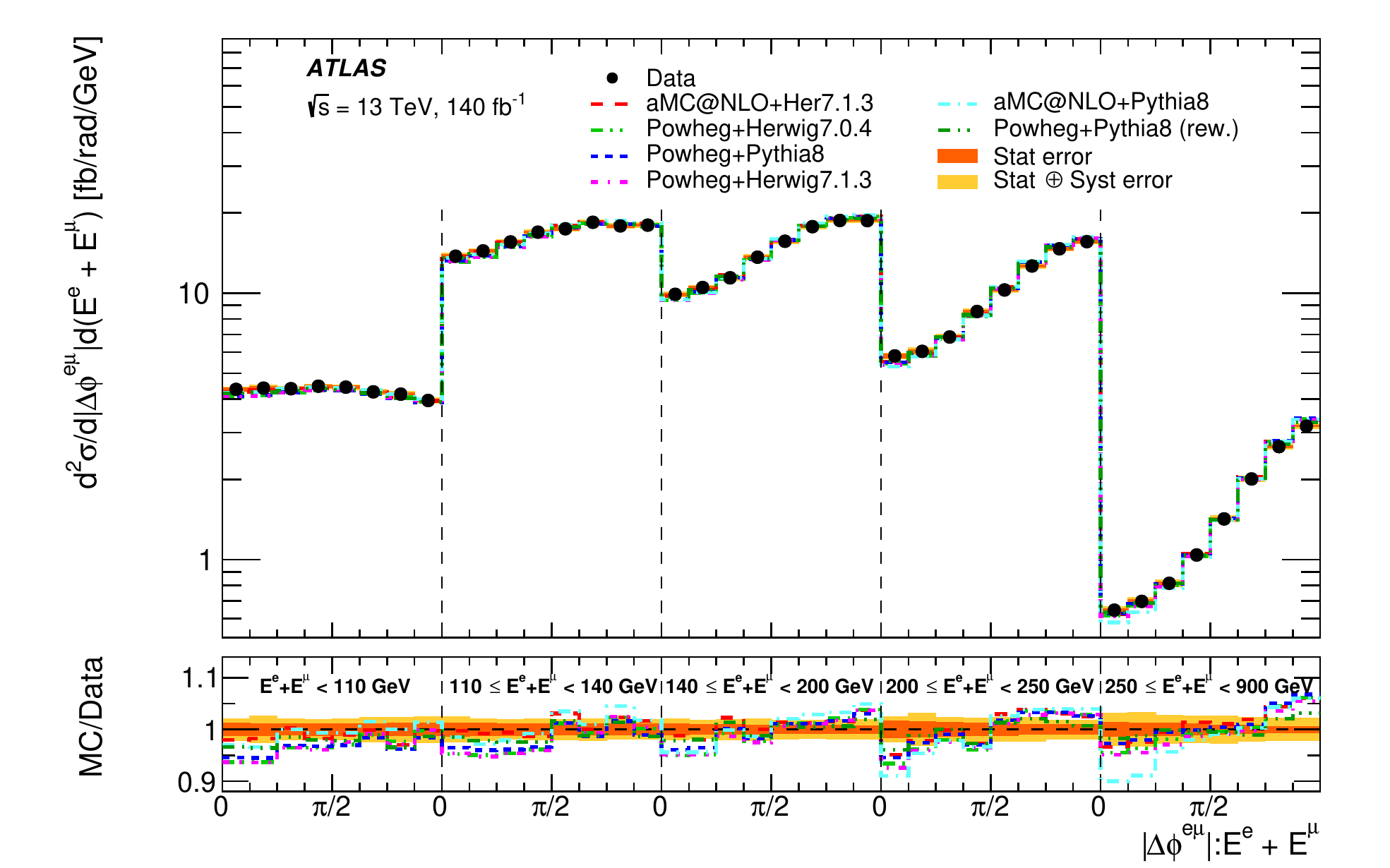}}
\caption[Absolute double-differential cross-sections with statistical and systematic uncertainties]{Absolute double-differential cross-sections as a function of (a) $|\Delta \phi^{e\mu}|$ in bins of $\pT^{e\mu}$ and (b) $|\Delta \phi^{e\mu}|$ in bins of $E^e + E^{\mu}$ with statistical (orange) and statistical plus systematic uncertainties (yellow). The data points are placed at the centre of each bin. The results are compared with the predictions from different Monte Carlo generators normalised to the \TOPpp prediction: the baseline \POWPY[8.230] \ttbar sample (blue), \MGNLOHER[7.1.3] (red), \POWHER[7.0.4] (green), \POWHER[7.1.3] (purple), \MGNLOPY[8.230] (cyan) and \POWPY[8.230] rew.\ (dark green), which refers to \POWPY[8.230] reweighted according to the top-quark \pT. The lower panel shows the ratios of the predictions to data, with the bands indicating the statistical and systematic uncertainties. The highest $E^{e}+E^{\mu}$ and \pT bin contains the overflows.}
\label{fig:double_xs_abs_predictions2}
\end{figure}

The differential cross-section is measured as a function of several lepton kinematic variables: $\pT^{\ell}$, $|\eta^{\ell}|$, $m^{e\mu}$, $\pT^{e\mu}$, $|y^{e\mu}|$, $E^e + E^{\mu}$, $\pT^{e}+\pT^{\mu}$ and $|\Delta \phi^{e\mu}|$.
The absolute differential cross-sections in the fiducial region are presented in Figures~\ref{fig:xs_abs_predictions} and \ref{fig:xs_abs_predictions2}, as well as in Tables~\ref{tab:abs_diff_xs_pTell}--\ref{tab:abs_diff_xs_y} in the Appendix. 
The absolute double-differential cross-sections are presented in Figures~\ref{fig:double_xs_abs_predictions1} and \ref{fig:double_xs_abs_predictions2}, as well as in Tables~\ref{tab:double_diff_xs_YvsM}--\ref{tab:double_diff_xs_PhivsE} in the Appendix.  
The luminosity gives the largest contribution to the cross-section uncertainty in most bins, resulting in a typical uncertainty of 1\% out of a total uncertainty around 2\%. An uncertainty of around 1\% is expected since it affects both the signal and background yields in Equation~\ref{eq:double_tagging} and the final impact on the measured cross-section is subject to background fluctuations, having a relative impact on the measurement ranging between 0.9\% and 1.1\%.
However, uncertainties related to the modelling of the \ttbar process, those affecting the reconstruction of leptons and those affecting the modelling of the background processes also have a significant contribution in all bins of the distributions.
The statistical uncertainty increases with increasing transverse momentum, combined mass or energy, but is overtaken by the uncertainty related to the interference between \ttbar and $\Wboson t$ amplitudes that dominates the uncertainty in the high mass or energy bins.
 
\begin{figure}[htbp]
\centering
\subfloat[]{\label{fig:norm_xs_pT} \includegraphics[width=0.49\linewidth]{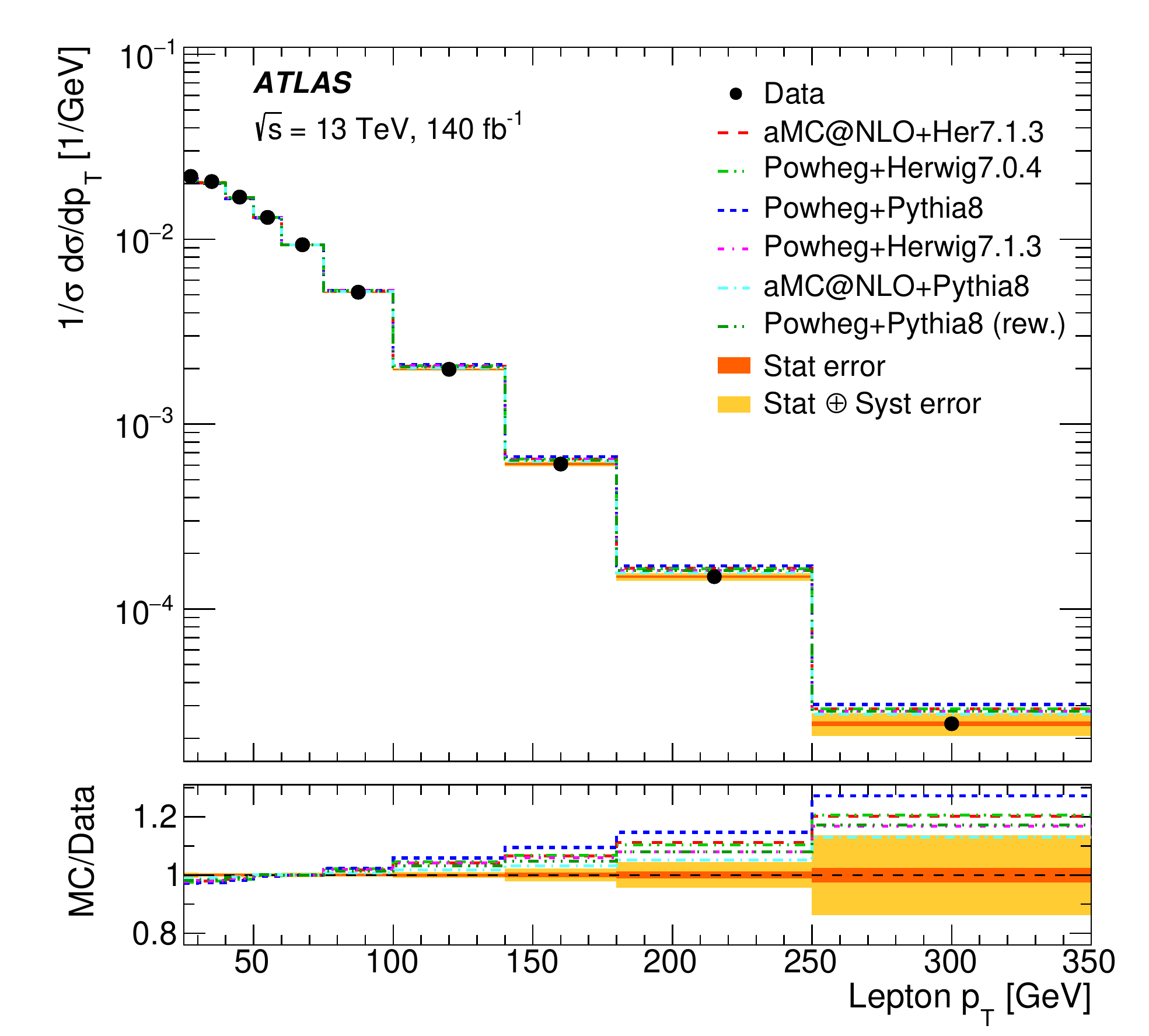}}
\subfloat[]{\label{fig:norm_xs_eta} \includegraphics[width=0.49\linewidth]{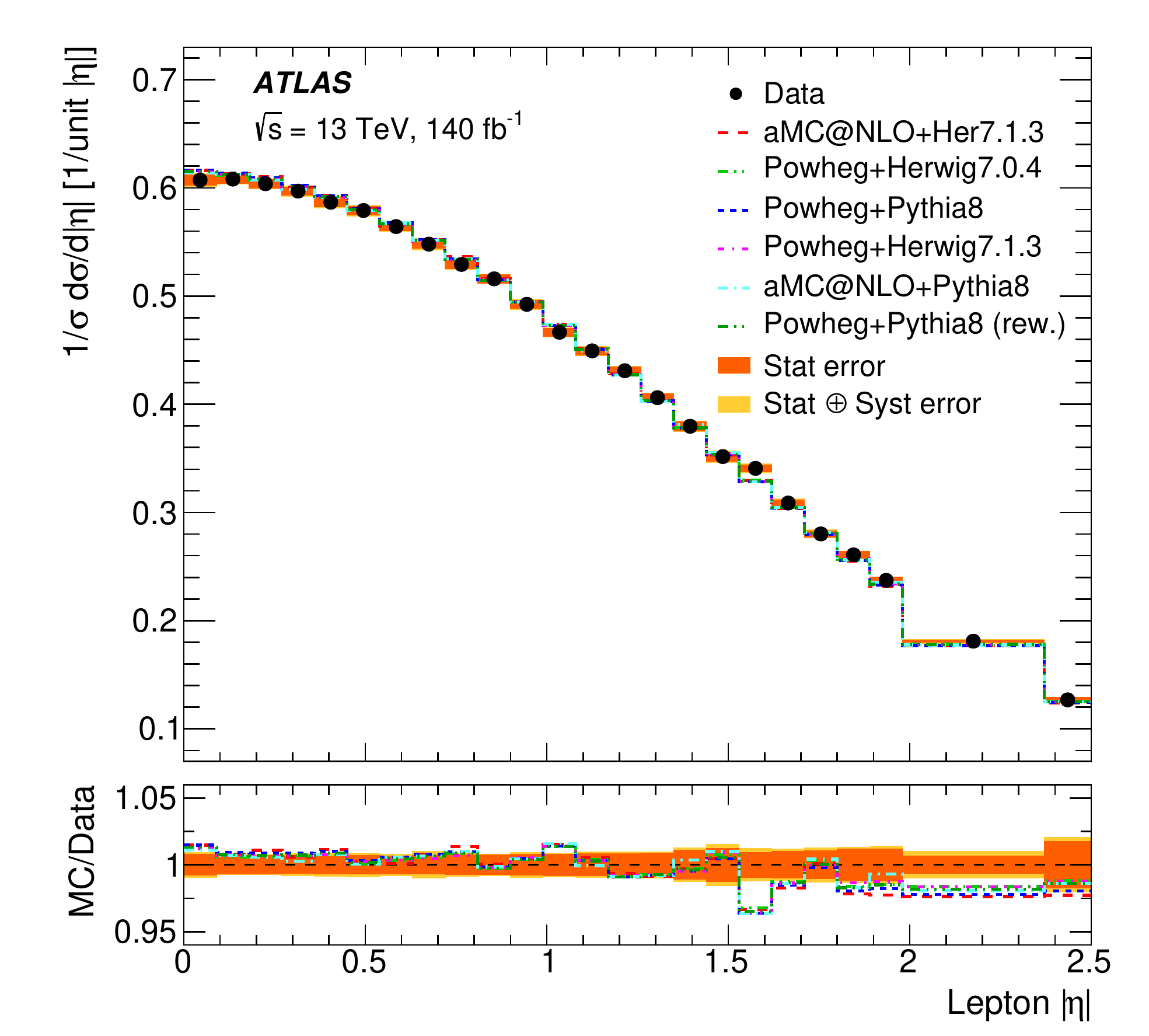}} \\
\subfloat[]{\label{fig:norm_xs_sumE} \includegraphics[width=0.49\linewidth]{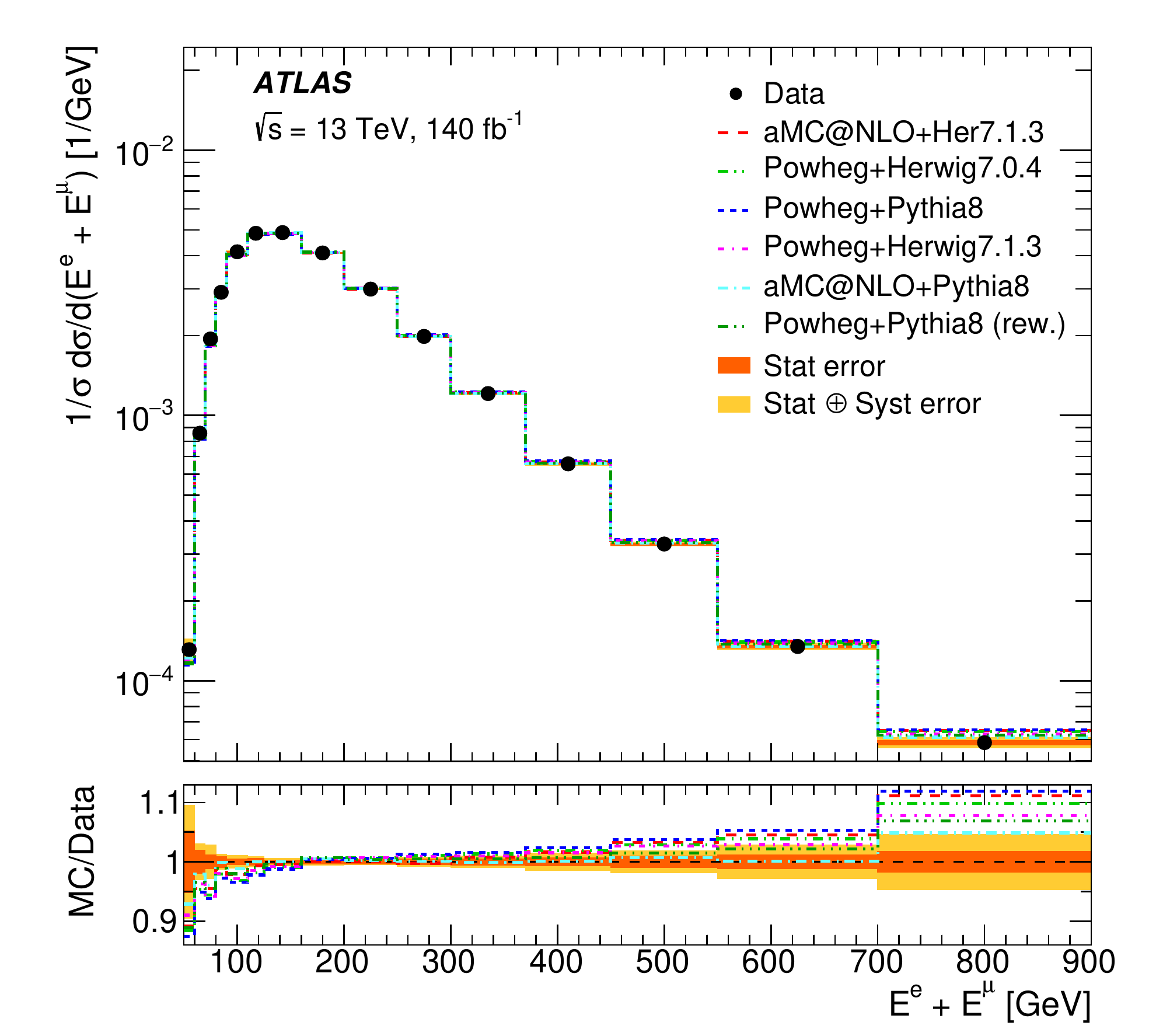}}
\subfloat[]{\label{fig:norm_xs_combinedM} \includegraphics[width=0.49\linewidth]{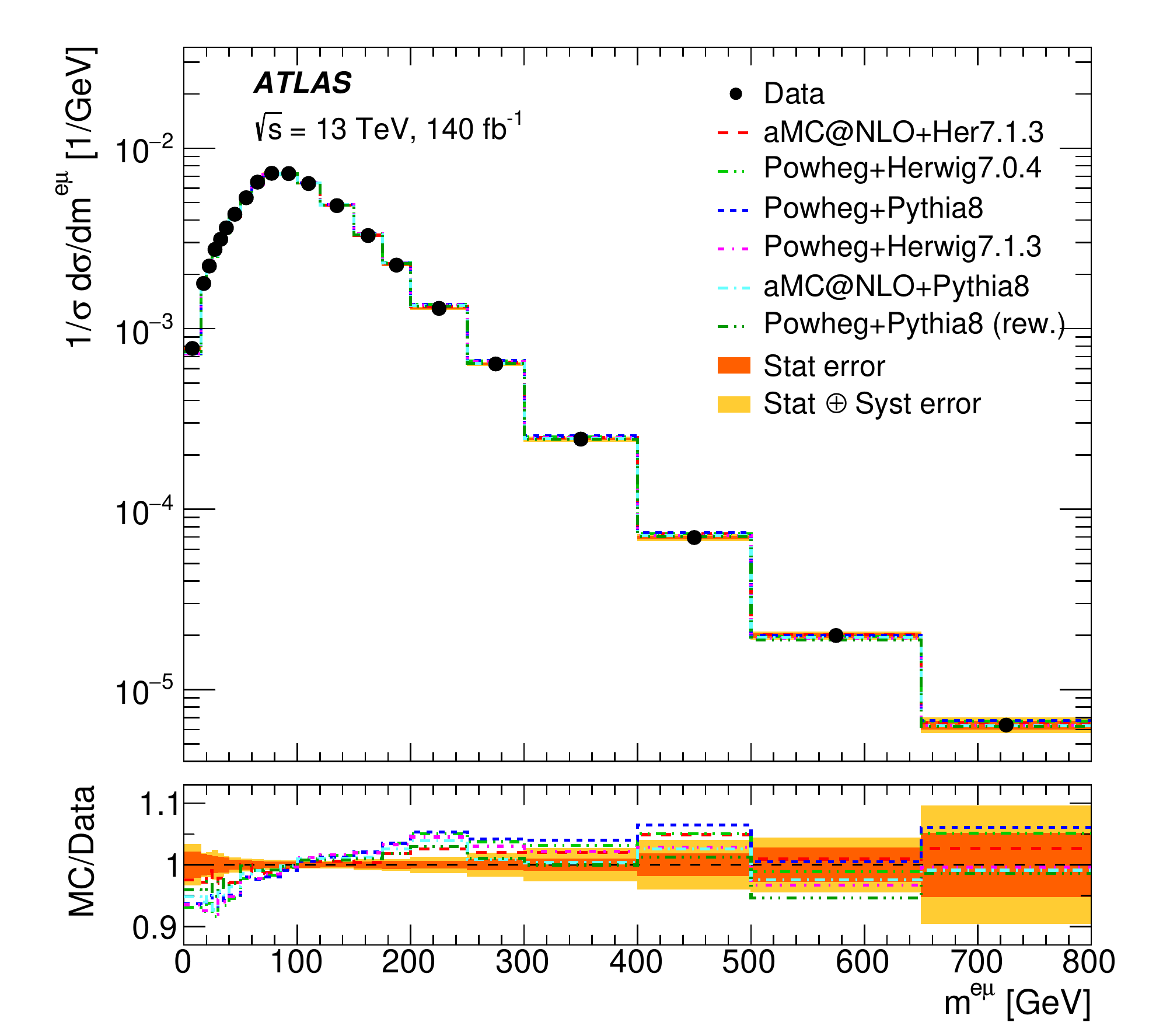}} \\
\caption[Normalised differential cross-sections with statistical and systematic uncertainties]{Normalised differential cross-sections as a function of (a) $\pT^{\ell}$, (b) $|\eta^{\ell}|$, (c) $E^e + E^{\mu}$ and (d) $m^{e\mu}$ with statistical (orange) and statistical plus systematic uncertainties (yellow). The data points are placed at the centre of each bin. The results are compared with the predictions from different Monte Carlo generators normalised to the \TOPpp prediction: the baseline \POWPY[8.230] \ttbar sample (blue), \MGNLOHER[7.1.3] (red), \POWHER[7.0.4] (green), \POWHER[7.1.3] (purple), \MGNLOPY[8.230] (cyan) and \POWPY[8.230] rew.\ (dark green), which refers to \POWPY[8.230] reweighted according to the top-quark \pT. The lower panel shows the ratios of the predictions to data, with the bands indicating the statistical and systematical uncertainties. The last bin in (a), (c) and (d) also contains overflow events.}
\label{fig:xs_norm_predictions}
\end{figure}

\begin{figure}[htbp]
\centering
\subfloat[]{\label{fig:norm_xs_sumpT} \includegraphics[width=0.49\linewidth]{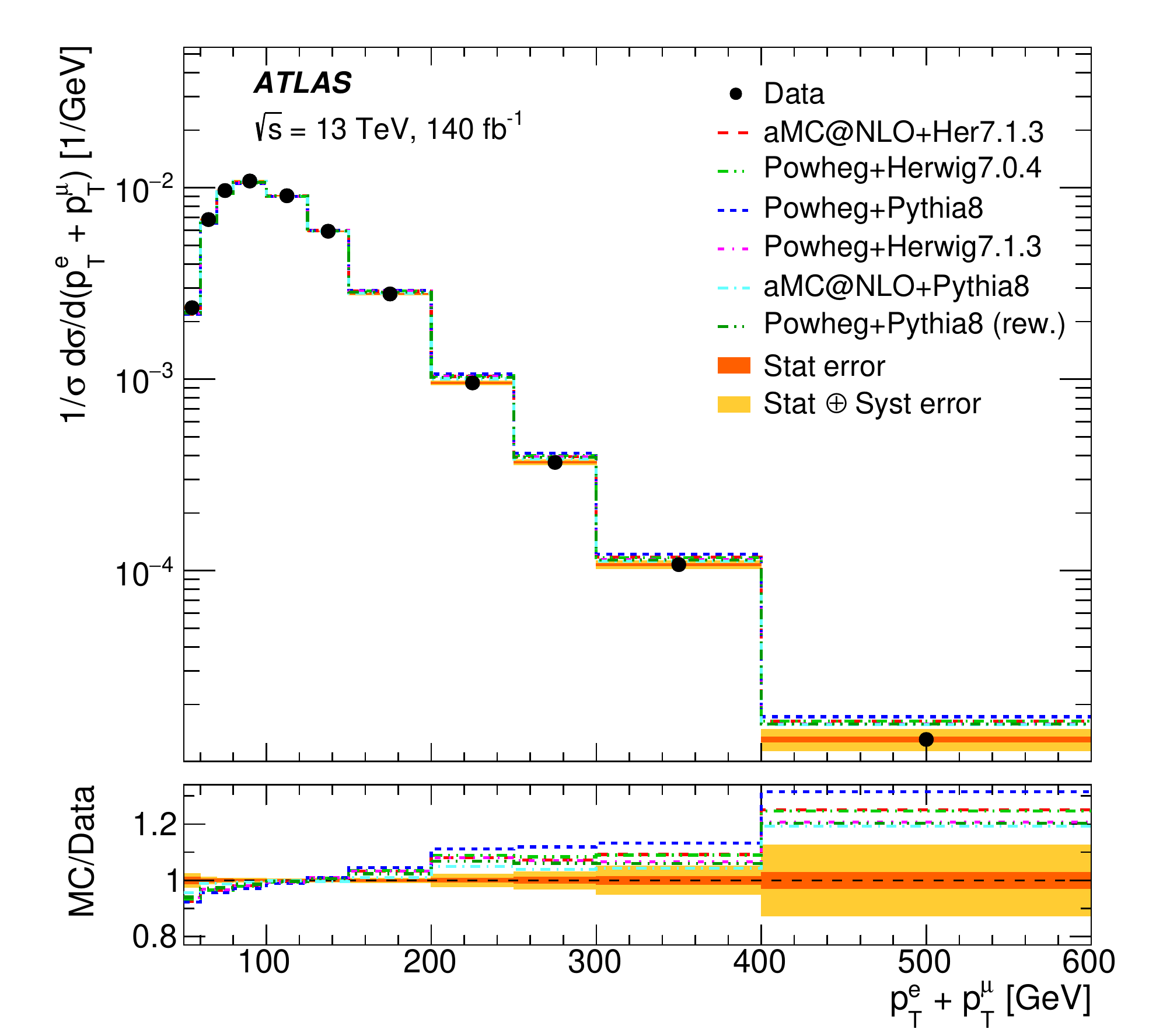}}
\subfloat[]{\label{fig:norm_xs_combinedpT} \includegraphics[width=0.49\linewidth]{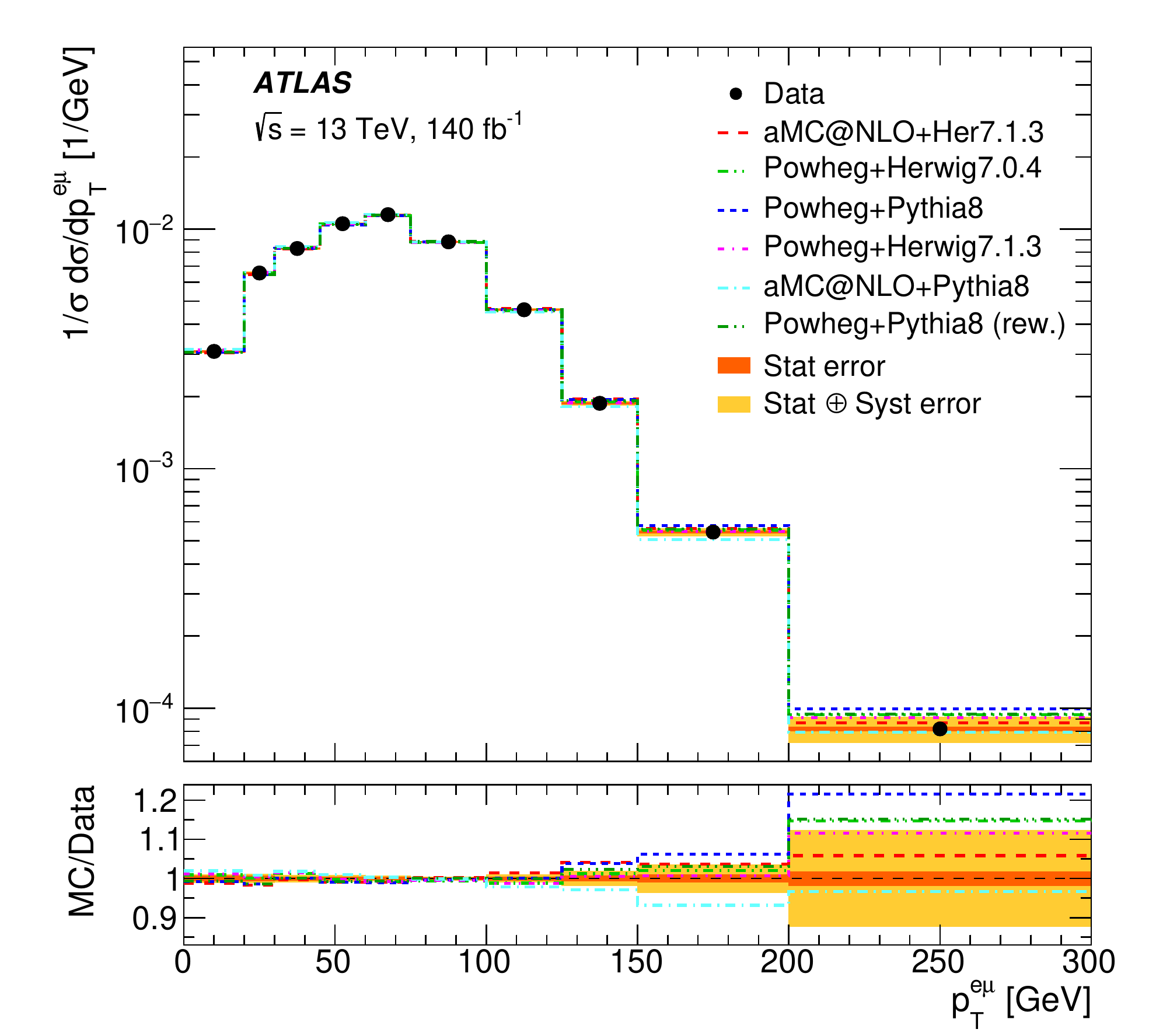}} \\
\subfloat[]{\label{fig:norm_xs_phi} \includegraphics[width=0.49\linewidth]{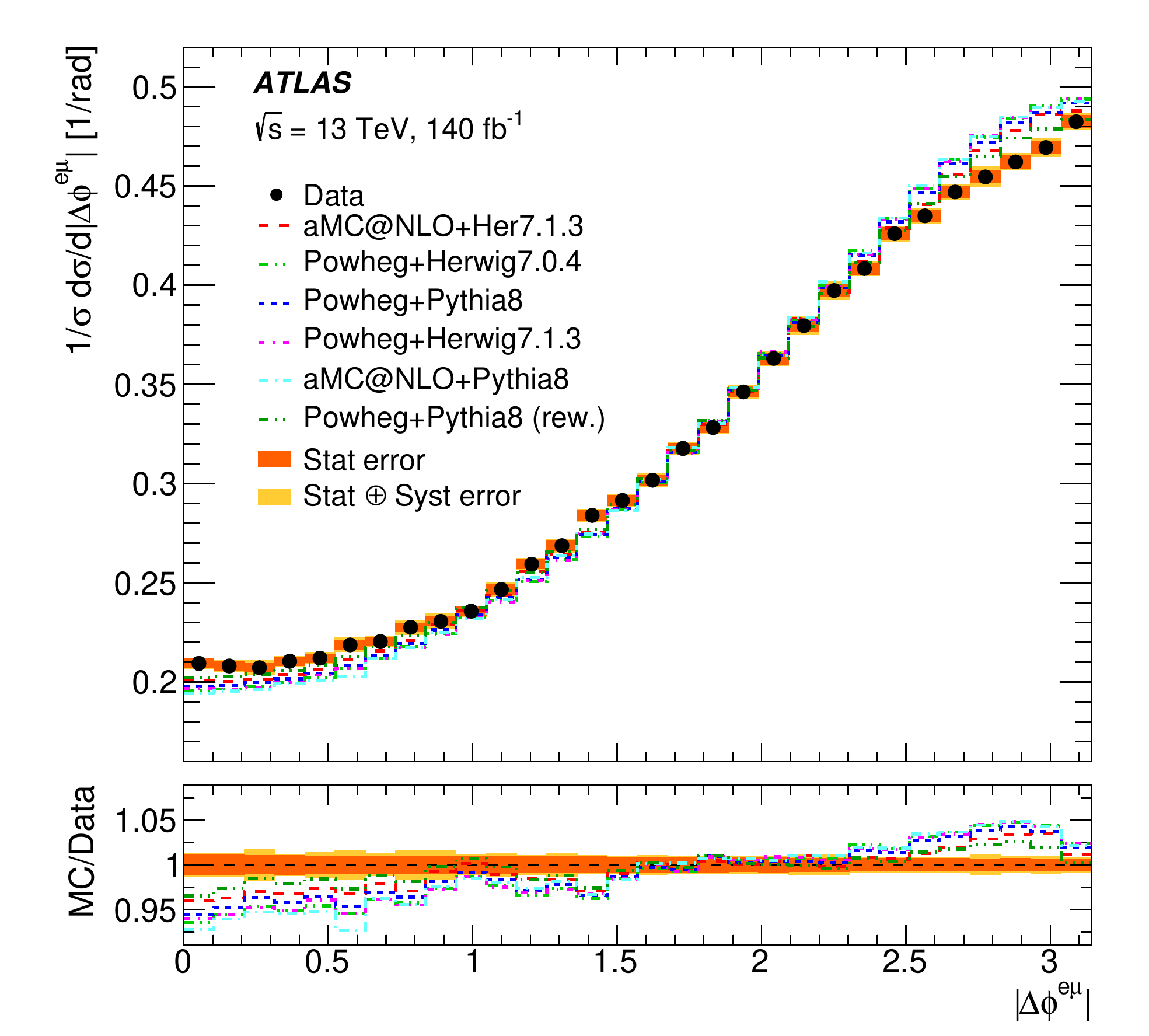}}
\subfloat[]{\label{fig:norm_xs_y} \includegraphics[width=0.49\linewidth]{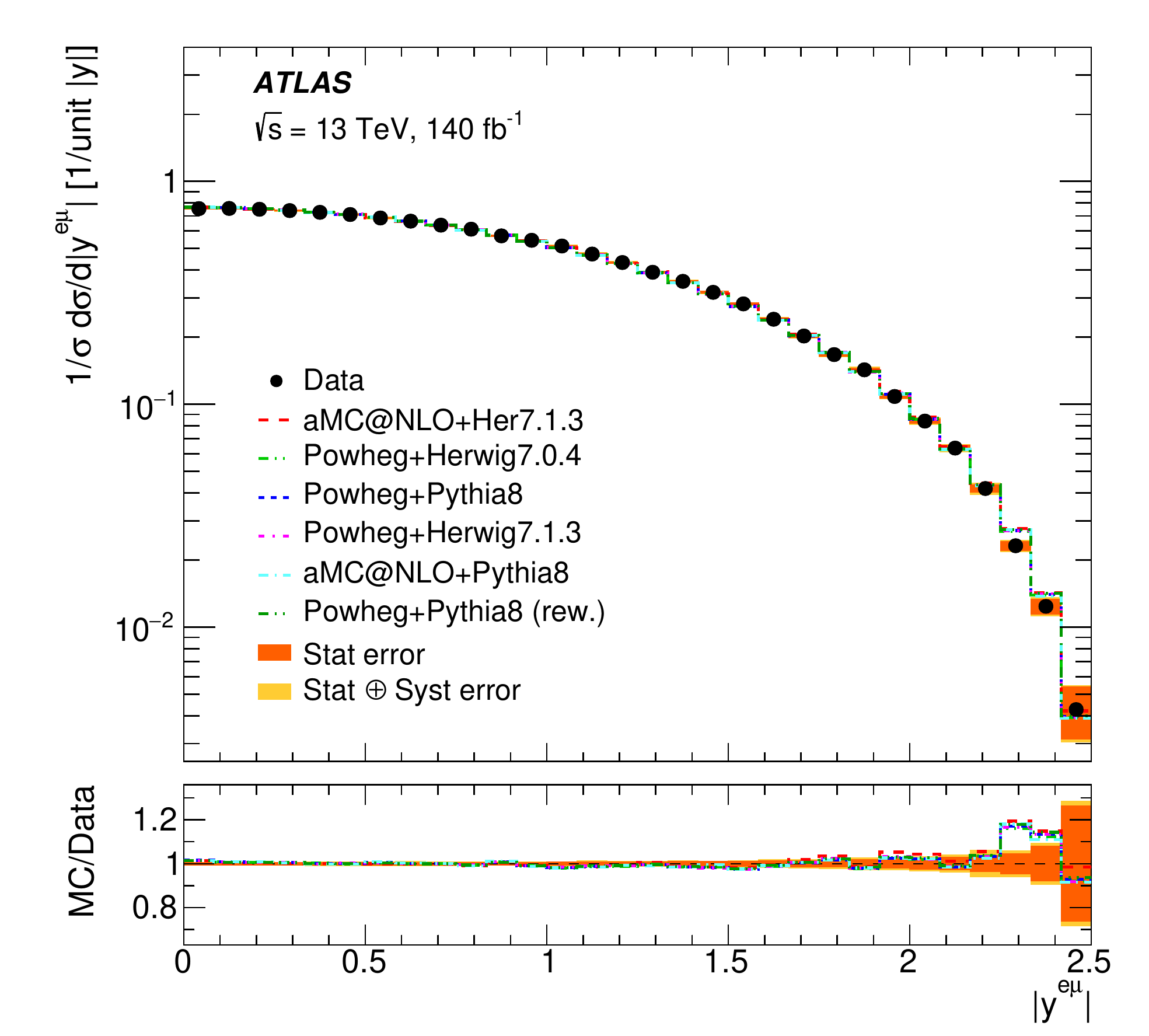}} \\
\caption[Normalised differential cross-sections with statistical and systematic uncertainties]{Normalised differential cross-sections as a function of (a) $\pT^{e}+\pT^{\mu}$, (b) $\pT^{e\mu}$, (c) $|\Delta \phi^{e\mu}|$ and (d) $|y^{e\mu}|$ with statistical (orange) and statistical plus systematic uncertainties (yellow). The data points are placed at the centre of each bin. The results are compared with the predictions from different Monte Carlo generators normalised to the \TOPpp prediction: the baseline \POWPY[8.230] \ttbar sample (blue), \MGNLOHER[7.1.3] (red), \POWHER[7.0.4] (green), \POWHER[7.1.3] (purple), \MGNLOPY[8.230] (cyan) and \POWPY[8.230] rew.\ (dark green), which refers to \POWPY[8.230] reweighted according to the top-quark \pT. The lower panel shows the ratios of the predictions to data, with the bands indicating the statistical and systematic uncertainties. The last bin in (a) and (b) also contains overflow events.}
\label{fig:xs_norm_predictions2}
\end{figure}

However, in the normalised differential cross-section the uncertainty due to the luminosity largely cancels out. This results in a typical uncertainty
of 1\%, except in the highest energy bins.
These results are presented
in Figures~\ref{fig:xs_norm_predictions} and \ref{fig:xs_norm_predictions2}, as well as in Tables~\ref{tab:norm_diff_xs_pTell}--\ref{tab:norm_diff_xs_y} in the Appendix. 
Instead of the luminosity, the interference between \ttbar and $\Wboson t$ amplitudes becomes the most important source of uncertainty,
especially for high values of
variables with dimensions of energy.
The \ttbar modelling uncertainties are also important, while other uncertainties are very small in the normalised distributions.

Due to differences in the fiducial region definition, these results are not directly comparable to the previous results from Ref.~\cite{TOPQ-2018-17}.
The lepton \pT requirement in this analysis, $\pT^{\ell}>27\ (25)$~\GeV \ for the leading (sub-leading) lepton, differs from that in the 36 fb$^{-1}$ analysis~\cite{TOPQ-2018-17}, in which the minimum lepton \pT was $20$~\GeV\ whilst requiring at least one lepton to be above the lepton trigger threshold of $21-27$ \GeV.

\begin{figure}[htbp]
\centering
\subfloat[]{\label{fig:double_norm_xs_double_YvsM} \includegraphics[width=0.85\linewidth]{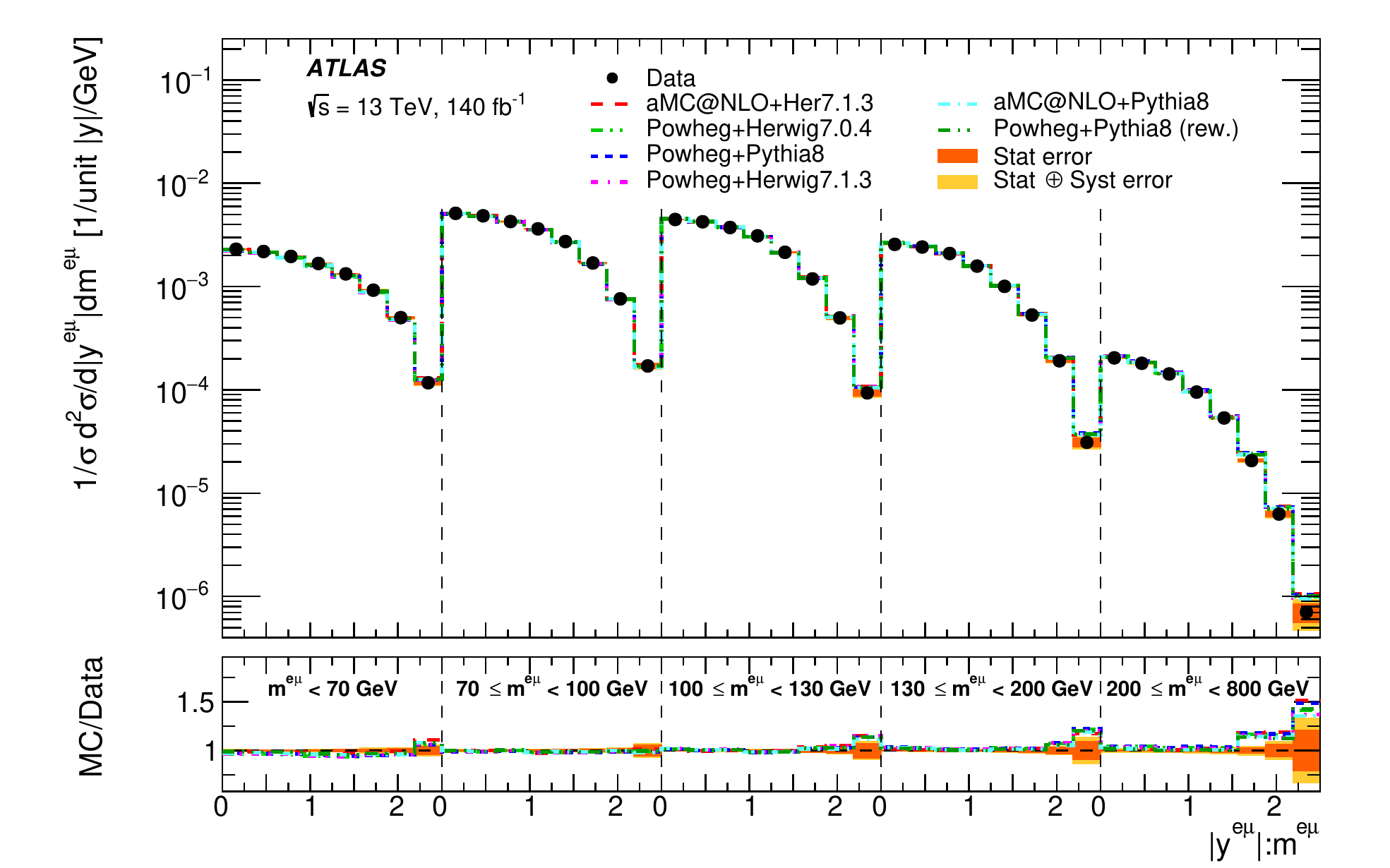}} \\
\subfloat[]{\label{fig:double_norm_xs_double_PhivsM} \includegraphics[width=0.85\linewidth]{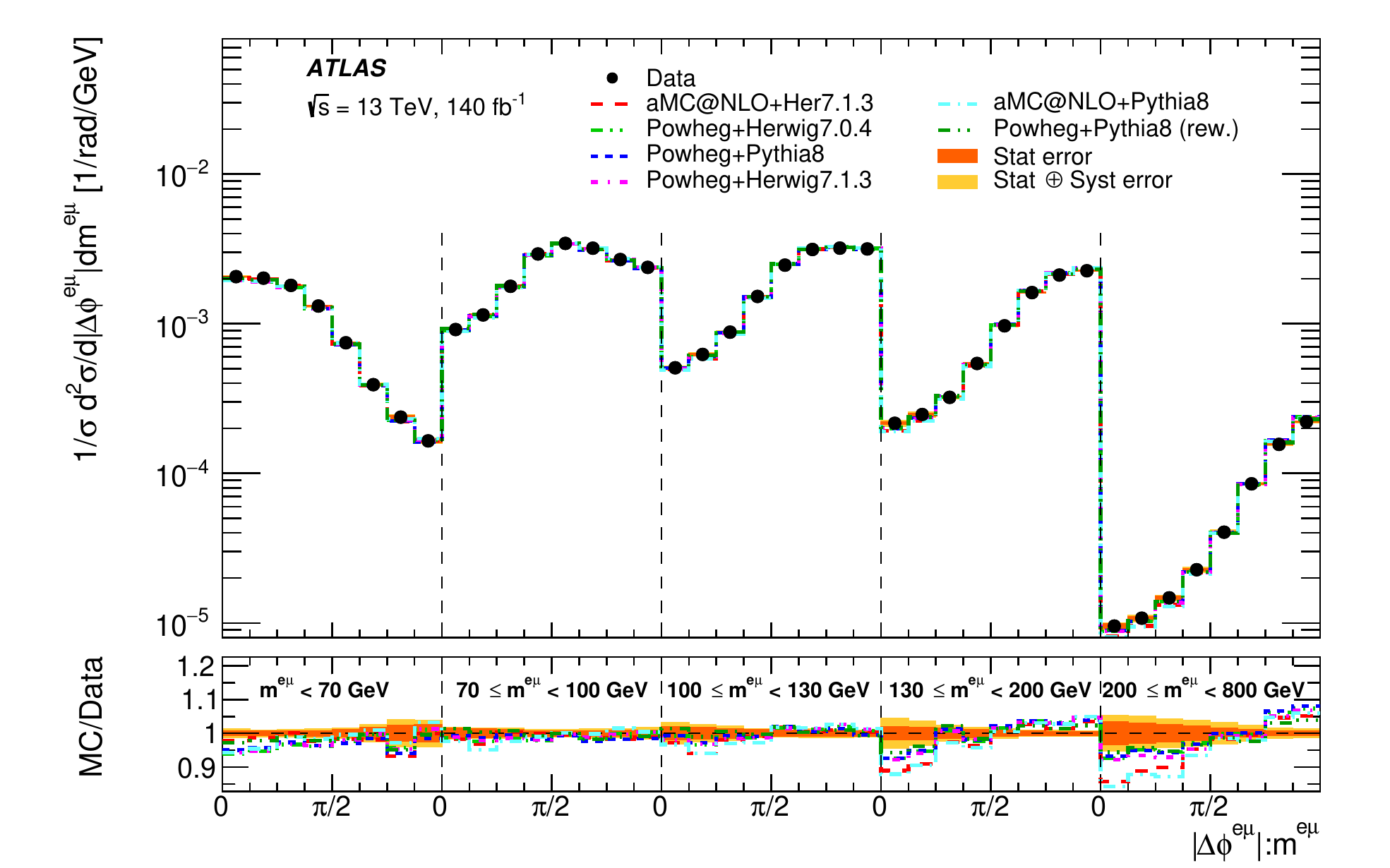}}
\caption[Normalised double-differential cross-sections with statistical and systematic uncertainties]{Normalised double-differential cross-sections as a function of (a) $|y^{e\mu}|$ in bins of $m^{e\mu}$ and (b) $|\Delta \phi^{e\mu}|$ in bins of $m^{e\mu}$ with statistical (orange) and statistical plus systematic uncertainties (yellow). The data points are placed at the centre of each bin. The results are compared with the predictions from different Monte Carlo generators normalised to the \TOPpp prediction: the baseline \POWPY[8.230] \ttbar sample (blue), \MGNLOHER[7.1.3] (red), \POWHER[7.0.4] (green), \POWHER[7.1.3] (purple), \MGNLOPY[8.230] (cyan) and \POWPY[8.230] rew.\ (dark green), which refers to \POWPY[8.230] reweighted according to the top-quark \pT. The lower panel shows the ratios of the predictions to data, with the bands indicating the statistical and systematic uncertainties. The highest invariant mass bin of the two distributions contains the overflows.}
\label{fig:double_xs_norm_predictions1}
\end{figure}

\begin{figure}[htbp]
\centering
\subfloat[]{\label{fig:double_norm_xs_double_PhivspTll} \includegraphics[width=0.85\linewidth]{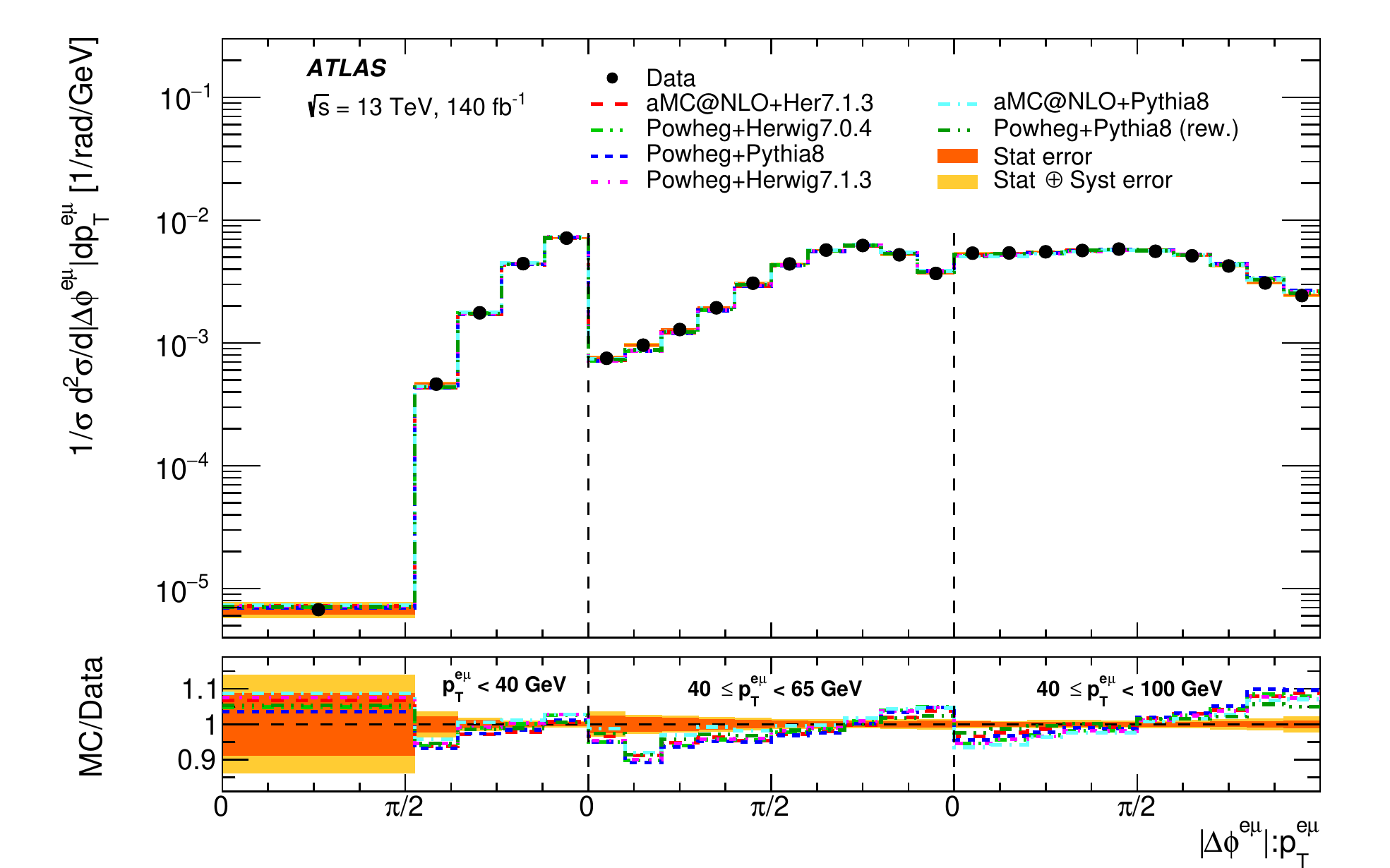}} \\
\subfloat[]{\label{fig:double_norm_xs_double_PhivsE+E} \includegraphics[width=0.85\linewidth]{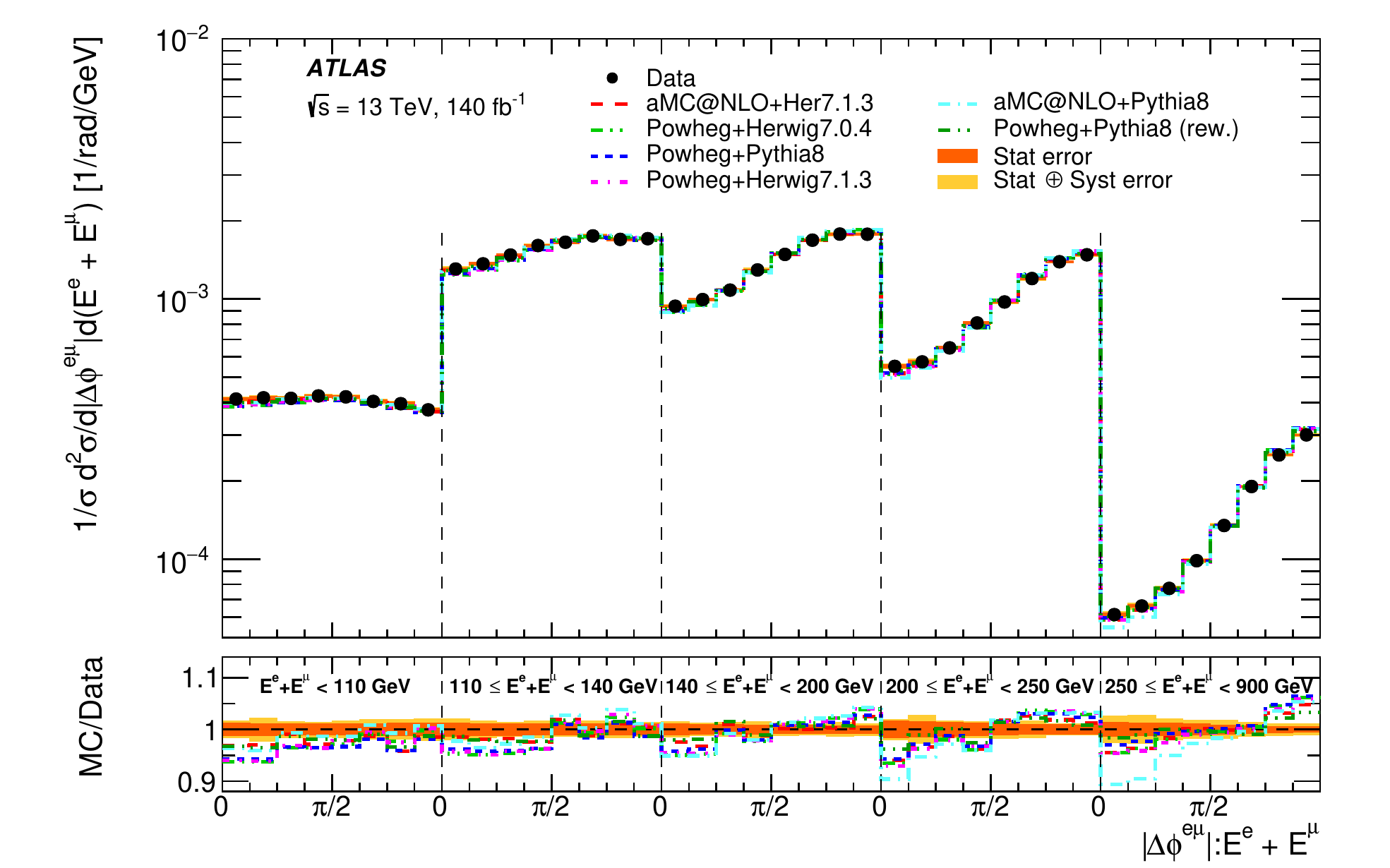}}
\caption[Normalised double-differential cross-sections with statistical and systematic uncertainties]{Normalised double-differential cross-sections as a function of (a) $|\Delta \phi^{e\mu}|$ in bins of $\pT^{e\mu}$ and (b) $|\Delta \phi^{e\mu}|$ in bins of $E^e + E^{\mu}$ with statistical (orange) and statistical plus systematic uncertainties (yellow). The data points are placed at the centre of each bin. The results are compared with the predictions from different Monte Carlo generators normalised to the \TOPpp prediction: the baseline \POWPY[8.230] \ttbar sample (blue), \MGNLOHER[7.1.3] (red), \POWHER[7.0.4] (green), \POWHER[7.1.3] (purple), \MGNLOPY[8.230] (cyan) and \POWPY[8.230] rew.\ (dark green), which refers to \POWPY[8.230] reweighted according to the top-quark \pT. The lower panel shows the ratios of the predictions to data, with the bands indicating the statistical and systematic uncertainties. The highest $E^{e}+E^{\mu}$ and \pT bin contains the overflows.}
\label{fig:double_xs_norm_predictions2}
\end{figure}

The gain in precision from using the full \RunTwo sample is especially significant in the normalised double-differential cross-sections,
allowing the binning to have finer granularity than in the previous analysis. The obtained double-differential cross-section measurements as a function of $|y^{e\mu}|$ in bins of $m^{e\mu}$, $|\Delta \phi^{e\mu}|$ in bins of $m^{e\mu}$, $|\Delta \phi^{e\mu}|$ in bins of $\pT^{e\mu}$, and $|\Delta \phi^{e\mu}|$ in bins of $E^{e} + E^{\mu}$ are presented in Figures~\ref{fig:double_xs_norm_predictions1} and \ref{fig:double_xs_norm_predictions2} and in Tables~\ref{tab:norm_double_diff_xs_YvsM}--\ref{tab:norm_double_diff_xs_PhivsE} in the Appendix.
 
\begin{figure}[htbp]
\centering
\subfloat[]{\label{fig:abs_xs_sumpT} \includegraphics[width=0.49\linewidth]{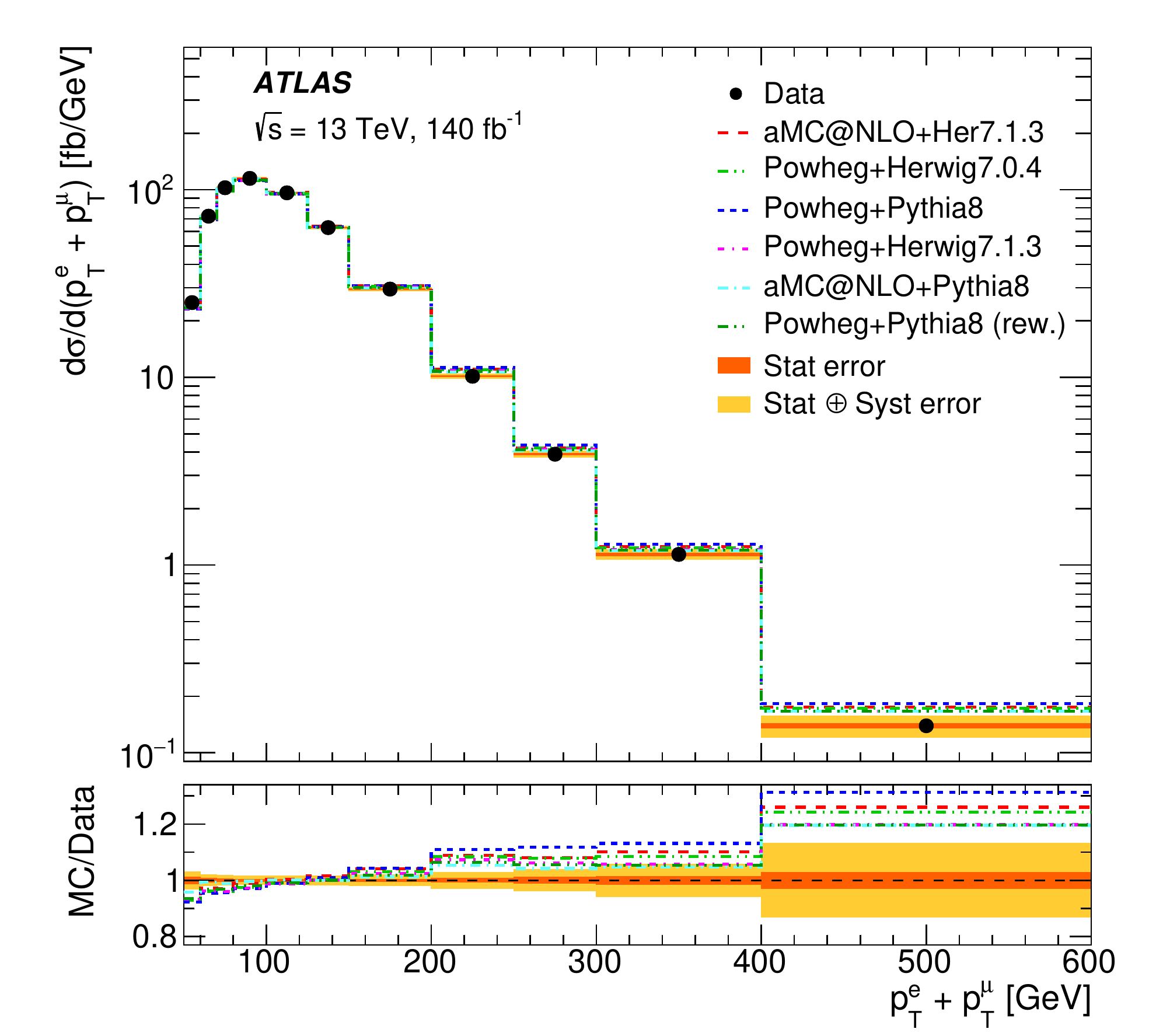}}
\subfloat[]{\label{fig:abs_xs_combinedpT} \includegraphics[width=0.49\linewidth]{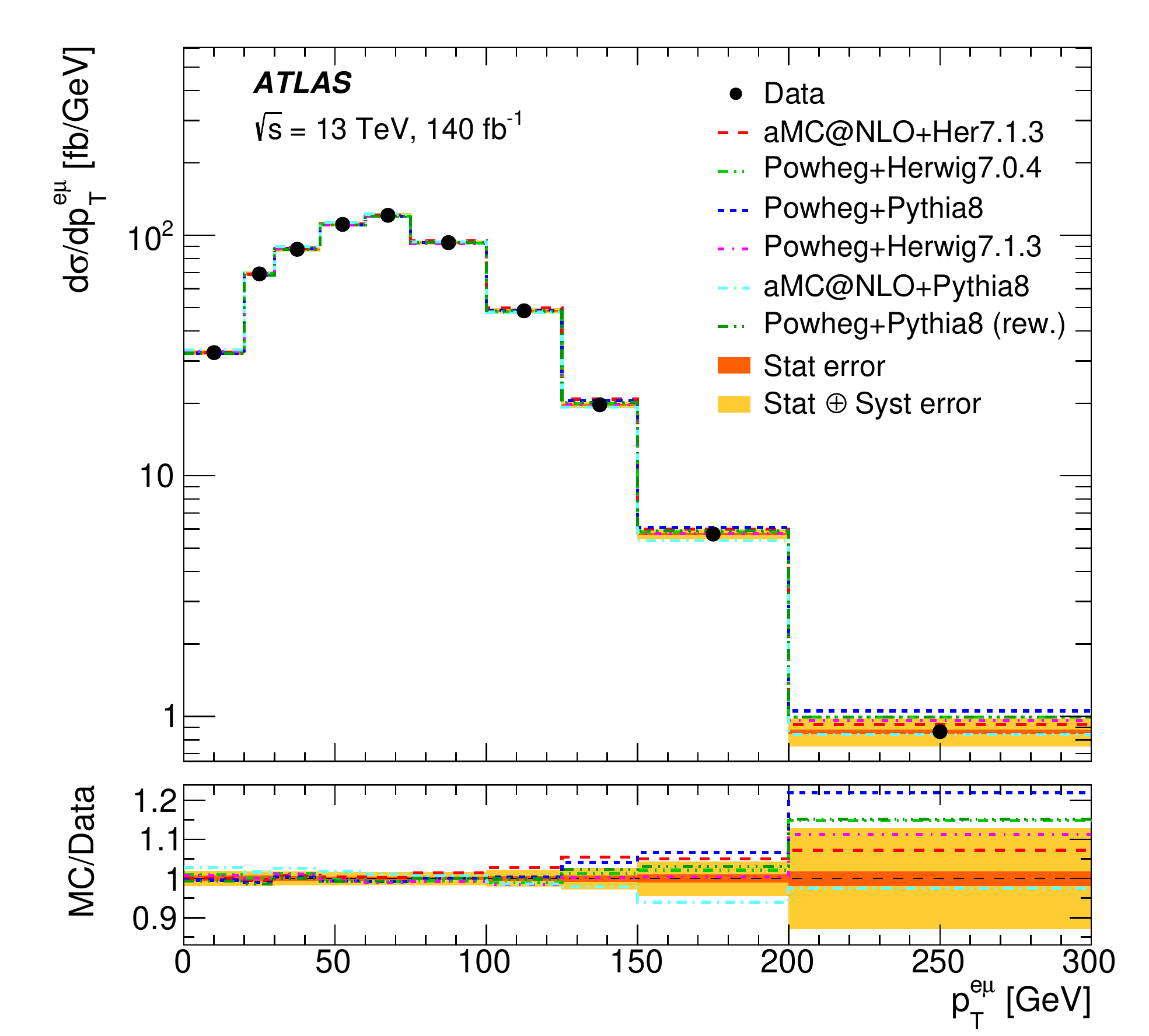}} \\
\subfloat[]{\label{fig:abs_xs_phi} \includegraphics[width=0.49\linewidth]{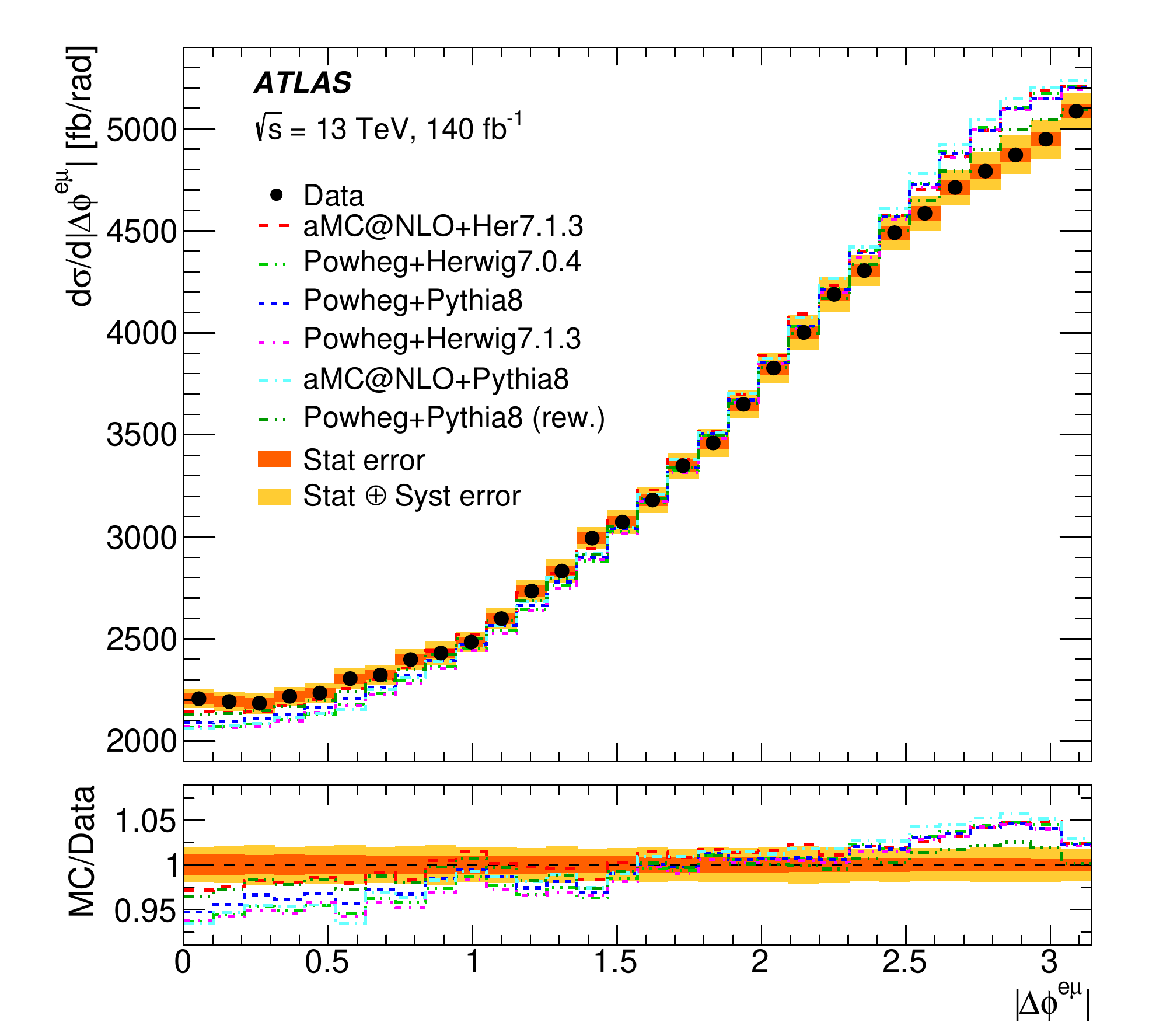}}
\subfloat[]{\label{fig:abs_xs_y} \includegraphics[width=0.49\linewidth]{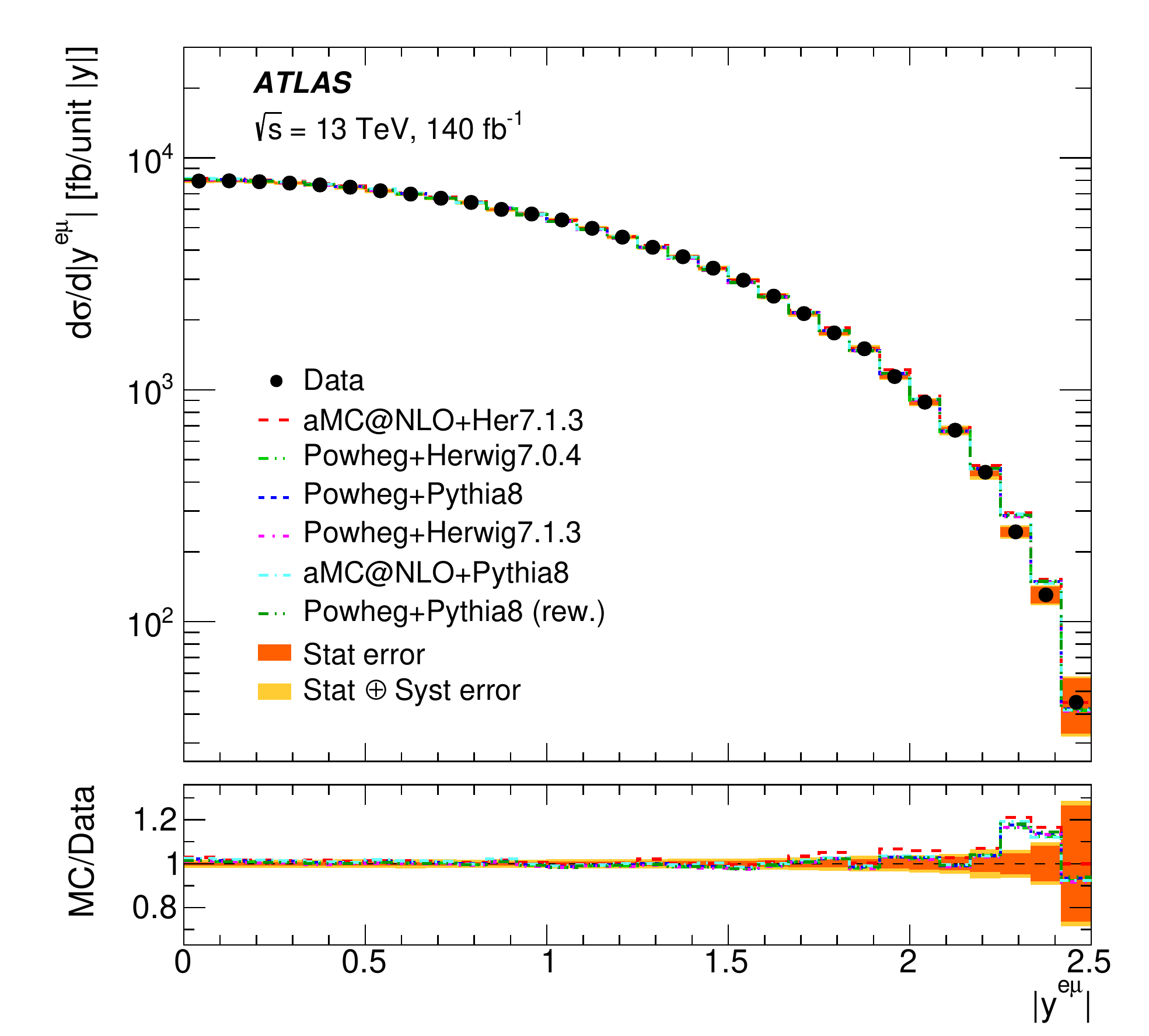}} \\
\caption[Absolute differential cross-sections with statistical and systematic uncertainties]{Absolute differential cross-sections as a function of (a) $\pT^{e}+\pT^{\mu}$, (b) $\pT^{e\mu}$, (c) $|\Delta \phi^{e\mu}|$ and (d) $|y^{e\mu}|$ with statistical (orange) and statistical plus systematic uncertainties (yellow). The data points are placed at the centre of each bin. The results are compared with the predictions from different Monte Carlo generators normalised to the \TOPpp prediction: the baseline \POWPY[8.230] \ttbar sample (blue), \MGNLOHER[7.1.3] (red), \POWHER[7.0.4] (green), \POWHER[7.1.3] (purple), \MGNLOPY[8.230] (cyan) and \POWPY[8.230] rew.\ (dark green), which refers to \POWPY[8.230] reweighted according to the top-quark \pT. The lower panel shows the ratios of the predictions to data, with the bands indicating the statistical and systematic uncertainties. The last bin in (a) and (b) also contains overflow events.}
\label{fig:xs_abs_predictions2}
\end{figure}
 
\FloatBarrier

\subsection{Comparison with predictions}
 
Defining a vector  $V_b$ as the difference between the measured and predicted values for an
absolute differential cross-section in $b$ kinematic bins, the compatibility of the data and predictions is assessed by using a $\chi^2$
test with $b$ degrees of freedom
\begin{equation}
\label{eq:chi2}
\chi^2 = V_{b}^{T} \cdot  \textrm{Cov}^{-1}_{b\times b} \cdot V_{b}
\end{equation}
The covariance matrix $\textrm{Cov}_{b\times b}$ is a sum of several terms. The statistical covariance matrices for both the data and the simulated event sample are calculated
with the method described in Section~\ref{sec:validation}. Each systematic uncertainty contributes another term,
where the uncertainties are assumed to be fully correlated between bins, except for the statistical uncertainties of the misidentified-lepton estimate.
No uncertainty was assigned to the theoretical prediction.
 
The normalised differential cross-sections have one less degree of freedom and the $\chi^2$ is calculated by simply dropping one bin in Eq.~(\ref{eq:chi2})
\begin{equation*}
\label{eq:chi2_norm}
\chi^2 = V_{(b-1)}^{T} \cdot \textrm{Cov}^{-1}_{(b-1) \times (b-1)} \cdot V_{(b-1)}
\end{equation*}
The statistical covariance matrix terms are constructed using the same method as for the absolute differential cross-sections, and the systematic covariance
matrix contributions are propagated to the normalised differential cross-sections.
The resulting combined $\chi^2$ for each variable and each \ttbar generator set-up are shown in Table~\ref{tab:chi2_compact}, while those for each double-differential distribution are shown in Table~\ref{tab:chi2_compact_double}.
The results show that no generator describes all distributions with a $\chi^2$ probability larger than 1\% .
However, some interesting features stand out and were also observed in the earlier ATLAS results at $\sqrt{s} = 13$~\TeV~\cite{TOPQ-2018-17}:
\begin{itemize}
\item All generators, except for the \MGNLOPY[8.230] sample, predict a spectrum for the variables $\pT^{\ell}$, $E^{e}+E^{\mu}$ and $\pT^{e}+\pT^{\mu}$ which is harder than in the data.
Among the various combinations of matrix element and shower
generators, \POWPY[8.230] gives the poorest agreement with these distributions, while \MGNLOPY[8.230] provides acceptable matches to the measured normalised distributions, in particular for the $E^{e}+E^{\mu}$ and $\pT^{e}+\pT^{\mu}$ differential cross-sections.
\item The $m^{e\mu}$ distribution is well represented by \MGNLOHER[7.1.3], while the $\pT^{e\mu}$ distribution is better represented
by \POWPY[8.230], especially the sample with the top-quark transverse momentum reweighted.
\item All generators fail to reproduce the $|\eta^{\ell}|$ distribution because of a 2\% deficit for $|\eta^{\ell}|>1.5$. The  $|y^{e\mu}|$ distribution is best represented by \POWPY[8.230] with the PDF4LHC15\_nnlo\_mc set, while other generators predict a surplus of events for $|y^{e\mu}|>2.2$, especially at high $m^{e\mu}$.
\item All generators predict a different trend than seen in data for the distribution of $|\Delta \phi^{e\mu}|$. The data tend to be higher than the predictions at low $|\Delta \phi^{e\mu}|$ , whereas they are generally lower in the high $|\Delta \phi^{e\mu}|$ region. This trend was observed in various previous measurements~\cite{CMS-TOP-17-014,TOPQ-2016-10,TOPQ-2018-17}.
\end{itemize}
 
\FloatBarrier

\begin{table}[htbp]
\centering
\caption{$\chi^2$ values for the comparison of the normalised measured differential cross-sections with different \ttbar simulation samples. $N_{\textrm{dof}}$ is the number of degrees of freedom. The $\chi^2$ values are displayed to one decimal place if the corresponding $\chi^2$ probability is greater than 1\%, and rounded to integers otherwise.}
\label{tab:chi2_compact}
\begin{tabular}{lcccccccc}
\toprule
Generator               &   $\pT^{\ell}$ &   $|\eta^{\ell}|$ &   $\pT^{e\mu}$ &   $\pT^{e} + \pT^{\mu}$ &   $E^e + E^{\mu}$ &   $m^{e\mu}$ &   $|\Delta \phi^{e\mu}|$ &   $|y^{e\mu}|$ \\
$N_{\textrm{dof}}$                    &     9   &       23 &       9    &           10    &         14    &     20    &          29 &     29    \\
\midrule
\POWPY[8]                 			&   196  			& 	132	& 12.0 			& 130			& 33    			& 102    			& 193 			& 47\phantom{.0} \\
\POWPY[8] - top \pT rew.			& \phantom{0}51  	&      114 	& \phantom{0}7.8	& \phantom{0}42    	& \phantom{0.}20.4 	& \phantom{0}53    	& \phantom{0}65 	& 45.2  \\
\POWPY[8] - $\hdamp \times 2$	&   228   			&      139 	& 26\phantom{.0}    	& 167    			& 38    			& \phantom{0}97   	& 121 			& 45.3  \\
\POWPY[8] - PDF4LHC       		&   186   			&      100 	& 11.5 			& 125   			& 32    			& \phantom{0}93    	& 185 			& 33.6 \\
\POWPY[8] - ISR up       			&   149   			&      111 	& 17.3  			& 120 			& 34    			& \phantom{0}79    	& \phantom{0}66 	& 50\phantom{.0} \\
\POWPY[8] - ISR down      		&   216   			&      159 	& 10.6 			& 131    			& 30    			& 113    			& 311 			& 44.5  \\
\POWPY[8] - Rad up        		&   164   			&      115 	& 27\phantom{.0}    	& 139    			& 38    			& \phantom{0}78    	& \phantom{0}49	& 47.6  \\
\POWPY[8] - Rad down      		&   216   			&      159 	& 10.6 			& 131    			& 30    			& 113    			& 311 			& 44.5  \\
\POWPY[8] - FSR up        		&   216   			&      132 	& 12.5 			& 143    			& 35    			& 106    			& 194 			& 46.8  \\
\POWPY[8] - FSR down      		&   171   			&      139 	& \phantom{0}9.5 	& 118    			& 30    			& \phantom{0}98    	& 185 			& 49\phantom{.0} \\
\POWPY[8] - MEC off       		& \phantom{0}42   	&      136 	& 41\phantom{.0}    	& \phantom{0}37    	& \phantom{0.}16.5	& \phantom{0}83    	& 181 			& 42.7  \\
$\AMCatNLO{+}\PYTHIA[8]$            	& \phantom{0.0}16.5 &      126 	& 48\phantom{.0}    	& \phantom{0.0}14.4	& \phantom{0.}14.3	& \phantom{0}89    	& 300 			& 50\phantom{.0} \\
$\AMCatNLO{+}\HERWIG[7.0.4]$       & \phantom{0}98   	&      137 	& 24\phantom{.0}    	& \phantom{0}74    	& \phantom{0.}24.1	& \phantom{0.0}29.1	& 110 			& 54\phantom{.0} \\
\POWHER[7.0.4]            			&   113   			&      104 	& 28\phantom{.0}    	& \phantom{0}82    	& 28    			& 135    			& 271 			& 45.8  \\
\POWHER[7.1.3]            			&   101   			&      107 	& 31\phantom{.0}    	& \phantom{0}75    	& \phantom{0.}25.5	& 138    			& 259 			& 45.5  \\
\bottomrule
\end{tabular}
\end{table}

\begin{table}[htbp]
\centering
\caption{$\chi^2$ values for the comparison of the normalised measured double-differential cross-sections with different \ttbar simulation samples. $N_{\textrm{dof}}$ is the number of degrees of freedom. The $\chi^2$ values are displayed to one decimal place if the corresponding $\chi^2$ probability is greater than 1\%, and rounded to integers otherwise.}
\label{tab:chi2_compact_double}
\begin{tabular}{lcccc}
\toprule
Generator               &   $|y^{e\mu}|:m^{e\mu}$ &   $|\Delta \phi^{e\mu}|:m^{e\mu}$ &   $|\Delta \phi^{e\mu}|:\pT^{e\mu}$ &   $|\Delta \phi^{e\mu}|:E^{e}+E^{\mu}$ \\
$N_{\textrm{dof}}$                     	&            39 		&              39 &                 24 	&                39 \\
\midrule
\POWPY[8]                 			&           131 		&             364 &                264 	&               263 \\
\POWPY[8] - top \pT rew.  		& \phantom{0}82 	&             140 & \phantom{0}81 	& \phantom{0}96 \\
\POWPY[8] - $\hdamp \times 2$	&           129 		&             250 &                182 	&               183 \\
\POWPY[8] - PDF4LHC       		&           114 		&             351 &                252 	&               253 \\
\POWPY[8] - ISR up        			&           108 		&             153 &                105 	&               112 \\
\POWPY[8] - ISR down      		&           143 		&             562 &                413 	&               409 \\
\POWPY[8] - Rad up        		&           109 		&             130 & \phantom{0}90 	&               104 \\
\POWPY[8] - Rad down      		&           143 		&             562 &                413 	&               409 \\
\POWPY[8] - FSR up        		&           137 		&             374 &                271 	&               268 \\
\POWPY[8] - FSR down      		&           122 		&             349 &                247 	&               255 \\
\POWPY[8] - MEC off       		&           107 		&             276 &                219 	&               237 \\
$\AMCatNLO{+}\PYTHIA[8]$            &           108 		&             436 &                363 	&               386 \\
$\AMCatNLO{+}\HERWIG[7.0.4]$	& \phantom{0}95 	&             270 &                154 	&               162 \\
\POWHER[7.0.4]           			&           151 		&             400 &                334 	&               345 \\
\POWHER[7.1.3]            			&           147 		&             392 &                318 	&               336 \\
\bottomrule
\end{tabular}
\end{table}

\FloatBarrier

\section{Conclusion}
\label{sec:conclusion}
 
The production of \ttbar pairs in \pp collisions at $\sqrt{s}$ = 13 \TeV\, is measured using opposite-sign
$e\mu$ events in association with one or two $b$-tagged jets in the
LHC \RunTwo (2015--2018) data collected by the
ATLAS experiment with an integrated luminosity of $140 \, \si{\per\fb}$.

The inclusive fiducial and total cross-sections for \ttbar production are measured.
The total inclusive cross-section is measured to be
\begin{equation*}
\sigma_{\ttbar} = 829 \pm 1\; (\textrm{stat}) \pm 13\; (\textrm{syst})  \pm 8\; (\textrm{lumi}) \pm 2\; (\textrm{beam})\ \textrm{pb},
\end{equation*}
where the uncertainties are due to statistics, theoretical and experimental systematic effects, the integrated luminosity and the beam energy.
This result is in excellent agreement with theoretical expectations.
It is the most precise measurement of the inclusive \ttbar cross-section to date.

The \ttbar absolute and normalised differential cross-sections are measured as functions of
eight different variables ($\pT^{e\mu}$, $\pT^{e} + \pT^{\mu}$, $\pT^{\ell}$, $E^{e} + E^{\mu}$, $m^{e\mu}$, $|\eta^{\ell}|$, $|\Delta \phi^{e\mu}|$
and $|y^{e\mu}|$). Furthermore, four double-differential cross-sections are measured as a function of
$|y^{e\mu}|$ in bins of $m^{e\mu}$, and as a function of $|\Delta \phi^{e\mu}|$ in bins of $m^{e\mu}$, $\pT^{e\mu}$ and $E^{e}+E^{\mu}$.
These distributions are confined to the fiducial region $\pT^{\ell}>27\ (25)$~\GeV \ for the leading (sub-leading) lepton and $|\eta^\ell |<2.5$
for both leptons.
 
The precision of the measurements is typically 2\% for the absolute differential cross-sections and at the 1\% level
for the normalised differential cross-sections, except in the highest energy bins where the \ttbar/$Wt$ interference uncertainty contribution increases.
The measurements are compared with a wide range of models for \ttbar production in \pp collisions. No model can describe all measured distributions within their uncertainties.
 
\section*{Acknowledgements}


We thank CERN for the very successful operation of the LHC, as well as the
support staff from our institutions without whom ATLAS could not be
operated efficiently.
 
We acknowledge the support of
ANPCyT, Argentina;
YerPhI, Armenia;
ARC, Australia;
BMWFW and FWF, Austria;
ANAS, Azerbaijan;
CNPq and FAPESP, Brazil;
NSERC, NRC and CFI, Canada;
CERN;
ANID, Chile;
CAS, MOST and NSFC, China;
Minciencias, Colombia;
MEYS CR, Czech Republic;
DNRF and DNSRC, Denmark;
IN2P3-CNRS and CEA-DRF/IRFU, France;
SRNSFG, Georgia;
BMBF, HGF and MPG, Germany;
GSRI, Greece;
RGC and Hong Kong SAR, China;
ISF and Benoziyo Center, Israel;
INFN, Italy;
MEXT and JSPS, Japan;
CNRST, Morocco;
NWO, Netherlands;
RCN, Norway;
MEiN, Poland;
FCT, Portugal;
MNE/IFA, Romania;
MESTD, Serbia;
MSSR, Slovakia;
ARRS and MIZ\v{S}, Slovenia;
DSI/NRF, South Africa;
MICINN, Spain;
SRC and Wallenberg Foundation, Sweden;
SERI, SNSF and Cantons of Bern and Geneva, Switzerland;
MOST, Taiwan;
TENMAK, T\"urkiye;
STFC, United Kingdom;
DOE and NSF, United States of America.
In addition, individual groups and members have received support from
BCKDF, CANARIE, Compute Canada and CRC, Canada;
PRIMUS 21/SCI/017 and UNCE SCI/013, Czech Republic;
COST, ERC, ERDF, Horizon 2020 and Marie Sk{\l}odowska-Curie Actions, European Union;
Investissements d'Avenir Labex, Investissements d'Avenir Idex and ANR, France;
DFG and AvH Foundation, Germany;
Herakleitos, Thales and Aristeia programmes co-financed by EU-ESF and the Greek NSRF, Greece;
BSF-NSF and MINERVA, Israel;
Norwegian Financial Mechanism 2014-2021, Norway;
NCN and NAWA, Poland;
La Caixa Banking Foundation, CERCA Programme Generalitat de Catalunya and PROMETEO and GenT Programmes Generalitat Valenciana, Spain;
G\"{o}ran Gustafssons Stiftelse, Sweden;
The Royal Society and Leverhulme Trust, United Kingdom.
 
The crucial computing support from all WLCG partners is acknowledged gratefully, in particular from CERN, the ATLAS Tier-1 facilities at TRIUMF (Canada), NDGF (Denmark, Norway, Sweden), CC-IN2P3 (France), KIT/GridKA (Germany), INFN-CNAF (Italy), NL-T1 (Netherlands), PIC (Spain), ASGC (Taiwan), RAL (UK) and BNL (USA), the Tier-2 facilities worldwide and large non-WLCG resource providers. Major contributors of computing resources are listed in Ref.~\cite{ATL-SOFT-PUB-2021-003}.


\clearpage
\appendix
\part*{Appendix}

\clearpage

\begin{table}[htbp]
\footnotesize
\caption[Absolute differential cross-section for $\pT^{\ell}$]{Absolute differential cross-section for $\pT^{\ell}$.}
\label{tab:abs_diff_xs_pTell}
\centering
\begin{tabular}{l|c|ccccccc|c}
\toprule
$\pT^{\ell}$ bins & $\dif\sigma/\dif\pT^{\ell}$ & Data & MC & $t\bar{t}$ & Lep. & Jets/ & Bkg. & Lumi + & Total  \\
$[\textrm{GeV}]$ &  [fb/GeV] & stat.\ [\%] & stat.\ [\%] & mod.\ [\%] & [\%] & $b$-tag.\ [\%] & [\%] & $E_{\textrm{beam}}$ [\%] & unc. [\%]\\
\midrule
\phantom{0}25.0--30.0 & 461.6 & 0.38 & 0.11 & 0.48 & 1.09 & 0.11 & 0.71 & 0.93 & 1.72  \\
\phantom{0}30.0--40.0 & 433.1 & 0.27 & 0.08 & 0.56 & 0.87 & 0.11 & 0.67 & 0.93 & 1.57  \\
\phantom{0}40.0--50.0 & 357.0 & 0.30 & 0.07 & 0.66 & 0.74 & 0.11 & 0.66 & 0.92 & 1.54  \\
\phantom{0}50.0--60.0 & 277.2 & 0.33 & 0.07 & 0.53 & 0.74 & 0.12 & 0.68 & 0.92 & 1.51  \\
\phantom{0}60.0--75.0 & 197.2 & 0.29 & 0.07 & 0.59 & 0.74 & 0.13 & 0.67 & 0.92 & 1.52  \\
\phantom{0}75.0--100.0 & 109.2 & 0.30 & 0.07 & 0.69 & 0.82 & 0.15 & 0.85 & 0.93 & 1.69  \\
100.0--140.0 & \phantom{00}41.79 & 0.40 & 0.09 & 0.65 & 0.96 & 0.17 & 1.25 & 0.94 & 2.00  \\
140.0--180.0 & \phantom{00}12.85 & 0.69 & 0.17 & 0.71 & 1.42 & 0.20 & 2.22 & 0.97 & 2.99  \\
180.0--250.0 & \phantom{000}3.16 & 1.02 & 0.28 & 1.04 & 1.86 & 0.26 & 4.25 & 1.01 & 4.98  \\
250.0--350.0 & \phantom{000}0.51 & 2.35 & 0.66 & 2.54 & 4.24 & 0.41 & 12.97\phantom{0} & 1.15 & 14.15\phantom{0}  \\
\bottomrule
 
\end{tabular}
\end{table}

\begin{table}[htbp]
\footnotesize
\caption[Absolute differential cross-section for $|\eta^{\ell}|$]{Absolute differential cross-section for $|\eta^{\ell}|$.}
\label{tab:abs_diff_xs_etaell}
\centering
\begin{tabular}{l|c|ccccccc|c}
\toprule
$|\eta^{\ell}|$ bins & $\dif\sigma/\dif|\eta^{\ell}|$ & Data & MC & $t\bar{t}$ & Lep. & Jets/ & Bkg. & Lumi + & Total  \\
- &  [pb/units of $\eta$] & stat.\ [\%] & stat.\ [\%] & mod.\ [\%] & [\%] & $b$-tag.\ [\%] & [\%] & $E_{\textrm{beam}}$ [\%] & unc.\ [\%]\\
\midrule
0.00--0.09 & 12.79 & 0.36 & 0.12 & 0.67 & 1.01 & 0.12 & 0.72 & 0.93 & 1.74  \\
0.09--0.18 & 12.81 & 0.32 & 0.12 & 0.49 & 0.81 & 0.13 & 0.77 & 0.93 & 1.58  \\
0.18--0.27 & 12.72 & 0.32 & 0.11 & 0.62 & 0.76 & 0.13 & 0.74 & 0.93 & 1.58  \\
0.27--0.36 & 12.57 & 0.30 & 0.12 & 0.81 & 0.76 & 0.13 & 0.76 & 0.93 & 1.67  \\
0.36--0.45 & 12.36 & 0.30 & 0.12 & 0.60 & 0.76 & 0.13 & 0.75 & 0.93 & 1.57  \\
0.45--0.54 & 12.19 & 0.32 & 0.11 & 0.84 & 0.76 & 0.14 & 0.68 & 0.93 & 1.65  \\
\bottomrule
 
\end{tabular}
\end{table}

\begin{table}[htbp]
\footnotesize
\caption[Absolute differential cross-section for $E^{e} + E^{\mu}$]{Absolute differential cross-section for $E^{e} + E^{\mu}$.}
\label{tab:abs_diff_xs_sumE}
\centering
\begin{tabular}{l|c|ccccccc|c}
\toprule
$E^{e} + E^{\mu}$ bins & $\dif\sigma/\dif(E^{e} + E^{\mu})$ & Data & MC & $t\bar{t}$ & Lep. & Jets/ & Bkg. & Lumi + & Total  \\
$[\textrm{GeV}]$ &  [fb/GeV] & stat.\ [\%] & stat.\ [\%] & mod.\ [\%] & [\%] & $b$-tag.\ [\%] & [\%] & $E_{\textrm{beam}}$ [\%] & unc.\ [\%]\\
\midrule
\phantom{0}50.0--60.0 & \phantom{0}1.38 & 5.20 & 3.78 & 4.20 & 1.38 & 1.08 & 5.61 & 0.88 & 9.71  \\
\phantom{0}60.0--70.0 & \phantom{0}9.04 & 1.93 & 0.89 & 1.82 & 1.10 & 0.38 & 1.03 & 0.94 & 3.33  \\
\phantom{0}70.0--80.0 & 20.47 & 1.16 & 0.38 & 2.43 & 0.96 & 0.14 & 0.97 & 0.92 & 3.18  \\
\phantom{0}80.0--90.0 & 30.69 & 0.95 & 0.29 & 0.79 & 0.88 & 0.17 & 0.65 & 0.92 & 1.92  \\
\phantom{0}90.0--110.0 & 43.64 & 0.52 & 0.14 & 0.74 & 0.80 & 0.13 & 0.65 & 0.92 & 1.66  \\
110.0--125.0 & 51.23 & 0.56 & 0.13 & 0.55 & 0.75 & 0.12 & 0.59 & 0.92 & 1.55  \\
125.0--160.0 & 51.53 & 0.37 & 0.09 & 0.46 & 0.73 & 0.12 & 0.59 & 0.92 & 1.45  \\
160.0--200.0 & 43.20 & 0.37 & 0.09 & 0.67 & 0.74 & 0.13 & 0.66 & 0.92 & 1.56  \\
200.0--250.0 & 31.56 & 0.38 & 0.09 & 0.65 & 0.78 & 0.11 & 0.76 & 0.93 & 1.62  \\
250.0--300.0 & 20.93 & 0.48 & 0.11 & 0.56 & 0.84 & 0.15 & 1.02 & 0.93 & 1.79  \\
300.0--370.0 & 12.74 & 0.54 & 0.13 & 0.79 & 0.93 & 0.13 & 1.25 & 0.94 & 2.06  \\
370.0--450.0 & \phantom{0}6.92 & 0.70 & 0.16 & 0.91 & 1.07 & 0.15 & 1.61 & 0.95 & 2.45  \\
450.0--550.0 & \phantom{0}3.45 & 0.97 & 0.22 & 0.70 & 1.29 & 0.17 & 1.92 & 0.96 & 2.79  \\
550.0--700.0 & \phantom{0}1.42 & 1.19 & 0.29 & 1.39 & 1.61 & 0.16 & 2.37 & 0.96 & 3.55  \\
700.0--900.0 & \phantom{0}0.62 & 1.78 & 0.44 & 2.25 & 2.48 & 0.27 & 3.77 & 1.00 & 5.46  \\
\bottomrule
 
\end{tabular}
\end{table}

\begin{table}[htbp]
\footnotesize
\caption[Absolute differential cross-section for $m^{e\mu}$]{Absolute differential cross-section for $m^{e\mu}$.}
\label{tab:abs_diff_xs_combinedM}
\centering
\begin{tabular}{l|c|ccccccc|c}
\toprule
$m^{e\mu}$ bins & $\dif\sigma/\dif m^{e\mu}$ & Data & MC & $t\bar{t}$ & Lep. & Jets/ & Bkg. & Lumi + & Total  \\
$[\textrm{GeV}]$ &  [fb/GeV] & stat.\ [\%] & stat.\ [\%] & mod.\ [\%] & [\%] & $b$-tag.\ [\%] & [\%] & $E_{\textrm{beam}}$ [\%] & unc.\ [\%]\\
\midrule
\phantom{00}0.0--15.0 & \phantom{0}8.21\phantom{0} & 2.08 & 0.47 & 2.38 & 0.90 & 0.14 & 1.31 & 0.93 & 3.69  \\
\phantom{0}15.0--20.0 & 18.81\phantom{0} & 1.71 & 0.36 & 0.59 & 0.85 & 0.12 & 0.76 & 0.92 & 2.36  \\
\phantom{0}20.0--25.0 & 23.48\phantom{0} & 1.54 & 0.33 & 1.19 & 0.84 & 0.12 & 0.73 & 0.92 & 2.45  \\
\phantom{0}25.0--30.0 & 29.02\phantom{0} & 1.33 & 0.30 & 1.72 & 0.84 & 0.09 & 1.07 & 0.92 & 2.75  \\
\phantom{0}30.0--35.0 & 33.04\phantom{0} & 1.24 & 0.29 & 1.17 & 0.85 & 0.09 & 0.79 & 0.92 & 2.28  \\
\phantom{0}35.0--40.0 & 38.24\phantom{0} & 1.19 & 0.26 & 0.37 & 0.84 & 0.09 & 0.74 & 0.93 & 1.93  \\
\bottomrule
 
\end{tabular}
\end{table}

\begin{table}[htbp]
\footnotesize
\caption[Absolute differential cross-section for $\pT^{e} + \pT^{\mu}$]{Absolute differential cross-section for $\pT^{e} + \pT^{\mu}$.}
\label{tab:abs_diff_xs_sumpT}
\centering
\begin{tabular}{l|c|ccccccc|c}
\toprule
$\pT^{e} + \pT^{\mu}$ bins & $\dif\sigma/\dif(\pT^{e} + \pT^{\mu})$ & Data & MC & $t\bar{t}$ & Lep. & Jets/ & Bkg. & Lumi + & Total  \\
$[\textrm{GeV}]$ &  [fb/GeV] & stat.\ [\%] & stat.\ [\%] & mod.\ [\%] & [\%] & $b$-tag.\ [\%] & [\%] & $E_{\textrm{beam}}$ [\%] & unc.\ [\%]\\
\midrule
\phantom{0}50.0--60.0 & 25.03 & 1.25 & 0.49 & 2.10 & 1.36 & 0.19 & 0.74 & 0.93 & 3.09  \\
\phantom{0}60.0--70.0 & 72.26 & 0.66 & 0.22 & 0.70 & 1.05 & 0.16 & 0.85 & 0.93 & 1.92  \\
\phantom{0}70.0--80.0 & 102.4\phantom{00} & 0.52 & 0.14 & 0.70 & 0.88 & 0.14 & 0.65 & 0.92 & 1.69  \\
\phantom{0}80.0--100.0 & 114.9\phantom{00} & 0.33 & 0.08 & 0.59 & 0.79 & 0.11 & 0.65 & 0.92 & 1.53  \\
100.0--125.0 & 96.23 & 0.33 & 0.07 & 0.55 & 0.74 & 0.12 & 0.63 & 0.92 & 1.49  \\
125.0--150.0 & 62.80 & 0.39 & 0.09 & 0.66 & 0.76 & 0.13 & 0.74 & 0.93 & 1.61  \\
150.0--200.0 & 29.54 & 0.41 & 0.10 & 0.79 & 0.87 & 0.16 & 1.11 & 0.94 & 1.92  \\
200.0--250.0 & 10.14 & 0.65 & 0.17 & 0.92 & 1.19 & 0.20 & 2.18 & 0.96 & 2.91  \\
250.0--300.0 & \phantom{0}3.90 & 1.09 & 0.26 & 1.09 & 1.52 & 0.22 & 2.97 & 0.98 & 3.82  \\
300.0--400.0 & \phantom{0}1.14 & 1.43 & 0.36 & 2.37 & 2.06 & 0.26 & 4.45 & 1.00 & 5.73  \\
400.0--600.0 & \phantom{0}0.14 & 2.84 & 0.82 & 2.70 & 3.73 & 0.44 & 11.87\phantom{0} & 1.11 & 13.13\phantom{0}  \\
\bottomrule
 
\end{tabular}
\end{table}

\begin{table}[htbp]
\footnotesize
\caption[Absolute differential cross-section for $\pT^{e\mu}$]{Absolute differential cross-section for $\pT^{e\mu}$.}
\label{tab:abs_diff_xs_combinedpT}
\centering
\begin{tabular}{l|c|ccccccc|c}
\toprule
$\pT^{e\mu}$ bins & $\dif\sigma/\dif\pT^{e\mu}$ & Data & MC & $t\bar{t}$ & Lep. & Jets/ & Bkg. & Lumi + & Total  \\
$[\textrm{GeV}]$ &  [fb/GeV] & stat.\ [\%] & stat.\ [\%] & mod.\ [\%] & [\%] & $b$-tag.\ [\%] & [\%] & $E_{\textrm{beam}}$ [\%] & unc.\ [\%]\\
\midrule
\phantom{00}0.0--20.0 & 32.51 & 0.66 & 0.21 & 0.87 & 0.79 & 0.19 & 0.67 & 0.92 & 1.79  \\
\phantom{0}20.0--30.0 & 69.08 & 0.64 & 0.17 & 0.87 & 0.78 & 0.22 & 0.68 & 0.92 & 1.78  \\
\phantom{0}30.0--45.0 & 87.41 & 0.46 & 0.11 & 0.90 & 0.78 & 0.18 & 0.66 & 0.92 & 1.72  \\
\phantom{0}45.0--60.0 & 111.0\phantom{00} & 0.41 & 0.09 & 0.71 & 0.79 & 0.15 & 0.64 & 0.92 & 1.61  \\
\phantom{0}60.0--75.0 & 121.0\phantom{00} & 0.38 & 0.09 & 0.66 & 0.79 & 0.14 & 0.63 & 0.92 & 1.57  \\
\phantom{0}75.0--100.0 & 93.20 & 0.33 & 0.07 & 0.70 & 0.78 & 0.12 & 0.66 & 0.92 & 1.59  \\
100.0--125.0 & 48.51 & 0.45 & 0.10 & 0.97 & 0.89 & 0.15 & 0.99 & 0.93 & 1.96  \\
125.0--150.0 & 19.74 & 0.70 & 0.17 & 0.38 & 1.20 & 0.21 & 1.98 & 0.95 & 2.64  \\
150.0--200.0 & \phantom{0}5.73 & 0.95 & 0.23 & 0.55 & 1.62 & 0.28 & 3.65 & 1.00 & 4.28  \\
200.0--300.0 & \phantom{0}0.86 & 1.78 & 0.49 & 3.39 & 2.64 & 0.35 & 11.88\phantom{0} & 1.14 & 12.82\phantom{0}  \\
\bottomrule
 
\end{tabular}
\end{table}

\begin{table}[htbp]
\footnotesize
\caption[Absolute differential cross-section for $\Delta \phi^{e\mu}$]{Absolute differential cross-section for $|\Delta \phi^{e\mu}|$.}
\label{tab:abs_diff_xs_phi}
\centering
\begin{tabular}{l|c|ccccccc|c}
\toprule
$|\Delta\phi^{e\mu}|$ bins & $\dif\sigma/\dif|\Delta\phi^{e\mu}|$ & Data & MC & $t\bar{t}$ & Lep. & Jets/ & Bkg. & Lumi + & Total  \\
$[\textrm{rad}]$ &  [pb/rad] & stat.\ [\%] & stat.\ [\%] & mod.\ [\%] & [\%] & $b$-tag.\ [\%] & [\%] & $E_{\textrm{beam}}$ [\%] & unc.\ [\%]\\
\midrule
0.00--0.10 & 2.21 & 1.12 & 0.24 & 0.46 & 0.84 & 0.11 & 0.84 & 0.92 & 1.95  \\
0.10--0.21 & 2.19 & 1.12 & 0.24 & 0.73 & 0.84 & 0.10 & 0.83 & 0.92 & 2.03  \\
0.21--0.31 & 2.18 & 1.05 & 0.24 & 1.26 & 0.84 & 0.10 & 0.79 & 0.92 & 2.22  \\
0.31--0.42 & 2.22 & 1.08 & 0.23 & 0.33 & 0.85 & 0.08 & 1.02 & 0.93 & 1.99  \\
0.42--0.52 & 2.23 & 1.03 & 0.23 & 0.93 & 0.84 & 0.10 & 0.78 & 0.92 & 2.04  \\
0.52--0.63 & 2.31 & 1.02 & 0.22 & 1.19 & 0.84 & 0.09 & 0.77 & 0.92 & 2.16  \\
\bottomrule

\end{tabular}
\end{table}

\begin{table}[htbp]
\footnotesize
\caption[Absolute differential cross-section for $|y^{e\mu}|$]{Absolute differential cross-section for $|y^{e\mu}|$.}
\label{tab:abs_diff_xs_y}
\centering
\begin{tabular}{l|c|ccccccc|c}
\toprule
$|y^{e\mu}|$ bins & $\dif\sigma/\dif|y^{e\mu}|$ & Data & MC & $t\bar{t}$ & Lep. & Jets/ & Bkg. & Lumi + & Total  \\
- &  [pb/units of y] & stat.\ [\%] & stat.\ [\%] & mod.\ [\%] & [\%] & $b$-tag.\ [\%] & [\%] & $E_{\textrm{beam}}$ [\%] & unc.\ [\%]\\
\midrule
0.00--0.08 & 7.94\phantom{0} & 0.59 & 0.15 & 0.58 & 0.77 & 0.13 & 0.75 & 0.92 & 1.65  \\
0.08--0.17 & 7.96\phantom{0} & 0.57 & 0.14 & 0.70 & 0.78 & 0.13 & 0.75 & 0.92 & 1.69  \\
0.17--0.25 & 7.89\phantom{0} & 0.61 & 0.15 & 0.43 & 0.77 & 0.13 & 0.69 & 0.93 & 1.59  \\
0.25--0.33 & 7.78\phantom{0} & 0.56 & 0.14 & 0.51 & 0.78 & 0.14 & 0.82 & 0.93 & 1.65  \\
0.33--0.42 & 7.64\phantom{0} & 0.60 & 0.14 & 0.65 & 0.77 & 0.12 & 0.80 & 0.93 & 1.71  \\
0.42--0.50 & 7.47\phantom{0} & 0.61 & 0.15 & 0.81 & 0.77 & 0.12 & 0.89 & 0.93 & 1.82  \\
\bottomrule
 
\end{tabular}
\end{table}
 
\FloatBarrier

\begin{table}[htbp]
\tiny
\caption[Double-differential cross-section for $|y^{e\mu}|:m^{e\mu}$]{Double-differential cross-section for $|y^{e\mu}|:m^{e\mu}$. The cross-section measured in the last region includes events with an invariant mass greater than 800~\GeV.}
\label{tab:double_diff_xs_YvsM}
\centering
\begin{tabular}{cc|c|ccccccc|c}
\toprule
\multicolumn{2}{l|}{$|y^{e\mu}| \times m^{e\mu}$ bins} & $\dif^{2}\sigma/\dif|y^{e\mu}|\dif m^{e\mu}$ & Data & MC & $t\bar{t}$ & Lep. & Jets/ & Bkg. & Lumi + & Total  \\
\multicolumn{2}{l|}{-} &  [fb/GeV] & stat.\ [\%] & stat.\ [\%] & mod.\ [\%] & [\%] & $b$-tag.\ [\%] & [\%] & $E_{\textrm{beam}}$ [\%] & unc.\ [\%]\\
\midrule
& 0.00--0.31 & 24.27 & 0.68 & 0.20 & 0.62 & 0.84 & 0.11 & 0.61 & 0.92 & 1.68  \\
& 0.31--0.62 & 23.06 & 0.68 & 0.18 & 0.47 & 0.80 & 0.09 & 0.64 & 0.92 & 1.62  \\
& 0.62--0.94 & 20.66 & 0.74 & 0.20 & 0.38 & 0.78 & 0.12 & 0.73 & 0.93 & 1.66  \\
$0.0 \leq m^{e\mu}$ & 0.94--1.25 & 17.66 & 0.85 & 0.22 & 0.62 & 0.81 & 0.09 & 0.87 & 0.93 & 1.85  \\
$< 70.0$ GeV & 1.25--1.56 & 14.01 & 1.07 & 0.29 & 0.65 & 0.93 & 0.10 & 0.81 & 0.93 & 2.01  \\
& 1.56--1.88 & \phantom{0}9.74 & 1.34 & 0.39 & 1.86 & 1.12 & 0.12 & 1.43 & 0.96 & 3.10  \\
\bottomrule
\end{tabular}
\end{table}

\begin{table}[htbp]
\tiny
\caption[Double-differential cross-section for $|\Delta \phi^{e\mu}|:m^{e\mu}$]{Double-differential cross-section for $|\Delta \phi^{e\mu}|:m^{e\mu}$. The cross-section measured in the last region includes events with an invariant mass greater than 800~\GeV.}
\label{tab:double_diff_xs_PhivsM}
\centering
\begin{tabular}{cc|c|ccccccc|c}
\toprule
\multicolumn{2}{l|}{$|\Delta\phi^{e\mu}| \times m^{e\mu}$ bins} & $\dif^{2}\sigma/\dif|\Delta\phi^{e\mu}|\dif m^{e\mu}$ & Data & MC & $t\bar{t}$ & Lep. & Jets/ & Bkg. & Lumi + & Total  \\
\multicolumn{2}{l|}{$[\textrm{rad}]$} &  [fb/rad GeV] & stat.\ [\%] & stat.\ [\%] & mod.\ [\%] & [\%] & $b$-tag.\ [\%] & [\%] & $E_{\textrm{beam}}$ [\%] & unc.\ [\%]\\
\midrule
& 0.00--0.39 & 21.80 & 0.69 & 0.15 & 1.11 & 0.83 & 0.09 & 0.74 & 0.92 & 1.96  \\
& 0.39--0.79 & 21.29 & 0.66 & 0.15 & 0.66 & 0.82 & 0.09 & 0.73 & 0.93 & 1.72  \\
& 0.79--1.18 & 19.05 & 0.71 & 0.16 & 0.67 & 0.82 & 0.10 & 0.76 & 0.93 & 1.76  \\
$0.0 \leq m^{e\mu}$ & 1.18--1.57 & 13.86 & 0.82 & 0.20 & 0.72 & 0.85 & 0.12 & 0.75 & 0.93 & 1.84  \\
$< 70.0$ GeV & 1.57--1.96 & \phantom{0}7.88 & 1.13 & 0.39 & 0.72 & 0.92 & 0.14 & 0.79 & 0.93 & 2.08  \\
& 1.96--2.36 & \phantom{0}4.13 & 1.74 & 0.82 & 1.49 & 1.04 & 0.30 & 1.04 & 0.96 & 3.01  \\
\bottomrule
\end{tabular}
\end{table}

\begin{table}[htbp]
\tiny
\caption[Double-differential cross-section for $|\Delta \phi^{e\mu}|:\pT^{e\mu}$]{Double-differential cross-section for $|\Delta \phi^{e\mu}|:\pT^{e\mu}$. The cross-section measured in the last region includes events with a \pT greater than 100~\GeV.}
\label{tab:double_diff_xs_PhivspTll}
\centering
\begin{tabular}{cc|c|ccccccc|c}
\toprule
\multicolumn{2}{l|}{$|\Delta\Phi^{e\mu}| \times \pT^{e\mu}$ bins} & $\dif^{2}\sigma/\dif|\Delta\phi^{e\mu}|\dif\pT^{e\mu}$ & Data & MC & $t\bar{t}$ & Lep. & Jets/ & Bkg. & Lumi + & Total  \\
\multicolumn{2}{l|}{$[\textrm{rad}]$} &  [fb/rad GeV] & stat. [\%] & stat.\ [\%] & mod.\ [\%] & [\%] & $b$-tag.\ [\%] & [\%] & $E_{\textrm{beam}}$ [\%] & unc.\ [\%]\\
\midrule
& 0.00--1.65 & \phantom{0}0.077 & 8.69 & 3.11 & 8.43 & 2.01 & 0.29 & 6.00 & 0.95 & 14.04\phantom{0}  \\
$0.0 \leq \pT^{e\mu}$ & 1.65--2.02 & \phantom{0}4.89\phantom{0} & 2.11 & 0.79 & 2.49 & 1.15 & 0.21 & 1.46 & 0.93 & 3.95  \\
$< 40.0$ GeV & 2.02--2.40 & 18.63\phantom{0} & 1.05 & 0.37 & 1.44 & 0.91 & 0.18 & 0.85 & 0.93 & 2.40  \\
& 2.40--2.77 & 46.83\phantom{0} & 0.63 & 0.18 & 0.52 & 0.78 & 0.19 & 0.67 & 0.92 & 1.63  \\
& 2.77--3.14 & 75.62\phantom{0} & 0.49 & 0.13 & 0.78 & 0.77 & 0.21 & 0.66 & 0.92 & 1.66  \\
\midrule
& 0.00--0.31 & \phantom{0}7.94\phantom{0} & 2.39 & 0.53 & 2.22 & 1.18 & 0.26 & 0.97 & 0.91 & 3.76  \\
\bottomrule
\end{tabular}
\end{table}

\begin{table}[htbp]
\tiny
\caption[Double-differential cross-section for $|\Delta \phi^{e\mu}|:E^{e} + E^{\mu}$]{Double-differential cross-section for $|\Delta \phi^{e\mu}|:E^{e} + E^{\mu}$. The cross-section measured in the last region includes events with a sum of the two energies greater than 900~\GeV.}
\label{tab:double_diff_xs_PhivsE}
\centering
\begin{tabular}{cc|c|ccccccc|c}
\toprule
\multicolumn{2}{l|}{$|\Delta\phi^{e\mu}| \times (E^{e} + E^{\mu})$ bins} & $\dif^{2}\sigma/\dif|\Delta\phi^{e\mu}|\dif(E^{e} + E^{\mu})$ & Data & MC & $t\bar{t}$ & Lep. & Jets/ & Bkg. & Lumi + & Total  \\
\multicolumn{2}{l|}{$[\textrm{rad}]$} &  [fb/rad GeV] & stat.\ [\%] & stat.\ [\%] & mod.\ [\%] & [\%] & $b$-tag.\ [\%] & [\%] & $E_{\textrm{beam}}$ [\%] & unc.\ [\%]\\
\midrule
& 0.00--0.39 & \phantom{0}4.35 & 1.21 & 0.27 & 0.90 & 0.85 & 0.12 & 0.69 & 0.91 & 2.09  \\
& 0.39--0.79 & \phantom{0}4.40 & 1.13 & 0.24 & 1.74 & 0.85 & 0.11 & 0.78 & 0.91 & 2.56  \\
& 0.79--1.18 & \phantom{0}4.38 & 1.08 & 0.26 & 0.76 & 0.85 & 0.13 & 0.80 & 0.91 & 2.01  \\
$0.0 \leq E^{e} + E^{\mu}$ & 1.18--1.57 & \phantom{0}4.48 & 1.13 & 0.28 & 0.47 & 0.85 & 0.12 & 0.61 & 0.92 & 1.88  \\
$< 110.0$ GeV & 1.57--1.96 & \phantom{0}4.44 & 1.12 & 0.32 & 0.91 & 0.85 & 0.15 & 0.65 & 0.92 & 2.05  \\
& 1.96--2.36 & \phantom{0}4.26 & 1.15 & 0.44 & 1.40 & 0.86 & 0.23 & 0.62 & 0.93 & 2.35  \\
\bottomrule
\end{tabular}
\end{table}
 
\FloatBarrier

\begin{table}[htbp]
\footnotesize
\caption[Normalised differential cross-section for $\pT^{\ell}$]{Normalised differential cross-section for $\pT^{\ell}$.}
\label{tab:norm_diff_xs_pTell}
\centering
\begin{tabular}{l|c|ccccccc|c}
\toprule
$\pT^{\ell}$ bins & $1/\sigma$ $\dif\sigma/\dif\pT^{\ell}$ & Data & MC & $t\bar{t}$ & Lep. & Jets/ & Bkg. & Lumi + & Total  \\
$[\textrm{GeV}]$ &  $\times 10^{-3}$ [1/GeV] & stat.\ [\%] & stat.\ [\%] & mod.\ [\%] & [\%] & $b$-tag.\ [\%] & [\%] & $E_{\textrm{beam}}$ [\%] & unc.\ [\%]\\
\midrule
\phantom{0}25.0--30.0 & 21.85\phantom{00} & 0.34 & 0.10 & 0.24 & 0.55 & 0.07 & 0.24 & 0.01 & 0.74  \\
\phantom{0}30.0--40.0 & 20.50\phantom{00} & 0.23 & 0.06 & 0.23 & 0.21 & 0.04 & 0.30 & 0.01 & 0.49  \\
\phantom{0}40.0--50.0 & 16.90\phantom{00} & 0.25 & 0.06 & 0.37 & 0.19 & 0.02 & 0.27 & 0.01 & 0.56  \\
\phantom{0}50.0--60.0 & 13.12\phantom{00} & 0.28 & 0.07 & 0.13 & 0.23 & 0.02 & 0.20 & 0.01 & 0.44  \\
\phantom{0}60.0--75.0 & 9.33\phantom{0} & 0.26 & 0.06 & 0.46 & 0.25 & 0.02 & 0.20 & 0.01 & 0.62  \\
\phantom{0}75.0--100.0 & 5.17\phantom{0} & 0.26 & 0.07 & 0.34 & 0.20 & 0.04 & 0.15 & 0.00 & 0.51  \\
100.0--140.0 & 1.98\phantom{0} & 0.37 & 0.09 & 0.31 & 0.43 & 0.08 & 0.64 & 0.02 & 0.92  \\
140.0--180.0 & 0.61\phantom{0} & 0.68 & 0.17 & 0.57 & 0.92 & 0.13 & 1.67 & 0.05 & 2.11  \\
180.0--250.0 & 0.15\phantom{0} & 1.01 & 0.28 & 0.86 & 1.42 & 0.20 & 3.74 & 0.09 & 4.23  \\
250.0--350.0 & 0.024 & 2.34 & 0.66 & 2.57 & 3.90 & 0.35 & 12.48\phantom{0} & 0.24 & 13.55\phantom{0}  \\
\bottomrule
\end{tabular}
\end{table}

\begin{table}[htbp]
\footnotesize
\caption[Normalised cross-section for $|\eta^{\ell}|$]{Normalised differential cross-section for $|\eta^{\ell}|$.}
\label{tab:norm_diff_xs_etaell}
\centering
\begin{tabular}{l|c|ccccccc|c}
\toprule
$|\eta^{\ell}|$ bins & $1/\sigma$ $\dif\sigma/\dif|\eta^{\ell}|$ & Data & MC & $t\bar{t}$ & Lep. & Jets/ & Bkg. & Lumi + & Total  \\
- &  $\times 10^{-2}$ [1/units of $\eta$] & stat.\ [\%] & stat.\ [\%] & mod.\ [\%] & [\%] & $b$-tag.\ [\%] & [\%] & $E_{\textrm{beam}}$ [\%] & unc.\ [\%]\\
\midrule
0.00--0.09 & 60.72 & 0.78 & 0.12 & 0.24 & 0.40 & 0.02 & 0.17 & 0.01 & 0.93  \\
0.09--0.18 & 60.81 & 0.69 & 0.11 & 0.09 & 0.11 & 0.02 & 0.14 & 0.01 & 0.73  \\
0.18--0.27 & 60.38 & 0.68 & 0.11 & 0.24 & 0.08 & 0.03 & 0.12 & 0.02 & 0.75  \\
0.27--0.36 & 59.70 & 0.65 & 0.11 & 0.46 & 0.08 & 0.03 & 0.13 & 0.01 & 0.82  \\
0.36--0.45 & 58.67 & 0.68 & 0.11 & 0.49 & 0.08 & 0.02 & 0.19 & 0.01 & 0.87  \\
0.45--0.54 & 57.91 & 0.69 & 0.11 & 0.49 & 0.08 & 0.02 & 0.22 & 0.01 & 0.88  \\
\bottomrule
 
\end{tabular}
\end{table}

\begin{table}[htbp]
\footnotesize
\caption[Normalised differential cross-section for $E^{e} + E^{\mu}$]{Normalised differential cross-section for $E^{e} + E^{\mu}$.}
\label{tab:norm_diff_xs_sumE}
\centering
\begin{tabular}{l|c|ccccccc|c}
\toprule
$E^{e} + E^{\mu}$ bins & $1/\sigma$ $\dif\sigma/\dif(E^{e} + E^{\mu})$ & Data & MC & $t\bar{t}$ & Lep. & Jets/ & Bkg. & Lumi + & Total  \\
$[\textrm{GeV}]$ &  $\times 10^{-3}$ [1/GeV] & stat.\ [\%] & stat.\ [\%] & mod.\ [\%] & [\%] & $b$-tag.\ [\%] & [\%] & $E_{\textrm{beam}}$ [\%] & unc.\ [\%]\\
\midrule
\phantom{0}50.0--60.0 & 0.13\phantom{0} & 5.19 & 3.77 & 3.95 & 0.92 & 1.07 & 5.68 & 0.06 & 9.54  \\
\phantom{0}60.0--70.0 & 0.86\phantom{0} & 1.92 & 0.88 & 1.84 & 0.52 & 0.40 & 1.02 & 0.03 & 3.05  \\
\phantom{0}70.0--80.0 & 1.94\phantom{0} & 1.15 & 0.37 & 2.25 & 0.42 & 0.13 & 1.14 & 0.03 & 2.83  \\
\phantom{0}80.0--90.0 & 2.91\phantom{0} & 0.93 & 0.28 & 0.45 & 0.35 & 0.11 & 0.63 & 0.03 & 1.29  \\
\phantom{0}90.0--110.0 & 4.14\phantom{0} & 0.50 & 0.13 & 0.54 & 0.31 & 0.07 & 0.69 & 0.03 & 1.07  \\
110.0--125.0 & 4.86\phantom{0} & 0.54 & 0.12 & 0.30 & 0.26 & 0.03 & 0.58 & 0.03 & 0.90  \\
125.0--160.0 & 4.89\phantom{0} & 0.33 & 0.08 & 0.26 & 0.22 & 0.02 & 0.39 & 0.02 & 0.62  \\
160.0--200.0 & 4.10\phantom{0} & 0.35 & 0.08 & 0.36 & 0.17 & 0.02 & 0.24 & 0.01 & 0.59  \\
200.0--250.0 & 2.99\phantom{0} & 0.35 & 0.08 & 0.30 & 0.13 & 0.02 & 0.06 & 0.00 & 0.49  \\
250.0--300.0 & 1.99\phantom{0} & 0.46 & 0.11 & 0.47 & 0.17 & 0.05 & 0.35 & 0.01 & 0.77  \\
300.0--370.0 & 1.21\phantom{0} & 0.52 & 0.13 & 0.45 & 0.27 & 0.04 & 0.60 & 0.03 & 0.96  \\
370.0--450.0 & 0.66\phantom{0} & 0.68 & 0.16 & 0.74 & 0.44 & 0.05 & 0.99 & 0.04 & 1.49  \\
450.0--550.0 & 0.33\phantom{0} & 0.95 & 0.22 & 0.43 & 0.73 & 0.08 & 1.33 & 0.06 & 1.85  \\
550.0--700.0 & 0.13\phantom{0} & 1.18 & 0.28 & 1.49 & 1.11 & 0.08 & 1.79 & 0.08 & 2.85  \\
700.0--900.0 & 0.058 & 1.75 & 0.43 & 1.97 & 2.06 & 0.21 & 3.18 & 0.14 & 4.65  \\
\bottomrule
 
\end{tabular}
\end{table}

\begin{table}[htbp]
\footnotesize
\caption[Normalised differential cross-section for $m^{e\mu}$]{Normalised differential cross-section for $m^{e\mu}$.}
\label{tab:norm_diff_xs_combinedM}
\centering
\begin{tabular}{l|c|ccccccc|c}
\toprule
$m^{e\mu}$ bins & $1/\sigma$ $\dif\sigma/\dif m^{e\mu}$ & Data & MC & $t\bar{t}$ & Lep. & Jets/ & Bkg. & Lumi + & Total  \\
$[\textrm{GeV}]$ &  $\times 10^{-3}$ [1/GeV] & stat.\ [\%] & stat.\ [\%] & mod.\ [\%] & [\%] & $b$-tag.\ [\%] & [\%] & $E_{\textrm{beam}}$ [\%] & unc.\ [\%]\\
\midrule
\phantom{00}0.0--15.0 & 0.78\phantom{0} & 2.06 & 0.47 & 2.18 & 0.32 & 0.16 & 1.21 & 0.01 & 3.29  \\
\phantom{0}15.0--20.0 & 1.78\phantom{0} & 1.70 & 0.36 & 0.45 & 0.18 & 0.15 & 0.60 & 0.01 & 1.91  \\
\phantom{0}20.0--25.0 & 2.22\phantom{0} & 1.54 & 0.32 & 1.10 & 0.19 & 0.14 & 0.43 & 0.02 & 1.98  \\
\phantom{0}25.0--30.0 & 2.75\phantom{0} & 1.32 & 0.30 & 1.52 & 0.16 & 0.10 & 1.15 & 0.02 & 2.35  \\
\phantom{0}30.0--35.0 & 3.13\phantom{0} & 1.23 & 0.28 & 1.20 & 0.17 & 0.07 & 0.53 & 0.01 & 1.83  \\
\phantom{0}35.0--40.0 & 3.62\phantom{0} & 1.18 & 0.25 & 0.28 & 0.17 & 0.08 & 0.49 & 0.01 & 1.35  \\
\bottomrule
 
\end{tabular}
\end{table}

\begin{table}[htbp]
\footnotesize
\caption[Normalised differential cross-section for $\pT^{e} + \pT^{\mu}$]{Normalised differential cross-section for $\pT^{e} + \pT^{\mu}$.}
\label{tab:norm_diff_xs_sumpT}
\centering
\begin{tabular}{l|c|ccccccc|c}
\toprule
$\pT^{e} + \pT^{\mu}$ bins & $1/\sigma$ $\dif\sigma/\dif(\pT^{e} + \pT^{\mu})$ & Data & MC & $t\bar{t}$ & Lep. & Jets/ & Bkg. & Lumi + & Total  \\
$[\textrm{GeV}]$ &  $\times 10^{-3}$ [1/GeV] & stat.\ [\%] & stat.\ [\%] & mod.\ [\%] & [\%] & $b$-tag.\ [\%] & [\%] & $E_{\textrm{beam}}$ [\%] & unc.\ [\%]\\
\midrule
\phantom{0}50.0--60.0 & 2.36\phantom{0} & 1.23 & 0.48 & 1.76 & 0.85 & 0.17 & 0.55 & 0.01 & 2.42  \\
\phantom{0}60.0--70.0 & 6.82\phantom{0} & 0.62 & 0.21 & 0.44 & 0.43 & 0.12 & 0.78 & 0.01 & 1.20  \\
\phantom{0}70.0--80.0 & 9.67\phantom{0} & 0.49 & 0.13 & 0.22 & 0.25 & 0.09 & 0.44 & 0.01 & 0.76  \\
\phantom{0}80.0--100.0 & 10.85\phantom{00} & 0.29 & 0.08 & 0.31 & 0.25 & 0.05 & 0.46 & 0.01 & 0.68  \\
100.0--125.0 & 9.09\phantom{0} & 0.29 & 0.06 & 0.20 & 0.22 & 0.01 & 0.32 & 0.01 & 0.53  \\
125.0--150.0 & 5.93\phantom{0} & 0.36 & 0.08 & 0.50 & 0.17 & 0.04 & 0.04 & 0.00 & 0.64  \\
150.0--200.0 & 2.79\phantom{0} & 0.37 & 0.09 & 0.34 & 0.32 & 0.08 & 0.53 & 0.01 & 0.81  \\
200.0--250.0 & 0.96\phantom{0} & 0.63 & 0.17 & 1.11 & 0.72 & 0.13 & 1.69 & 0.04 & 2.25  \\
250.0--300.0 & 0.37\phantom{0} & 1.09 & 0.26 & 0.76 & 1.07 & 0.14 & 2.49 & 0.07 & 3.03  \\
300.0--400.0 & 0.11\phantom{0} & 1.42 & 0.36 & 2.08 & 1.64 & 0.17 & 4.00 & 0.09 & 5.02  \\
400.0--600.0 & 0.013 & 2.84 & 0.83 & 2.97 & 3.38 & 0.37 & 11.44\phantom{0} & 0.20 & 12.65\phantom{0}  \\
\bottomrule
 
\end{tabular}
\end{table}

\begin{table}[htbp]
\footnotesize
\caption[Normalised differential cross-section for $\pT^{e\mu}$]{Normalised differential cross-section for $\pT^{e\mu}$.}
\label{tab:norm_diff_xs_combinedpT}
\centering
\begin{tabular}{l|c|ccccccc|c}
\toprule
$\pT^{e\mu}$ bins & $1/\sigma$ $\dif\sigma/\dif\pT^{e\mu}$ & Data & MC & $t\bar{t}$ & Lep. & Jets/ & Bkg. & Lumi + & Total  \\
$[\textrm{GeV}]$ &  $\times 10^{-3}$ [1/GeV] & stat.\ [\%] & stat.\ [\%] & mod.\ [\%] & [\%] & $b$-tag.\ [\%] & [\%] & $E_{\textrm{beam}}$ [\%] & unc.\ [\%]\\
\midrule
\phantom{00}0.0--20.0 & 3.08\phantom{0} & 0.64 & 0.20 & 0.47 & 0.29 & 0.12 & 0.47 & 0.01 & 0.99  \\
\phantom{0}20.0--30.0 & 6.55\phantom{0} & 0.61 & 0.16 & 0.77 & 0.27 & 0.17 & 0.42 & 0.01 & 1.13  \\
\phantom{0}30.0--45.0 & 8.29\phantom{0} & 0.42 & 0.10 & 0.47 & 0.25 & 0.09 & 0.47 & 0.01 & 0.84  \\
\phantom{0}45.0--60.0 & 10.53\phantom{00} & 0.37 & 0.09 & 0.55 & 0.17 & 0.06 & 0.42 & 0.01 & 0.81  \\
\phantom{0}60.0--75.0 & 11.48\phantom{00} & 0.35 & 0.08 & 0.23 & 0.13 & 0.03 & 0.40 & 0.01 & 0.60  \\
\phantom{0}75.0--100.0 & 8.84\phantom{0} & 0.29 & 0.07 & 0.24 & 0.05 & 0.04 & 0.19 & 0.00 & 0.43  \\
100.0--125.0 & 4.60\phantom{0} & 0.43 & 0.09 & 0.66 & 0.30 & 0.12 & 0.37 & 0.00 & 0.94  \\
125.0--150.0 & 1.87\phantom{0} & 0.67 & 0.16 & 0.53 & 0.73 & 0.21 & 1.46 & 0.03 & 1.86  \\
150.0--200.0 & 0.54\phantom{0} & 0.93 & 0.22 & 0.73 & 1.20 & 0.28 & 3.13 & 0.08 & 3.57  \\
200.0--300.0 & 0.082 & 1.78 & 0.48 & 3.71 & 2.24 & 0.33 & 11.38\phantom{0} & 0.22 & 12.32\phantom{0}  \\
\bottomrule
 
\end{tabular}
\end{table}

\begin{table}[htbp]
\footnotesize
\caption[Normalised differential cross-section for $|\Delta \phi^{e\mu}|$]{Normalised differential cross-section for $|\Delta \phi^{e\mu}|$.}
\label{tab:norm_diff_xs_phi}
\centering
\begin{tabular}{l|c|ccccccc|c}
\toprule
$|\Delta\phi^{e\mu}|$ bins & $1/\sigma$ $\dif\sigma/\dif|\Delta\phi^{e\mu}|$ & Data & MC & $t\bar{t}$ & Lep. & Jets/ & Bkg. & Lumi + & Total  \\
$[\textrm{rad}]$ &  $\times 10^{-2}$ [1/rad] & stat.\ [\%] & stat.\ [\%] & mod.\ [\%] & [\%] & $b$-tag.\ [\%] & [\%] & $E_{\textrm{beam}}$ [\%] & unc.\ [\%]\\
\midrule
0.00--0.10 & 20.94 & 1.11 & 0.24 & 0.61 & 0.14 & 0.13 & 0.15 & 0.01 & 1.31  \\
0.10--0.21 & 20.81 & 1.11 & 0.24 & 0.59 & 0.13 & 0.11 & 0.26 & 0.01 & 1.32  \\
0.21--0.31 & 20.72 & 1.04 & 0.24 & 1.39 & 0.13 & 0.12 & 0.32 & 0.01 & 1.79  \\
0.31--0.42 & 21.04 & 1.07 & 0.22 & 0.25 & 0.14 & 0.10 & 0.35 & 0.01 & 1.19  \\
0.42--0.52 & 21.20 & 1.02 & 0.22 & 0.90 & 0.13 & 0.08 & 0.12 & 0.01 & 1.39  \\
0.52--0.63 & 21.87 & 1.01 & 0.22 & 1.19 & 0.12 & 0.09 & 0.22 & 0.01 & 1.60  \\
\bottomrule
 
\end{tabular}
\end{table}

\begin{table}[htbp]
\footnotesize
\caption[Normalised differential cross-section for $|y^{e\mu}|$]{Normalised differential cross-section for $|y^{e\mu}|$.}
\label{tab:norm_diff_xs_y}
\centering
\begin{tabular}{l|c|ccccccc|c}
\toprule
$|y^{e\mu}|$ bins & $1/\sigma$ $\dif\sigma/\dif|y^{e\mu}|$ & Data & MC & $t\bar{t}$ & Lep. & Jets/ & Bkg. & Lumi + & Total  \\
- &  $\times 10^{-3}$ [1/units of y] & stat.\ [\%] & stat.\ [\%] & mod.\ [\%] & [\%] & $b$-tag.\ [\%] & [\%] & $E_{\textrm{beam}}$ [\%] & unc.\ [\%]\\
\midrule
0.00--0.08 & 754.8 & 0.56 & 0.14 & 0.27 & 0.15 & 0.02 & 0.18 & 0.02 & 0.68  \\
0.08--0.17 & 756.2 & 0.55 & 0.14 & 0.33 & 0.15 & 0.03 & 0.18 & 0.02 & 0.69  \\
0.17--0.25 & 749.7 & 0.60 & 0.14 & 0.39 & 0.14 & 0.02 & 0.18 & 0.02 & 0.76  \\
0.25--0.33 & 739.6 & 0.55 & 0.13 & 0.14 & 0.14 & 0.03 & 0.26 & 0.02 & 0.65  \\
0.33--0.42 & 726.2 & 0.59 & 0.14 & 0.57 & 0.13 & 0.02 & 0.20 & 0.02 & 0.86  \\
0.42--0.50 & 710.1 & 0.59 & 0.14 & 0.65 & 0.12 & 0.03 & 0.25 & 0.01 & 0.93  \\
\bottomrule
 
\end{tabular}
\end{table}
 
\FloatBarrier

\begin{table}[htbp]
\tiny
\caption[Normalised double-differential cross-section for $|y^{e\mu}|:m^{e\mu}$]{Normalised double-differential cross-section for $|y^{e\mu}|:m^{e\mu}$. The cross-section measured in the last region includes events with an invariant mass greater than 800~\GeV.}
\label{tab:norm_double_diff_xs_YvsM}
\centering
\begin{tabular}{cc|c|ccccccc|c}
\toprule
\multicolumn{2}{l|}{$|y^{e\mu}| \times m^{e\mu}$ bins} & $1/\sigma$ $\dif^{2}\sigma/\dif|y^{e\mu}|dm^{e\mu}$ & Data & MC & $t\bar{t}$ & Lep. & Jets/ & Bkg. & Lumi + & Total  \\
\multicolumn{2}{l|}{-} &  $\times 10^{-3}$ [1/GeV] & stat.\ [\%] & stat.\ [\%] & mod.\ [\%] & [\%] & $b$-tag.\ [\%] & [\%] & $E_{\textrm{beam}}$ [\%] & unc.\ [\%]\\
\midrule
& 0.00--0.31 & 2.30\phantom{0} & 0.66 & 0.19 & 0.48 & 0.22 & 0.08 & 0.57 & 0.03 & 1.04  \\
& 0.31--0.62 & 2.19\phantom{0} & 0.66 & 0.17 & 0.27 & 0.21 & 0.07 & 0.52 & 0.03 & 0.93  \\
& 0.62--0.94 & 1.96\phantom{0} & 0.73 & 0.20 & 0.30 & 0.21 & 0.06 & 0.54 & 0.02 & 1.00  \\
$0.0 \leq m^{e\mu}$ & 0.94--1.25 & 1.67\phantom{0} & 0.82 & 0.21 & 0.58 & 0.19 & 0.08 & 0.79 & 0.01 & 1.31  \\
$< 70.0$ GeV & 1.25--1.56 & 1.33\phantom{0} & 1.05 & 0.29 & 0.36 & 0.29 & 0.07 & 0.54 & 0.01 & 1.31  \\
& 1.56--1.88 & 0.92\phantom{0} & 1.33 & 0.38 & 1.72 & 0.55 & 0.09 & 1.04 & 0.04 & 2.50  \\
\bottomrule
\end{tabular}
\end{table}

\begin{table}[htbp]
\tiny
\caption[Normalised double-differential cross-section for $|\Delta \phi^{e\mu}|:m^{e\mu}$]{Normalised double-differential cross-section for $|\Delta \phi^{e\mu}|:m^{e\mu}$. The cross-section measured in the last region includes events with an invariant mass greater than 800~\GeV.}
\label{tab:norm_double_diff_xs_PhivsM}
\centering
\begin{tabular}{cc|c|ccccccc|c}
\toprule
\multicolumn{2}{l|}{$|\Delta\phi^{e\mu}| \times m^{e\mu}$ bins} & $1/\sigma$ $\dif^{2}\sigma/\dif|\Delta\phi^{e\mu}|\dif m^{e\mu}$ & Data & MC & $t\bar{t}$ & Lep. & Jets/ & Bkg. & Lumi + & Total  \\
\multicolumn{2}{l|}{$[\textrm{rad}]$} &  $\times 10^{-3}$ [1/rad GeV] & stat.\ [\%] & stat.\ [\%] & mod.\ [\%] & [\%] & $b$-tag.\ [\%] & [\%] & $E_{\textrm{beam}}$ [\%] & unc.\ [\%]\\
\midrule
& 0.00--0.39 & 2.06\phantom{00} & 0.66 & 0.15 & 1.02 & 0.12 & 0.10 & 0.40 & 0.01 & 1.30  \\
& 0.39--0.79 & 2.02\phantom{00} & 0.64 & 0.14 & 0.54 & 0.12 & 0.08 & 0.47 & 0.01 & 0.98  \\
& 0.79--1.18 & 1.80\phantom{00} & 0.69 & 0.16 & 0.66 & 0.15 & 0.05 & 0.61 & 0.01 & 1.16  \\
$0.0 \leq m^{e\mu}$ & 1.18--1.57 & 1.31\phantom{00} & 0.81 & 0.20 & 0.63 & 0.28 & 0.05 & 0.68 & 0.01 & 1.28  \\
$< 70.0$ GeV & 1.57--1.96 & 0.75\phantom{00} & 1.12 & 0.39 & 0.50 & 0.33 & 0.09 & 0.63 & 0.01 & 1.47  \\
& 1.96--2.36 & 0.39\phantom{00} & 1.72 & 0.81 & 1.24 & 0.47 & 0.27 & 0.96 & 0.03 & 2.53  \\
\bottomrule
\end{tabular}
\end{table}

\begin{table}[htbp]
\tiny
\caption[Normalised double-differential cross-section for $|\Delta \phi^{e\mu}|:\pT^{e\mu}$]{Normalised double-differential cross-section for $|\Delta \phi^{e\mu}|:\pT^{e\mu}$. The cross-section measured in the last region includes events with a \pT greater than 100~\GeV.}
\label{tab:norm_double_diff_xs_PhivspTll}
\centering
\begin{tabular}{cc|c|ccccccc|c}
\toprule
\multicolumn{2}{l|}{$|\Delta\phi^{e\mu}| \times \pT^{e\mu}$ bins} & $1/\sigma$ $\dif^{2}\sigma/\dif|\Delta\phi^{e\mu}|\dif\pT^{e\mu}$ & Data & MC & $t\bar{t}$ & Lep. & Jets/ & Bkg. & Lumi + & Total  \\
\multicolumn{2}{l|}{$[\textrm{rad}]$} &  $\times 10^{-3}$ [1/rad GeV] & stat.\ [\%] & stat.\ [\%] & mod.\ [\%] & [\%] & $b$-tag.\ [\%] & [\%] & $E_{\textrm{beam}}$ [\%] & unc.\ [\%]\\
\midrule
& 0.00--1.65 & \phantom{0}0.0070 & 8.68 & 3.10 & 8.35 & 1.61 & 0.27 & 5.85 & 0.02 & 13.84\phantom{0}  \\
$0.0 \leq \pT^{e\mu}$ & 1.65--2.02 & 0.46\phantom{0} & 2.10 & 0.79 & 2.35 & 0.58 & 0.17 & 1.36 & 0.01 & 3.57  \\
$< 40.0$ GeV & 2.02--2.40 & 1.76\phantom{0} & 1.02 & 0.36 & 1.18 & 0.33 & 0.12 & 0.76 & 0.01 & 1.81  \\
& 2.40--2.77 & 4.43\phantom{0} & 0.60 & 0.17 & 0.28 & 0.32 & 0.12 & 0.48 & 0.01 & 0.90  \\
& 2.77--3.14 & 7.16\phantom{0} & 0.46 & 0.12 & 0.36 & 0.28 & 0.14 & 0.40 & 0.01 & 0.78  \\
\midrule
& 0.00--0.31 & 0.75\phantom{0} & 2.37 & 0.52 & 2.16 & 0.66 & 0.24 & 0.97 & 0.02 & 3.46  \\
\bottomrule
\end{tabular}
\end{table}

\begin{table}[htbp]
\tiny
\caption[Normalised double-differential  cross-section for $|\Delta \phi^{e\mu}|:E^{e} + E^{\mu}$]{Normalised double-differential cross-section for $|\Delta \phi^{e\mu}|:E^{e} + E^{\mu}$. The cross-section measured in the last region includes events with a sum of the two energies greater than 900~\GeV.}
\label{tab:norm_double_diff_xs_PhivsE}
\centering
\begin{tabular}{cc|c|ccccccc|c}
\toprule
\multicolumn{2}{l|}{$|\Delta\phi^{e\mu}| \times (E^{e} + E^{\mu})$ bins} & $1/\sigma$ $\dif^{2}\sigma/\dif|\Delta\phi^{e\mu}|\dif(E^{e} + E^{\mu})$ & Data & MC & $t\bar{t}$ & Lep. & Jets/ & Bkg. & Lumi + & Total  \\
\multicolumn{2}{l|}{$[\textrm{rad}]$} &  $\times 10^{-3}$ [1/rad GeV] & stat.\ [\%] & stat.\ [\%] & mod.\ [\%] & [\%] & $b$-tag.\ [\%] & [\%] & $E_{\textrm{beam}}$ [\%] & unc.\ [\%]\\
\midrule
& 0.00--0.39 & 0.41\phantom{0} & 1.19 & 0.27 & 0.72 & 0.30 & 0.11 & 0.78 & 0.04 & 1.65  \\
& 0.39--0.79 & 0.42\phantom{0} & 1.12 & 0.24 & 1.68 & 0.30 & 0.08 & 0.83 & 0.03 & 2.22  \\
& 0.79--1.18 & 0.42\phantom{0} & 1.07 & 0.26 & 0.49 & 0.30 & 0.10 & 0.94 & 0.03 & 1.56  \\
$0.0 \leq E^{e} + E^{\mu}$ & 1.18--1.57 & 0.42\phantom{0} & 1.11 & 0.27 & 0.40 & 0.33 & 0.07 & 0.62 & 0.03 & 1.40  \\
$< 110.0$ GeV & 1.57--1.96 & 0.42\phantom{0} & 1.11 & 0.31 & 0.76 & 0.34 & 0.07 & 0.67 & 0.03 & 1.57  \\
& 1.96--2.36 & 0.40\phantom{0} & 1.15 & 0.44 & 1.23 & 0.35 & 0.19 & 0.53 & 0.03 & 1.86  \\
\bottomrule
\end{tabular}
\end{table}
 
\FloatBarrier

\clearpage

\printbibliography

 
\begin{flushleft}
\hypersetup{urlcolor=black}
{\Large The ATLAS Collaboration}

\bigskip

\AtlasOrcid[0000-0002-6665-4934]{G.~Aad}$^\textrm{\scriptsize 102}$,
\AtlasOrcid[0000-0002-5888-2734]{B.~Abbott}$^\textrm{\scriptsize 120}$,
\AtlasOrcid[0000-0002-7248-3203]{D.C.~Abbott}$^\textrm{\scriptsize 103}$,
\AtlasOrcid[0000-0002-1002-1652]{K.~Abeling}$^\textrm{\scriptsize 55}$,
\AtlasOrcid[0000-0002-8496-9294]{S.H.~Abidi}$^\textrm{\scriptsize 29}$,
\AtlasOrcid[0000-0002-9987-2292]{A.~Aboulhorma}$^\textrm{\scriptsize 35e}$,
\AtlasOrcid[0000-0001-5329-6640]{H.~Abramowicz}$^\textrm{\scriptsize 151}$,
\AtlasOrcid[0000-0002-1599-2896]{H.~Abreu}$^\textrm{\scriptsize 150}$,
\AtlasOrcid[0000-0003-0403-3697]{Y.~Abulaiti}$^\textrm{\scriptsize 117}$,
\AtlasOrcid[0000-0003-0762-7204]{A.C.~Abusleme~Hoffman}$^\textrm{\scriptsize 137a}$
% ...
\bigskip
\\

% ...
$^{29}$Physics Department, Brookhaven National Laboratory, Upton NY; United States of America.\\
% ...
$^{35}$$^{(a)}$Facult\'e des Sciences Ain Chock, R\'eseau Universitaire de Physique des Hautes Energies - Universit\'e Hassan II, Casablanca;$^{(b)}$Facult\'{e} des Sciences, Universit\'{e} Ibn-Tofail, K\'{e}nitra;$^{(c)}$Facult\'e des Sciences Semlalia, Universit\'e Cadi Ayyad, LPHEA-Marrakech;$^{(d)}$LPMR, Facult\'e des Sciences, Universit\'e Mohamed Premier, Oujda;$^{(e)}$Facult\'e des sciences, Universit\'e Mohammed V, Rabat;$^{(f)}$Institute of Applied Physics, Mohammed VI Polytechnic University, Ben Guerir; Morocco.\\
% ...
$^{55}$II. Physikalisches Institut, Georg-August-Universit\"{a}t G\"ottingen, G\"ottingen; Germany.\\
% ...
$^{102}$CPPM, Aix-Marseille Universit\'e, CNRS/IN2P3, Marseille; France.\\
$^{103}$Department of Physics, University of Massachusetts, Amherst MA; United States of America.\\
% ...
$^{117}$Department of Physics, New York University, New York NY; United States of America.\\
% ...
$^{120}$Homer L. Dodge Department of Physics and Astronomy, University of Oklahoma, Norman OK; United States of America.\\
% ...
$^{137}$$^{(a)}$Departamento de F\'isica, Pontificia Universidad Cat\'olica de Chile, Santiago;$^{(b)}$Millennium Institute for Subatomic physics at high energy frontier (SAPHIR), Santiago;$^{(c)}$Instituto de Investigaci\'on Multidisciplinario en Ciencia y Tecnolog\'ia, y Departamento de F\'isica, Universidad de La Serena;$^{(d)}$Universidad Andres Bello, Department of Physics, Santiago;$^{(e)}$Instituto de Alta Investigaci\'on, Universidad de Tarapac\'a, Arica;$^{(f)}$Departamento de F\'isica, Universidad T\'ecnica Federico Santa Mar\'ia, Valpara\'iso; Chile.\\
% ...
$^{150}$Department of Physics, Technion, Israel Institute of Technology, Haifa; Israel.\\
$^{151}$Raymond and Beverly Sackler School of Physics and Astronomy, Tel Aviv University, Tel Aviv; Israel.\\
% ...
$^{a}$ Also Affiliated with an institute covered by a cooperation agreement with CERN.\\
$^{b}$ Also at An-Najah National University, Nablus; Palestine.\\
$^{c}$ Also at Borough of Manhattan Community College, City University of New York, New York NY; United States of America.\\
$^{d}$ Also at Bruno Kessler Foundation, Trento; Italy.\\
$^{e}$ Also at Center for High Energy Physics, Peking University; China.\\
$^{f}$ Also at Center for Interdisciplinary Research and Innovation (CIRI-AUTH), Thessaloniki ; Greece.\\
% ...

\end{flushleft}


\end{document}